\documentclass{emulateapj}

\newcommand{\vc}[1]{\textbf{\em #1}}

\newcommand{\pder}[2]{\frac{\partial #1}{\partial #2}}

\shorttitle{RECONNECTION OF WEAKLY STOCHASTIC B-FIELD}
\shortauthors{KOWAL ET AL.}

\begin{document}

\title{Numerical Tests of Fast Reconnection in Weakly Stochastic Magnetic Fields}

\author{Grzegorz Kowal}
\affil{Department of Astronomy, University of Wisconsin, 475 North Charter Street, Madison, WI 53706, USA}
\affil{Astronomical Observatory, Jagiellonian University, Orla 171, 30-244 Krak\'ow, POLAND}
\email{kowal@astro.wisc.edu}
\and
\author{A. Lazarian}
\affil{Department of Astronomy, University of Wisconsin, 475 North Charter Street, Madison, WI 53706, USA}
\email{lazarian@astro.wisc.edu}
\and
\author{E.~T. Vishniac}
\affil{Department of Physics and Astronomy, McMaster University, 1280 Main Street West, Hamilton, ON L8S 4M1, CANADA}
\email{ethan@mcmaster.ca}
\and
\author{K. Otmianowska-Mazur}
\affil{Astronomical Observatory, Jagiellonian University, Orla 171, 30-244 Krak\'ow, POLAND}
\email{otmian@astro.wisc.edu}

\begin{abstract}
We study the effects of turbulence on magnetic reconnection using
three-dimensional direct numerical simulations.  This is the first attempt to
test a model of fast magnetic reconnection in the presence of weak turbulence
proposed by Lazarian \& Vishniac (1999).  This model predicts that weak
turbulence, which is generically present in most of astrophysical systems,
enhances the rate of reconnection by reducing the transverse scale for
reconnection events and by allowing many independent flux reconnection events to
occur simultaneously.  As a result the reconnection speed becomes independent of
Ohmic resistivity and is determined by the magnetic field wandering induced by
turbulence.  We test the dependence of the reconnection speed on turbulent
power, the energy injection scale and resistivity.  We study the reconnection
model with the open and experiment with the outflow boundary conditions and
discuss the advantages and drawbacks of various setups.  To test our results, we
also perform simulations of turbulence with the same outflow boundaries but
without a large scale field reversal, thus without large scale reconnection.  To
quantify the reconnection speed we use both an intuitive definition, i.e. the
speed of the reconnected flux inflow, as well as a more sophisticated definition
based on a formally derived analytical expression.  Our results confirm the
predictions of the Lazarian \& Vishniac model.  In particular, we find that the
reconnection speed is proportional to the square root of the injected power, as
predicted by the model.  The dependence on the injection scale for some of our
models is a bit weaker than expected, i.e. $l_{inj}^{3/4}$, compared to the
predicted linear dependence on the injection scale, which may require some
refinement of the model or may be due to the effects like finite size of the
excitation region, which are not a part of the model.  The reconnection speed
was found to depend on the expected rate of magnetic field wandering and not on
the magnitude of the guide field. In our models, we see no dependence on the
guide field when its strength is comparable to the reconnected component.  More
importantly, while in the absence of turbulence we successfully reproduce the
Sweet-Parker scaling of reconnection, in the presence of turbulence we do not
observe any dependence on Ohmic resistivity, confirming that the reconnection of
weakly stochastic field is fast.  We also do not observe a dependence on
anomalous resistivity, which suggests that the presence of anomalous effects,
e.g. Hall MHD effects, may be irrelevant for astrophysical systems with weakly
stochastic magnetic fields.
\end{abstract}


\section{Introduction}
\label{sec:intro}

Magnetic fields play a key role in astrophysical processes such as star
formation, the transport and acceleration of cosmic rays, accretion disks, solar
phenomena, etc. \citep{crutcher99,beck02,schlickeiser04,elmegreen04}.  Typically
magnetic diffusion is very slow on astrophysical scales, so to a good
approximation we can treat magnetic fields as being purely advected with the
flow, which is frequently referred to in the literature as the "frozen in"
condition for the plasma \citep[see][]{moffat78}.

Do we expect the frozen in condition to be violated in typical astrophysical
conditions?  The answer to this question is a qualified yes.  Indeed, most
astrophysical flows are chaotic; when adjacent parcels of fluids do not move in
the same direction, the frozen-in magnetic fields become tangled.  This can be
easily visualized by viewing magnetic fields lines as threads moving with the
fluid.  When the distance between magnetic bundles of different direction
becomes small, the finite resistivity of fluids starts to be important.  In a
generic situation of 3D flows, the bundles of magnetic fields come into contact
with their neighbors at an angle of the order unity.  Over the small scales at
which fluid resistivity is important, magnetic field lines change their
topology, or reconnect.  However, once magnetic field energies become large,
bending magnetic fields on small scales requires energies much larger than the
turbulent energies on that scale.  The magnetic field lines become stiff, even
in the presence of strong turbulence, and the scales associated with contact
between regions with very different magnetic fields become large.

What is the resulting reconnection speed?  This is a vital question for many
areas of astrophysics
\cite[see][]{biskamp00,priest00,bhattacharjee04,zweibel09}.  A first guess could
be that magnetic reconnection is generically slow in astrophysical
circumstances.  There is a large disparity in the scales involved.  The scale of
the magnetic flux bundle is astronomically large.  The microphysical scale over
which the Ohmic dissipation is important is relatively small.  In this case the
crossing magnetic bundles will create ubiquitous unresolved intersections or
"knots" in the fluid.  Magnetic tension that arises from those intersections as
the magnetic bundles press against each other should dramatically change the
properties of magnetized fluids\footnote{Don Cox (private communication) refers
to this state of entangled magnetic fields as astrophysical Jello or
astrophysical felt, to reflect the peculiar non-fluid properties of the
hypothetical substance.}.  This would be devastating news for most numerical MHD
simulations, as numerical diffusion in the simulations is high and magnetic
bundles in simulations easily reconnect.

To understand the difference between astrophysical reconnection and the one in
numerical simulations, one should recall that the dimensionless combination that
controls the reconnection rate is the Lundquist number\footnote{The magnetic
Reynolds number, which is the ratio of the magnetic field decay time to the eddy
turnover time, is defined using the injection velocity $v_l$ as a characteristic
speed instead of the Alfv\'en speed $V_A$, which is taken in the Lundquist
number.}, defined as $S = L V_A / \eta$, where $L$ is the length of the
reconnection layer (see Figure~\ref{fig:lv99model}, upper panel), $V_A$ is the
Alfv\'en velocity, and $\eta$ is Ohmic diffusivity. Because of huge
astrophysical sizes $L$ involved, the astrophysical Lundquist numbers are huge,
e.g. for the ISM they are about $10^{16}$, while present-day MHD simulations
correspond to $S<10^4$. As the numerical efforts scale as $L_x^4$, where $L_x$
is the size of the box, it is not feasible at present and will not be feasible
in the foreseeable future to have simulations with the sufficiently high
Lundquist numbers. Incidentally, this also presents a problem for numerical
simulations of magnetic reconnection unless one has theoretically-derived
scaling relations to test. Even with the limited resolution, numerical
simulations are a good tool to study scaling relations, the point that has been
proved by successful numerical studies of MHD turbulence.

Due to huge values of astrophysical Lundquist numbers, any rate of reconnection
that depends on $S$ is extremely slow\footnote{The exception is the case when
the dependence on $S$ is logarithmic.}.  Fast reconnection is reconnection that
does not depend on resistivity.

What are the conditions that can make magnetic reconnection fast?  Does it rely
on special initial or boundary conditions for the flow, or does it require
particular plasma effects?  These are burning astrophysical questions, which,
for example, define the extent that we can rely on numerical simulations of
magnetized fluids as models of astrophysical phenomena.  To understand processes
of magnetic field generation associated with dynamos, the dynamics of the
interstellar medium and accretion disks, or other related phenomena, it is
important to understand magnetic reconnection.  In most cases, astrophysical
reconnection is difficult to observe, with the notable exception of solar flares
\cite[see][]{sturrock66,masuda94} and gamma ray bursts \cite[see][]{galama98}.
This sometimes creates an illusion that the importance of reconnection is
limited to those selected phenomena.

A famous model of magnetic reconnection was suggested by \citet{parker57} and
\citet{sweet58}.  Unfortunately, this model, usually referred to as Sweet-Parker
reconnection, provides very slow reconnection speeds.  In this model the
reconnection rate is inversely proportional to the square root of the Lundquist
number.  This very slow speed comes from a geometrical constraint.  The current
sheet dividing two magnetized regions must be very thin in order for Ohmic
resistivity to be important, but the plasma trapped in the current sheet must
follow the local magnetic field lines in order to escape.  A narrow current
sheet implies a highly restricted outflow.  Naturally, if the Lundquist number
is large, e.g. $S=10^{16}$, the Sweet-Parker reconnection speed, $V_{SP}\approx
V_A S^{-1/2}$, is negligible.

Fast reconnection has been investigated for many decades \citep[see][for
reviews]{biskamp00,priest00}.  In 1964 Petschek introduced the first fast
magnetic reconnection model \citep[see][]{petschek64}.  He proposed that
extended magnetic bundles come into contact over a tiny area determined by the
Ohmic diffusivity.  This configuration, called an X-point configuration, differs
dramatically from the expected generic configuration when magnetic bundles try
to press their way through each other.  Thus the first introduction of this
model raised questions of dynamical self-consistency.  An X-point configuration
has to persist in the face of compressive bulk forces.  However, numerical
simulations have shown that an initial X-point configuration of magnetic field
reconnection is unstable in the MHD limit for small values of the Ohmic
diffusivity \citep{biskamp96} and the magnetic field will relax to a
Sweet-Parker configuration.  The physical explanation for this effect is simple.
 In the Petschek model shocks are required in order to maintain the geometry of
the X-point.  These shocks must persist and be supported by the flows driven by
fast reconnection.  The simulations showed that the shocks fade away and the
contact region spontaneously increases.

More recently there has been progress in understanding the role of collisionless
modes in stabilizing X-point reconnection.  In particular, simulations using
Hall MHD have produced stable X-point reconnection \citep{shay98,shay04}.  This
was the first numerical demonstration of fast reconnection, which raised the
hope in astrophysical community that the problem of astrophysical reconnection
can be solved.  There is continuing discussion of whether this kind of fast
reconnection persists at scales substantially larger than the ion inertial scale
\cite[see][]{bhattacharjee03} with a number of numerical studies
\citep{wang01,smith04,fitzpatrick04} yielding reconnection rates that are not
fast, but depend on resistivity.  More importantly, one may also wonder to what
extent Petscheck-type collisionless reconnection solves the problem of
describing magnetic field dynamics in astrophysical settings.  Apart from the
issue of how natural it is to produce conditions that lead to Petscheck
reconnection, one should note that the requirement that the upper limit on the
collision rate required by this process is extremely restrictive.  For instance,
estimates in \cite{yamada07} show that the scale of the reconnection current
sheet should not exceed approximately 40 electron mean free paths.  This is not
satisfied in many astrophysical environments including the interstellar medium
\citep{yamada07}.

What happens in a collisional medium?  Is reconnection in the interstellar
medium slow?  The latter, as mentioned above, would have catastrophic
implications for the entire current crop of MHD simulations of interstellar
processes.

Nearly simultaneously with the discovery of collisionless X-point reconnection,
a model of magnetic reconnection in the presence of a weakly stochastic magnetic
field was proposed by \citet[][henceforth LV99]{lazarian99}.  They claimed that
the laminar magnetic fields considered in both Petscheck and Sweet-Parker
reconnection models are exceptional in astrophysics.  If we consider the
interstellar medium as an example, a so-called "Big Power Law in the Sky"
indicates the presence of turbulence on scales from tens of parsecs to thousands
of kilometers \citep{armstrong95}.  Thus, the fact that reconnection may be slow
in a collisional fluid with laminar field lines does not necessarily imply that
reconnection is generically slow.  The model in LV99 predicted extended current
sheets and wide outflows limited by the diffusive spread of magnetic field
lines.  In the limit of very weak turbulence or very large resistivities the
model converges to the Sweet-Parker model.

For many years the quantitative predictions of the stochastic reconnection model
in LV99 have not been tested explicitly.  There is some implicit evidence in the
favor of this model, e.g. observations of the thick reconnection current outflow
regions observed in the Solar flares \citep{ciaravella08}.  Such outflow regions
are incompatible with Petscheck reconnection.  This is suggestive, but falls far
short of constituting a definitive proof of the LV99 model.

It is, in general, extremely difficult to use analytic tools to test turbulent
processes.  However, \cite{eyink06} have studied the necessary conditions for
reconnection in a turbulent conducting medium in the limit of infinitesimal
resistivity.  They list three separate necessary conditions, and prove that the
flux conservation can be violated at an instant of time for an arbitrarily small
scale, if at least one of the conditions is satisfied.  The one that concerns us
here is that there must be overlapping vortex and current sheets in the fluid.
This is trivially satisfied if turbulent eddies on all scales give rise to
current sheets with inflow/outflow properties of the kind we study here.  While
this shows that the LV99 model is consistent with the physical requirements for
fast reconnection, it does not prove that the model is viable.

Numerical simulations have provided the main testing ground for collisionless
X-point reconnection.  These simulations are typically performed in 2D with
periodic boundary conditions and cannot be continued for longer than one
dynamical time.  This approach is inappropriate for stochastic reconnection
where the goal is to obtain a stationary reconnection rate averaged over many
dynamical times.  For this reason we have followed \cite{daughton06} in using
open outflow boundary conditions.  Moreover, the LV99 model of reconnection is
intrinsically three dimensional.  A simplification, however, arises from the
fact that LV99 predicts that fast reconnection takes place in the MHD
approximation, making plasma effects irrelevant.

In this paper we describe the choice of boundary conditions as well as an
appropriate measure of reconnection.  Indeed, while the rate of reconnection can
be trivially measured in the case of laminar magnetic fields, additional care is
required in the estimation of the reconnection rate in the presence of a
stochastic magnetic field.

In \S\ref{sec:lv99model} we review the LV99 model of reconnection and its
theoretical predictions.  In \S\ref{sec:setup} we describe in detail the
numerical model we have built and studied in this paper.  In
\S\ref{sec:rec_rate} we present methods used throughout the paper to measure the
reconnection rate and we introduce a new, more general, method of estimating the
reconnection rate which includes all processes contributing to the change of
magnetic flux.  In \S\ref{sec:results} we present an extensive description of
results obtained from studying our numeric model, which we discuss later in
\S\ref{sec:discussion}.  In \S\ref{sec:summary} we set forth our main
conclusions.

\section{Lazarian-Vishniac (1999) Model}
\label{sec:lv99model}

\begin{figure}
\includegraphics[width=0.45\textwidth]{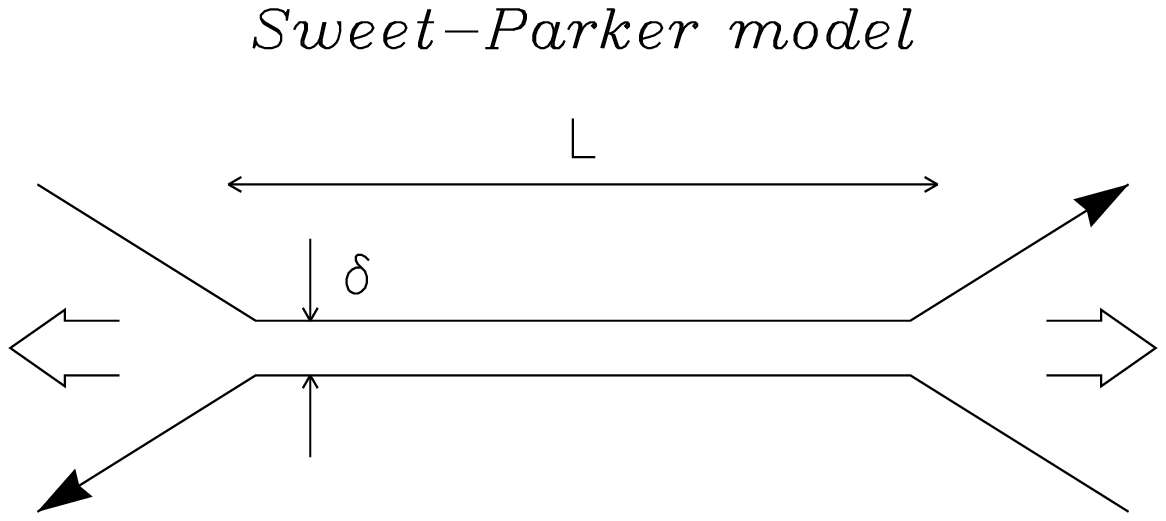}
\includegraphics[width=0.45\textwidth]{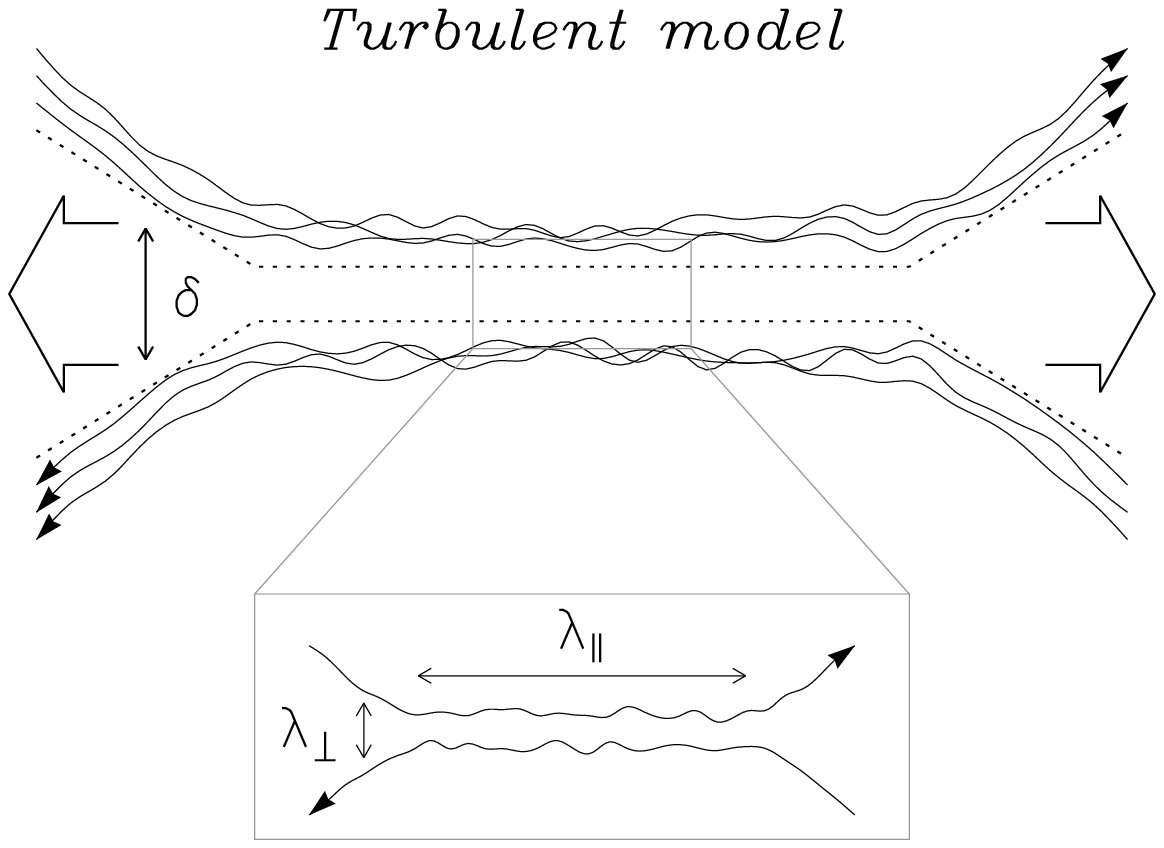}
\caption{{\it Upper plot}: Sweet-Parker model of reconnection.  The outflow is
limited to a thin width $\delta$, which is determined by Ohmic diffusivity.  The
other scale is an astrophysical scale $L \gg \delta$.  Magnetic field lines are
assumed to be laminar.
{\it Middle plot}: Turbulent reconnection model that accounts for the
stochasticity of magnetic field lines.  The stochasticity introduced by
turbulence is weak and the direction of the mean field is clearly defined.  The
outflow is limited by the diffusion of magnetic field lines, which depends on
macroscopic field line wandering rather than on microscales determined by
resistivity.
{\it Low plot}: An individual small scale reconnection region.  The reconnection
over small patches of magnetic field determines the local reconnection rate. The
global reconnection rate is substantially larger as many independent patches
reconnect simultaneously.  Conservatively, the LV99 model assumes that the small
scale events happen at a slow Sweet-Parker rate.  Following \cite{lazarian04}.}
\label{fig:lv99model}
\end{figure}

We begin by reviewing the basic features of the stochastic reconnection model in
\cite{lazarian99}.  This model might be seen as a generalization of the
Sweet-Parker model (see Fig.~\ref{fig:lv99model}) for the case of turbulent
fields.  Like the Sweet-Parker model it deals with an extremely generic
configuration, which should arise naturally as magnetic flux tubes try to make
their way one through another.  This avoids the problems related to the
preservation of outflow conditions which plague attempts to explain magnetic
reconnection via Petscheck-type solutions.  For example, if the outflow of
reconnected flux and entrained matter is temporarily impeded through a
fluctuation in the outflow, reconnection will slow down, but will not
permanently change its nature.

The essential difference between the Sweet-Parker model and the model in LV99 is
that while in the Sweet-Parker model the outflow is limited by microphysical
Ohmic diffusivity, in the LV99 model the large-scale magnetic field wandering
determines the thickness of outflow.  For extremely weak turbulence, when the
range of magnetic field wandering becomes smaller than the width of the
Sweet-Parker layer $L S^{-1/2}$, the reconnection rate reduces to the
Sweet-Parker rate\footnote{For the sake of simplicity, we do not raise here
issues related to the stability of the Sweet-Parker current sheet.  Estimates in
LV99 suggest that laminar current sheets subject to tearing instabilities have
reconnection rates which are just a bit faster than in the original Sweet-Parker
model.  These instabilities may, however, be important for increasing the 3D
stochasticity of magnetic field lines and thus initiating fast reconnection.  In
this paper we do not address this effect either.}.

LV99 consider the case of a large scale, well-ordered magnetic field, of the
kind that is normally used as a starting point for discussions of reconnection.
In the presence of turbulence we anticipate that the field will have some small
scale `wandering'. On any given scale the typical angle by which field lines
differ from their neighbors is $\phi\ll1$, and this angle persists for a
distance along the field lines $\lambda_{\|}$ with a correlation distance
$\lambda_{\perp}$ across field lines (see Fig.~\ref{fig:lv99model}).

LV99 suggested that the presence of a random magnetic field component
substantially enhances the reconnection rate, leading to fast reconnection.
There are two phenomena mainly responsible for this:
\begin{itemize}
\item only a small fraction of any magnetic field line is subject to direct
Ohmic annihilation.  The fraction of magnetic energy that goes directly into
heating the fluid drops down to zero as the fluid resistivity vanishes, and
\item the presence of turbulence enables many magnetic field lines to enter the
reconnection zone simultaneously.
\end{itemize}

In order to do quantitative estimates one has to adopt a model of MHD
turbulence.  Attempt to construct such a model can be traced back to
\cite{iroshnikov63} and \cite{kraichnan65} papers as well as to later
fundamental works \cite[e.g.][see also book by Biskamp
2003]{montgomery81,shebalin83,higdon84}. A model for realistic compressible
fluids may be constructed on the basis of the \cite[][henceforth
GS95]{goldreich95} model of incompressible turbulence\footnote{The exact
scalings of the MHD turbulence are still subject to debate and different
corrections to the original GS95 scalings have been proposed to account for
dynamical alignment, polarization intermittency and non-locality of the cascade
\cite[see][]{boldyrev05,boldyrev06,beresnyak06,beresnyak08b,gogoberidze07}.
However, calculations in LV99 show that reconnection rates are only weakly
dependent on the exact model of turbulence.  Subtle refinements of the scalings
that are currently debated can only be of marginal significance for this paper.
For the sake of simplicity, we adopt in what follows the original GS95 scaling.
We also do not consider imbalanced MHD turbulence
\cite[see][]{lithwick07,chandran08,beresnyak08a,beresnyak09}, although for the
modest degrees of imbalance between the turbulent energy fluxes moving in the
opposite directions, we do not expect substantial changes in our results}.
Reasonable alternatives are expected to have only marginal impact on the nature
of field line stochasticity \cite[see
also][]{cho00,maron01,lithwick01,cho02a,cho02b,cho03}, provided that the GS95
scalings are generalized to include the case of weak MHD turbulence.  The
equations in LV99 reflect the fact that when the turbulence is injected at
subAlfv\'enic velocities, the cascading is initially weak and the perturbations
can be reasonably well represented by a collection of Alfv\'en waves undergoing
occasional interactions.  However, the strength of the non-linear interactions
increases as the perpendicular scale decreases and at the scale $l_{\perp} = l
(v_l / V_A)^2$ the cascade becomes strong, i.e. with the GS95 critical balance
between motions at parallel and perpendicular scales $l_{\perp} / v_{\perp}
\approx l / V_A$ satisfied.  The corresponding velocity at $l_{\perp}$ is
$v_{\perp}\sim V_A (v_l / V_A)^2$ where $v_l$ is the velocity at which strong
turbulence is being injected.  For $v_l \ll V_A$ this scale may be small and the
field wandering induced by the turbulence is reduced\footnote{These comments
ignore the compressible modes, but their role in field wandering is marginal
anyhow (LV99).}.  The characteristics of the strong turbulent cascade dominate
transport phenomena and predictions of reconnection speeds.  In contrast the
weak turbulent cascade has very little effect on field line stochasticity, since
in this regime motions are largely periodic.

The modification of the global constraint induced by mass conservation in the
presence of a stochastic magnetic field component is self-evident.  Instead of
being squeezed from a layer whose width is determined by Ohmic diffusion, the
plasma diffuses through a much broader layer, $\delta \sim \langle y^2
\rangle^{1/2}$ (see Fig.~\ref{fig:lv99model}), determined by the diffusion of
magnetic field lines.  This suggests an upper limit on the reconnection speed of
$\sim V_A \left( \langle y^2 \rangle^{1/2} / L \right)$.  This will be the
actual speed of reconnection if the progress of reconnection in the current
sheet does not impose a smaller limit.  The value of $\langle y^2\rangle^{1/2}$
can be determined once a particular model of turbulence is adopted, but it is
obvious from the very beginning that this value is determined by field wandering
rather than Ohmic diffusion.

Following \citet[][hereafter LVC04]{lazarian04} we can generalize this upper
limit to include the whole range of scales between the large scale eddy size and
the current sheet thickness.  Consider two points initially separated by the
thickness of the current sheet.  Translating both points along the field lines
we can define their rms separation as a function of distance along the field
lines, $\delta ( \lambda_{\|} )$.  This gives an upper limit on the local
reconnection speed which is just $\sim ( \delta / \lambda_\| ) V_A$.  However,
the large scale current sheet contains many such reconnection regions, each
involving the reconnection of independent field lines.  In order to estimate the
global reconnection rate we need to multiply each local reconnection rate with
the number of simultaneous reconnection events happening on that scale.  Taking
$L/\lambda_\|$ as our best estimate we get a global limit of
\begin{equation}
V_{rec}\le\hbox{min} \left[{\delta(\lambda_\|)\over \lambda_\|}{L\over \lambda_\|} V_A\right], \label{eq:glimit}
\end{equation}
which should be evaluated for all $\lambda_\|$ between the length of the current
sheet and the length of an individual piece of the current sheet such that
translation along that piece will increase the rms separation by the thickness
of the current sheet.  As long as we are in the regime of strong turbulence this
expression has a rather simple form.  Each eddy scatters field lines by roughly
its own width, so an eddy of parallel length $\lambda_\|$ will have a
corresponding $\delta \sim \lambda_\perp(\lambda_\|)$.  With this in mind
Eq.~(\ref{eq:glimit}) becomes
\begin{equation}
V_{rec}\le\hbox{min}\left[ {\lambda_\perp L\over \lambda_\|^2} V_A\right]. \label{eq:glimit2}
\end{equation}
At its minimum Eq.~(\ref{eq:glimit2}) for GS95 model provides $V_{rec} < V_A$,
which is a natural limit of the reconnection speed and not a constraint.  In the
Goldreich-Sridhar model of turbulence $\lambda_\|\propto \lambda_\perp^{2/3}$,
so the minimum value of this expression corresponds to the largest scales.
(Once $\lambda_\|$ exceeds the scale of the largest eddies we change models to a
random walk with a fixed step size.)

This should make it clear why we are insensitive to the exact model of strong
turbulence used in our calculation.  In order to get a different answer we would
need $\lambda_\|\propto \lambda_\perp^{1/2}$ or less, which is not supported by
any of the available numerical evidence.  On the other hand, how one models the
weak turbulence does have an impact on the reconnection rate, since it will
define the length of the largest scale eddies.  Also, the presence of strong
viscous damping can complicate our argument, by creating a laminar field line
regime on very small scales and reducing $\delta$ far below its value in the
strong turbulent regime.  This point is discussed at length in LVC04 where it
was argued that in some of the colder and denser phases of the interstellar
medium we should expect a reduction in the reconnection speed by an order of
magnitude or more.  The numerical work presented here includes only a minimal
amount of viscosity, so testing these arguments is beyond the scope of this
paper.

The basic hypothesis of LV99 is that the the speed of magnetic reconnection in
the presence of 3D stochasticity of magnetic field lines is given by the most
stringent constraint imposed by the bottleneck condition expressed in
Eq.~(\ref{eq:glimit}) \footnote{In LV99 other processes that can impede
reconnection were found to be less restrictive.  For instance, the tangle of
reconnection field lines crossing the current sheet will need to reconnect
repeatedly before individual flux elements can leave the current sheet behind.
The rate at which this occurs can be estimated by assuming that it constitutes
the real bottleneck in reconnection events, and then analyzing each flux element
reconnection as part of a self-similar system of such events.   This turns out
not to impede reconnection.}.  With the GS95 model of turbulence LV99 obtained:
\begin{equation}
V_{rec}=V_A \min\left[\left({L\over l}\right)^{1/2}, \left({l\over L}\right)^{1/2}\right] \left({v_l\over V_A}\right)^{2}, \label{eq:constraint}
\end{equation}
where $l$ and $v_l$ are the energy injection scale and turbulent velocity at
this scale respectively.  Note that the combination $V_A(v_l/V_A)^2$ is the
velocity $v_{trans}$, i.e. the velocity at which the cascade transfers to the
strong regime.  This reflects the importance of strong turbulence for magnetic
field wandering and magnetic reconnection.  The term $(L/l)^{1/2}$ reflects the
random walk which takes place for $L<l$  Evidently, the goal of our paper is to
test formula (\ref{eq:constraint}).

There are several important characteristic features of
Eq.~(\ref{eq:constraint}).  First, and most important, there is no dependence on
resistivity.  Second, in general we expect reconnection to be fast since in most
cases the parameter ratios that enter the expression, i.e. the length of the
reconnection layer $L$ divided by the injection scale $l$, and the injection
velocity $v_l$ divided by the Alfv\'en velocity $V_A$ are of order unity.
Finally, we note that in particular situations when turbulence is extremely weak
the reconnection speed may be small.

Given the limited dynamical range of the simulations, we are forced to inject
turbulent energy on scales less than $L$.  Also, it is easier to control not
$v_l$, but the energy injection power $P$.  The power in the turbulent cascade
is $P \sim v_{turb}^2 (V_A/l)$ or $v_l^4/(lV_A)$, which reflects the fact that
the turbulence is weak at the injection scale $l$ (see LV99).  Using this in
Eq.~ (\ref{eq:constraint}) we get
\begin{equation}
V_{rec}\sim l P^{1/2},
\label{eq:scaling}
\end{equation}
which is the prediction we will test here.  In what follows we refer to the
injection power and scale using $P_{inj}$ and $l_{inj}$, respectively.

\section{Numerical Setup}
\label{sec:setup}

\subsection{Governing Equations}
\label{sec:equations}

We use a higher-order shock-capturing Godunov-type scheme based on the
essentially non oscillatory (ENO) spacial reconstruction and Runge-Kutta (RK)
time integration \citep[see][e.g.]{londrillo00, delzanna03} to solve isothermal
non-ideal MHD equations,
\begin{eqnarray}
 \pder{\rho}{t} + \nabla \cdot \left( \rho \vc{v} \right) & = & 0, \label{eq:mass} \\
 \pder{\rho \vc{v}}{t} + \nabla \cdot \left[ \rho \vc{v} \vc{v} + \left( a^2 \rho + \frac{B^2}{8 \pi} \right) I - \frac{1}{4 \pi} \vc{B} \vc{B} \right] & = & \vc{f}, \label{eq:momentum} \\
 \pder{\vc{A}}{t} + \vc{E} & = & 0, \label{eq:induction}
\end{eqnarray}
where $\rho$ and $\vc{v}$ are plasma density and velocity, respectively,
$\vc{A}$ is the vector potential, $\vc{E} = - \vc{v} \times \vc{B} + \eta \,
\vc{j}$ is the electric field, $\vc{B} \equiv \nabla \times \vc{A}$ is the
magnetic field, $\vc{j} = \nabla \times \vc{B}$ is the current density, $a$ is
the isothermal speed of sound, $\eta$ is the resistivity coefficient, and
$\vc{f}$ represents the forcing term.  We used Harten-Lax-van Leer
\cite[HLL,][]{harten83} and HLLD Riemann solvers \citep{mignone07} for solving
the isothermal MHD equations.  These schemes have widely different dissipation
properties.  The HLL solver was initially developed for solving Euler equations
and later adopted for solution of general set of time dependent differential
equations by averaging the Riemann fan over a region bound by the minimum and
maximum local characteristic speeds. In this way one intermediate state was
constructed.  The HLLD Riemann solver takes into account the discontinuities
resulting from the presence of magnetic field as well, separating this
intermediate state into multiple intermediate states, resolving e.g. Alfv\'en
waves with much less dissipation.  In this paper we are considering the
quasi-incompressible regime, where most of energy is transported by Alfv\'en
waves, therefore the application of the HLLD solver seems to be a better choice.
 We incorporated the field interpolated constrained transport (CT) scheme based
on a staggered mesh \citep[see][]{londrillo00} into the integration of the
induction equation (Eq.~\ref{eq:induction}) to maintain the $\nabla \cdot \vc{B}
= 0$ constraint numerically.

Some selected simulations that we perform include anomalous resistivity modeled
as
\begin{equation}
\eta = \eta_u + \eta_a \left( \frac{| \vc{j} |}{j_{crit}} - 1 \right) H \left( \frac{| \vc{j} |}{j_{crit}} \right),
\end{equation}
where $\eta_u$ and $\eta_a$ describe uniform and anomalous resistivity
coefficients, respectively, $j_{crit}$ is the critical level of the absolute
value of current density above which the anomalous effects start to work, and
$H$ is a step function.  However, for most of our simulations $\eta_a = 0$.

\subsection{Model Description and Initial Conditions}
\label{sec:initial}

\begin{figure}
\center
\includegraphics[width=0.45\textwidth]{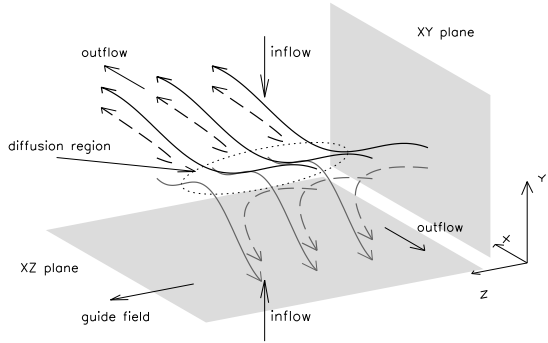}
\caption{Schematic 3D visualization of the reconnection problem setup.  Upper
and lower parts of the domain contain incoming oppositely directed magnetic
field lines (black and grey solid lines, respectively).  The inflow bends the
lines toward the center where they enter the diffusion region.  In the diffusion
region the lines reconnect and their product (dashed lines) is ejected along the
X direction. \label{fig:setup}}
\end{figure}

\begin{figure*}
\center
\includegraphics[width=0.45\textwidth]{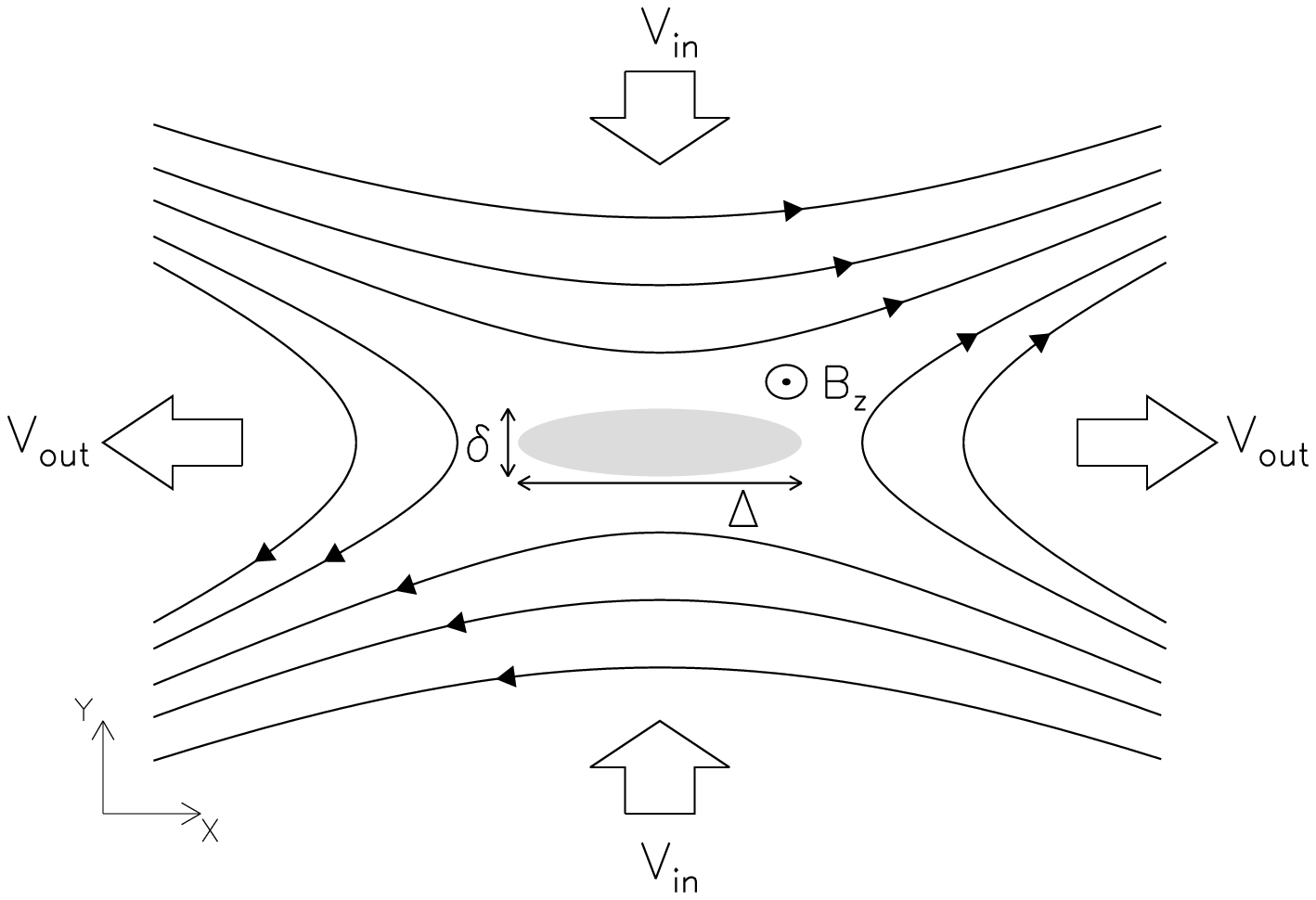}
\includegraphics[width=0.45\textwidth]{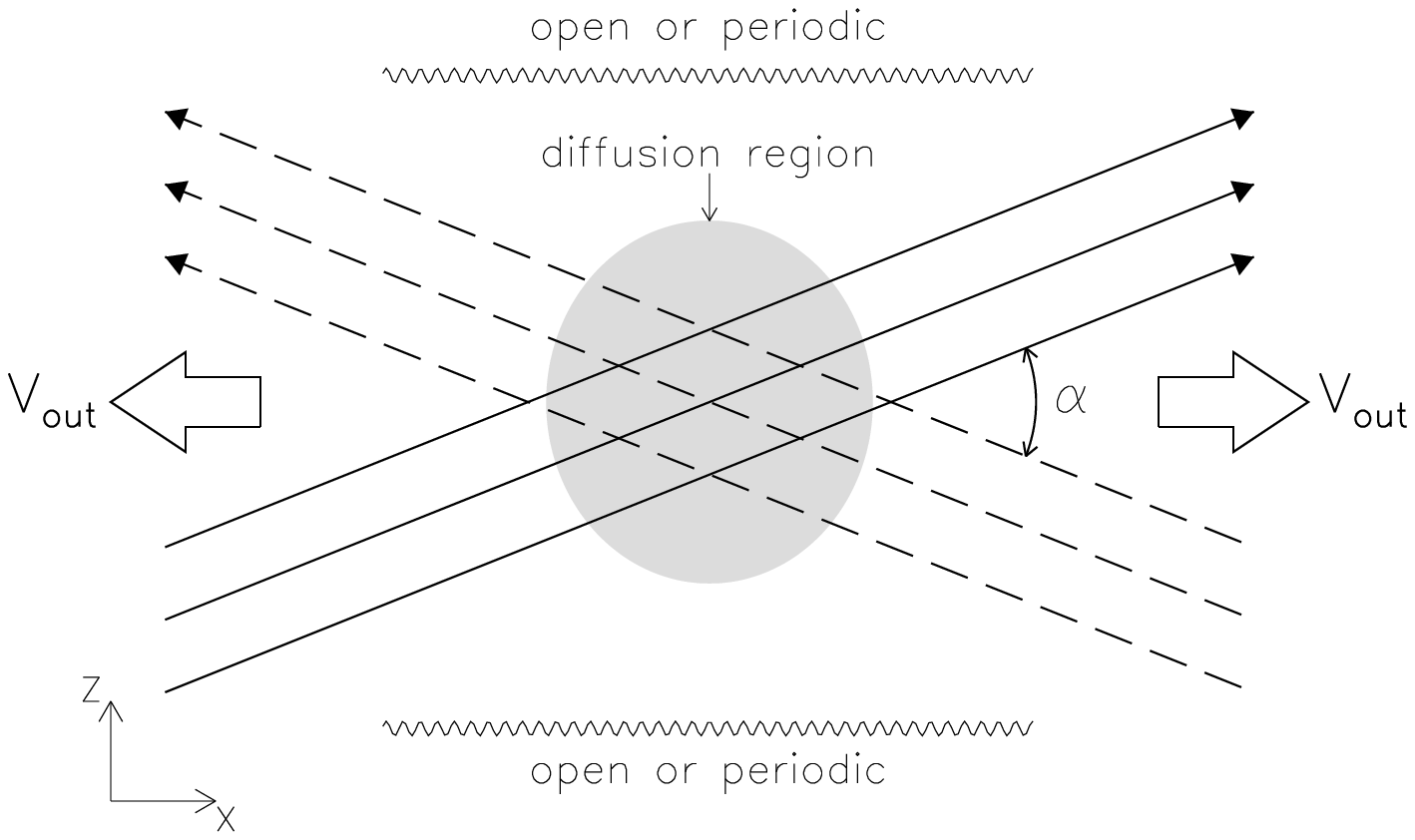}
\caption{3D magnetic field configuration in the studied problem projected on the
XY (left) and XZ (right) planes.
{\em Left:} XY projection of the magnetic field lines.  The grey area describes
the diffusion region where the incoming field lines reconnect.  The longitudinal
and transverse scales of the diffusion region are described by the parameters
$\Delta$ and $\delta$, respectively.  $B_z$ represents the direction of the
guide field with respect to the plane of projection.  We use inflow and outflow
boundary conditions at X and Y directions, respectively.
{\em Right:} XZ projection of the magnetic field lines as seen from the top.
Solid and dashed lines show the incoming field lines from the upper and lower
parts of the domain, respectively.  We see that the oppositely directed field
lines are not perfectly antiparallel but they create an angle $\alpha$
determined by the strength of the shared component $B_{z}$.  The Z boundary
conditions are open or periodic, depending on the model.
\label{fig:setup_planes}}
\end{figure*}

Figures~\ref{fig:setup} and \ref{fig:setup_planes} show a 3D visualization and
2D projections of the reconnection problem setup.  The domain contains two
regions of oppositely directed magnetic field lines (see Fig.~\ref{fig:setup}
and the left panel of Fig.~\ref{fig:setup_planes}).  The incoming lines (solid
lines in Fig.~\ref{fig:setup}) are bent by the inflow $V_{in}$ and enter the
diffusion region.  The diffusion region is characterized by the longitudinal
scale $\Delta$ and its thickness $\delta$ (see the left panel of
Fig.~\ref{fig:setup_planes}).  The diffusion region extends along the full Z
direction of the system.  The incoming magnetic lines are not perfectly
antiparallel.  The projection of the magnetic topology on the XZ plane shows
that the lines in the upper region (solid lines in the right panel of
Fig.~\ref{fig:setup_planes}) and in the lower region (dashed lines) create an
angle $\alpha$ determined by the strength of the shared component $B_{0z}$.
Once the incoming magnetic lines enter the diffusion region, they are
reconnected and the product of this process is ejected along X direction with a
speed $V_{out}$ (see Fig.~\ref{fig:setup} and the left panel of
Fig.~\ref{fig:setup_planes}).

Our initial magnetic field is a Harris current sheet of the form $B_x(x,y,z) =
B_{0x} \tanh (y/\theta)$ initialized using the magnetic vector potential
$A_z(x,y,z) = \ln | \cosh(y / \theta) |$.  In addition, we use a uniform guide
field $B_z(x,y,z) = B_{0z} = \mathrm{const}$.  The initial setup is completed by
setting the density profile from the condition of the uniform total (thermal
plus magnetic) pressure $p_T(t=0,x,y,z) = \mathrm{const}$ and setting the
initial velocity to zero everywhere.

In order to initiate magnetic reconnection we add a small initial perturbation
of vector potential $\delta A_z(x,y,z) = \delta B_{0x} \cos(2 \pi x)
\exp[-(y/d)^2]$ to the initial configuration of $A_z(t=0,x,y,z)$.  Parameters
$\delta B_{0x}$ and $d$ describe the strength of the initial perturbation and
thickness of the perturbed region, respectively.

We use dimensionless equations, so that the strength of the magnetic field is
expressed in terms of the Alfv\'en velocity (defined by the antiparallel
component of magnetic field) and the unperturbed density $\rho_0=1$.  All other
velocities are expressed as fractions of the fiducial Alfv\'en speed.  The
length of the box in X direction defines the unit of distance and time is
measured in units of $L_x/V_A$.  Initially, we set the strength of antiparallel
magnetic field component $B_{0x}$ to 1.0 and we vary the guide field $B_{0z}$
between 0.0 and 1.0, which corresponds to the range of angle $\alpha \in
(0^\circ, 90^\circ)$.  For a particular run we use $B_{0z}=4.0$, which
corresponds to $\alpha \approx 152^\circ$.  The speed of sound is set to 4.0 to
suppress compressibility in the system.  In order to study the importance of
resistivity in the reconnection process we vary the resistivity coefficient
$\eta_u$ between values $5\cdot10^{-4}$ and $5\cdot10^{-3}$ which are expressed
in the dimensionless units.  In the models where we include anomalous effects,
we vary the anomalous resistivity coefficient $\eta_a$ between 0.0 and
$2\cdot10^{-3}$.  The parameters describing the initial perturbation are set to
$\delta B_{0x} = 0.05$ and $d = 0.1$.

\subsection{Boundary Conditions}
\label{sec:boundaries}

Our numerical model of LV99 reconnection evolves in a box with dimensions $L_x =
L_z = 1$ and $L_y=2$ with resolution 256x512x256.  It is extended in the Y
direction in order to move the inflow boundaries far from the injection region.
This minimizes the influence of the injected turbulence on the inflow.

We use three different types of boundary conditions in our models, each of them
set along different direction: outflow boundary conditions along X direction,
inflow boundary conditions along Y direction and periodic boundary conditions
along Z direction.

We tested different types of open boundary conditions, including perfectly
permeable wave-type boundary conditions derived from the linearized MHD
equations similar to the method of characteristics \citep{hedstrom79}, which
seem to be the right choice. However, they are expensive since they require to
track the linear MHD waves crossing the boundaries and they have a relatively
small advantage over the simple "zero-gradient" boundary conditions in the
application to our models, which we found the simplest and most robust for our
purposes. In this type of open boundary conditions we set the normal derivatives
of the fluid variables (density and momentum) to zero.  This guarantees that all
waves generated in the system are free to leave the box without significant
reflections.  In the presence of turbulence, however, this type of boundary does
not enforce a constant mean value of density at the inflow boundary, which can
result in a small loss of the total mass in the system.  We will show later that
the total mass loss is very small and does not influence our results.

As for the vector potential, we have tested Daughton's approach for open
boundary conditions \cite[see][]{daughton06} in which the authors solved the
properly set wave equations for each component.  However, this approach was
limited to the 2.5 dimensional case and its extension to the full 3D case
requires a significant increase of complexity and, in our opinion, results in an
unacceptably expensive method.  Moreover, since \cite{daughton06} studied the
problem of reconnection using a kinetic approach, they used the speed of light
to evolve traveling electromagnetic waves.  In the MHD approximation, this would
imply a very large reduction in the time step, since the stability of the scheme
depends on the Courant-Friedrichs-Lewy \cite[CFL,][]{courant28} condition, which
in turn depends on the maximum speed in the system.  This issue excluded this
kind of approach from our consideration.

In the end we decided to use the simplest approach. We evolve the induction
equation for the vector potential $\vc{A}$ based on a staggered mesh, which
means that the vector potential components are located in the edge centers, i.e.
$A_x(i,j+1/2,k+1/2)$, $A_y(i+1/2,j,k+1/2)$, and $A_z(i+1/2,j+1/2,k)$, where
$(i,j,k)$ describes the cell center position. Thus, using the method described
in \cite{londrillo00} we calculate magnetic field components located at the
interface centers, $B_x(i+1/2,j,k)$, $B_y(i,j+1/2,k)$, and $B_z(i,j,k+1/2)$.
These locations of collocation points for the vector potential completed by a CT
integration of the induction equation guarantees the divergence-free evolution
of magnetic field.

In the treatment of vector potential $\vc{A}$ at the boundary we set its
components transverse to the considered boundary using the first order
extrapolation, while the normal derivative of the normal component is set to
zero. In this way the normal derivatives of transverse components of magnetic
field are zero, while the normal component of magnetic field is calculated from
the zero-divergence condition $\nabla \cdot \vc{B} = 0$.

This approach prevents from the generation of divergence of $\vc{B}$ at the
boundaries, however, it creates a small jump of the fluxes in the momentum
equation across the boundary resulting from the presence of non-zero terms
$\left( - B_x, B_y, B_z \right) \partial_x B_x$ at the X outflow boundary and
$\left( B_x, - B_y, B_z \right) \partial_y B_y$ at the Y inflow boundary.  In
order to justify the importance of these terms we have estimated directly from
our models the velocity increment they produce at each time step.  In models
with the strongest turbulence the magnitudes of these terms were of order of
$10^{-6}$ and $10^{-8}$ of $V_A$ at the X and Y boundaries, respectively,
signifying their negligible importance in the presence of strong inflow and
outflow which are of order of Alfv\'en speed in our models.

There is another constrain on the boundary conditions applying to the induction
equations resulting from its non-ideality, i.e. the presence of resistive term
and the treatment of current density $\vc{j}$ at the boundaries.  When we set
normal derivatives of transverse components of magnetic field to zero, we
violate the continuity of current density.  In the ideal case, $\eta=0$, the
resistivity term is unimportant even if the current density is not continuous.
However, once we introduce a non-zero resistivity coefficient, the resistive
term $\eta \vc{j}$ in the induction equation becomes discontinuous at the
boundary and we introduce a jump of electromotive force ${\cal E} = \vc{v}
\times \vc{B} - \eta \vc{j}$ across the boundary.  For example, considering the
Z component of electromotive force ${\cal E}_z = v_x B_y - v_y B_x - \eta \left(
\partial_x B_y - \partial_y B_x \right)$ we obtain a jump of ${\cal E}_z$ across
the X boundary equal to $\Delta_x {\cal E}_z = \left( {\cal E}_z^b - {\cal
E}_z^d\right) = \eta \partial_x B_y^d$, where indices $d$ (domain) and $b$
(boundary) stand for the limit values from the left and right sides of the
considered boundary, respectively.  This results in a constant generation of
$B_y$ at the boundary if the second derivative $\partial^2_x B_y$ is not zero at
that location.  One way to remove this undesired effect is by setting the second
order normal derivative of the field components to zero.  However, there is no
straightforward way to do this since it introduces another constraint on the
boundary values and requires complex and expensive methods.  We use another way
of diminishing the importance of resistive terms and current density jump at the
boundary.  Instead of modifying the boundary conditions, we introduce a zone of
decaying resistivity close to the boundary.  This means that the resistivity is
constant in most of the computational domain.  Only in a thin zone near the
boundary, the value of resistivity $\eta_u$ decays down to a very small value,
which is estimated to be around the numerical resistivity $\eta_n$ of our code.
In our models we adopt the value of $\eta_n = 3 \cdot 10^{-4}$.  Naturally, by
introducing such resistivity we do not change the reconnection speeds.  There
are two advantages to this approach.  First, the outflow in our models is not
constrained by the jump of resistive terms growing with the increasing value of
resistivity $\eta_u$, and is determined solely by the evolution in the central
part of the domain, where the resistivity is constant and uniform.  Second, we
control the importance of numerical effects by introducing a small, but larger
than numerical resistivity value of $\eta_n$ at the boundary.  The validation of
this method is presented in \S\ref{sec:resistivity}.

The above boundary conditions represent open boundaries, which adjust during the
evolution of the system.  This means that we do not set fixed values of the
fluid and magnetic variables and do not drive the flow at the boundaries in
order to achieve a stationary reconnection process in the system.

\subsection{Model of Turbulence}
\label{sec:forcing}

In our models we drive turbulence using a method described by \cite{alvelius99}.
The forcing is implemented in spectral space where it is concentrated around a
wave vector $k_{inj}$ corresponding to the injection scale $l_{inj}$.  We
perturb a number $N_f$ of discrete Fourier components of velocity in a shell
extending from $k_{inj}-\Delta k_{inj}$ to $k_{inj}+\Delta k_{inj}$ with a
Gaussian profile of the half width $k_c$ and the peak amplitude $\tilde{v}_f$ at
the injection scale (see Tab.~\ref{tab:models} for exact values in all models).
In all models $k_c = 0.4$.  Since we can control the scale of injection, the
power input is introduced into the flow at an arbitrary scale.  The amplitude of
driving is solely determined by its power $P_{inj}$, the number of driven
Fourier components and the time step of driving $\Delta t_f$, which in2 all our
models is equal to $10^{-5}$. The parameters describing our forcing do not
change during the evolution of the system. Because we perturb discrete number
$N_f$ of Fourier components which depends on the injection scale $k_{inj}$ and
the thickness of perturbed shell $\Delta k_{inj}$, the amplitude $\tilde{v}_f$
varies with the injection scale in order to keep the same power input $P_{inj}$.
In our models $\tilde{v}_f$ is always a small fraction of Alfv\'en speed (see
Tab.~\ref{tab:models}).

The randomness in time makes the force neutral in the sense that it does not
directly correlate with any of the time scales of the turbulent flow, and it
also determines the power input solely by the force-force correlation.  This
means that it is possible to generate different desirable turbulence states,
such as axisymmetric turbulence, where the degree of anisotropy of the forcing
can be chosen {\em a priori} through the forcing parameters.  In the models
presented in this paper we use isotropic forcing only.

In particular, the total amount of power input from the forcing can be set to
balance a desired dissipation at a statistically stationary state.  In order to
contribute to the input power in the discrete equations from the force-force
correlation only, the force is determined so that the velocity-force correlation
vanishes for each Fourier mode.  The procedure of reducing the velocity-force
correlation is described in \cite{alvelius99}.

We drive turbulence in a subvolume of the domain.  The size of the subvolume is
determined by two scales, the radius $r_d$ on the XZ plane around the center of
the domain and the height $h_d$ describing the thickness of the driving region
from the midplane.  In this way we avoid driving turbulence at the boundary and
reduce the influence of driving on the inflow or outflow.  All models are
evolved without turbulence for several dynamical times in order to allow the
system to achieve laminar stationary reconnection.  Then, at a given time $t_b$
we start driving turbulence, increasing its amplitude to the desired level,
until time $t_e$.  In this way we let the system to adjust to a new state.  From
time $t_e$ the turbulence is driven with the full power $P_{inj}$.

On the right hand side of Eq.~(\ref{eq:momentum}), the forcing is represented by
a function $\vc{f} = \rho \vc{a}$, where $\rho$ is local density and $\vc{a}$ is
random acceleration calculated using the method described above.

We do not set the viscosity coefficient explicitly in our models.  The scale at
which the dissipation starts to be important is defined by the numerical
diffusivity of the scheme.  The ENO-type schemes are considered relatively low
diffusion \cite[see][e.g.]{liu98,levy99}.  The numerical diffusion depends not
only on the adopted numerical scheme but also on the ``smoothness'' of the
solution, so it changes locally in the system.  In addition, it is also a
time-varying quantity.  All these problems make its estimation difficult and
incomparable between different applications.  However, the dissipation scales
can be estimated approximately from the velocity spectra.  Supported by our
studies of turbulence \cite[see][]{kowal07a,kowal07b} where we used similar
code, we estimated the dissipation scale $k_{\nu} \approx 30$ for the resolution
of models presented here.

\subsection{Table of Simulated Models}
\label{sec:models}

\begin{deluxetable*}{c|ccccccccccc}
\tablecaption{List of models. \label{tab:models}}
\tablehead{ \colhead{Name} & \colhead{$B_{0z}$} & \colhead{$\eta_u$ [$10^{-3}$]}
& \colhead{$\eta_a$ [$10^{-3}$]} & \colhead{$P_{inj}$} & \colhead{$k_{inj}$} & \colhead{$\Delta k_{inj}$} & \colhead{$N_f$} & \colhead{$\tilde{v}_f$} & \colhead{$v_l$ (t=12)} & \colhead{$\tilde{V}_A$} & \colhead{$\beta\equiv p/p_{mag}$} }
\startdata
PD & 0.1 & 1.0 & 0.0 & 0.1 &  8 & 0.5 &  96 & 0.0015 & 0.031 & 1.05 & 31.7 \\
{} & 0.1 & 1.0 & 0.0 & 0.2 &  8 & 0.5 &  96 & 0.0022 & 0.041 & 1.05 & 31.7 \\
{} & 0.1 & 1.0 & 0.0 & 0.5 &  8 & 0.5 &  96 & 0.0035 & 0.051 & 1.05 & 31.7 \\
{} & 0.1 & 1.0 & 0.0 & 1.0 &  8 & 0.5 &  96 & 0.0049 & 0.065 & 1.05 & 31.7 \\
{} & 0.1 & 1.0 & 0.0 & 2.0 &  8 & 0.5 &  96 & 0.0069 & 0.084 & 1.05 & 31.7 \\
{} & 1.0 & 1.0 & 0.0 & 0.1 &  8 & 0.5 &  96 & 0.0015 & 0.042 & 1.41 & 16.0 \\
{} & 1.0 & 1.0 & 0.0 & 0.2 &  8 & 0.5 &  96 & 0.0022 & 0.043 & 1.41 & 16.0 \\
{} & 1.0 & 1.0 & 0.0 & 0.5 &  8 & 0.5 &  96 & 0.0035 & 0.056 & 1.41 & 16.0 \\
{} & 1.0 & 1.0 & 0.0 & 1.0 &  8 & 0.5 &  96 & 0.0049 & 0.071 & 1.41 & 16.0 \\
{} & 1.0 & 1.0 & 0.0 & 2.0 &  8 & 0.5 &  96 & 0.0069 & 0.083 & 1.41 & 16.0 \\
\hline
SD & 0.1 & 1.0 & 0.0 & 1.0 &  5 & 0.5 &  24 & 0.0028 & 0.092 & 1.05 & 31.7 \\
{} & 0.1 & 1.0 & 0.0 & 1.0 &  8 & 0.5 &  96 & 0.0049 & 0.063 & 1.05 & 31.7 \\
{} & 0.1 & 1.0 & 0.0 & 1.0 & 12 & 1.0 &  48 & 0.0018 & 0.059 & 1.05 & 31.7 \\
{} & 0.1 & 1.0 & 0.0 & 1.0 & 16 & 1.0 &  96 & 0.0049 & 0.035 & 1.05 & 31.7 \\
{} & 0.1 & 1.0 & 0.0 & 1.0 & 25 & 1.0 & 265 & 0.0014 & 0.030 & 1.05 & 31.7 \\
{} & 1.0 & 1.0 & 0.0 & 1.0 &  5 & 0.5 &  24 & 0.0028 & 0.092 & 1.41 & 16.0 \\
{} & 1.0 & 1.0 & 0.0 & 1.0 &  8 & 0.5 &  96 & 0.0049 & 0.068 & 1.41 & 16.0 \\
{} & 1.0 & 1.0 & 0.0 & 1.0 & 12 & 1.0 &  48 & 0.0018 & 0.059 & 1.41 & 16.0 \\
{} & 1.0 & 1.0 & 0.0 & 1.0 & 16 & 1.0 &  96 & 0.0049 & 0.034 & 1.41 & 16.0 \\
{} & 1.0 & 1.0 & 0.0 & 1.0 & 25 & 1.0 & 265 & 0.0014 & 0.027 & 1.41 & 16.0 \\
\hline
RD & 0.1 & 0.5 & 0.0 & 1.0 &  8 & 0.5 &  96 & 0.0049 & 0.066 & 1.05 & 31.7 \\
{} & 0.1 & 1.0 & 0.0 & 1.0 &  8 & 0.5 &  96 & 0.0049 & 0.065 & 1.05 & 31.7 \\
{} & 0.1 & 2.0 & 0.0 & 1.0 &  8 & 0.5 &  96 & 0.0049 & 0.067 & 1.05 & 31.7 \\
{} & 0.1 & 4.0 & 0.0 & 1.0 &  8 & 0.5 &  96 & 0.0049 & 0.066 & 1.05 & 31.7 \\
{} & 0.1 & 5.0 & 0.0 & 1.0 &  8 & 0.5 &  96 & 0.0049 & 0.065 & 1.05 & 31.7 \\
{} & 1.0 & 1.0 & 0.0 & 1.0 &  8 & 0.5 &  96 & 0.0049 & 0.068 & 1.41 & 16.0 \\
{} & 1.0 & 2.0 & 0.0 & 1.0 &  8 & 0.5 &  96 & 0.0049 & 0.067 & 1.41 & 16.0 \\
{} & 1.0 & 4.0 & 0.0 & 1.0 &  8 & 0.5 &  96 & 0.0049 & 0.067 & 1.41 & 16.0 \\
\hline
AD & 0.2 & 0.5 & 0.0 & 0.5 &  5 & 0.5 &  24 & 0.0020 & 0.051 & 1.10 & 30.8 \\
{} & 0.2 & 0.5 & 0.5 & 0.5 &  5 & 0.5 &  24 & 0.0020 & 0.051 & 1.10 & 30.8 \\
{} & 0.2 & 0.5 & 1.0 & 0.5 &  5 & 0.5 &  24 & 0.0020 & 0.051 & 1.10 & 30.8 \\
{} & 0.2 & 0.5 & 2.0 & 0.5 &  5 & 0.5 &  24 & 0.0020 & 0.051 & 1.10 & 30.8 \\
\hline
BD & 0.1 & 1.0 & 0.0 & 1.0 &  8 & 0.5 &  96 & 0.0049 & 0.065 & 1.05 & 31.7 \\
{} & 0.5 & 1.0 & 0.0 & 1.0 &  8 & 0.5 &  96 & 0.0049 & 0.066 & 1.22 & 25.6 \\
{} & 1.0 & 1.0 & 0.0 & 1.0 &  8 & 0.5 &  96 & 0.0049 & 0.068 & 1.41 & 16.0 \\
{} & 4.0 & 1.0 & 0.0 & 1.0 &  8 & 0.5 &  96 & 0.0049 & 0.067 & 2.24 &  1.9 \\
\enddata
\end{deluxetable*}

In Table~\ref{tab:models} we list parameters of all the models presented in this
paper.  We divided them into several groups.  In each group we calculated models
in order to study the dependence of the reconnection rate on a characteristic
parameter of turbulence or resistivity.  We have studied the dependence of
reconnection on the power of turbulence (models ''PD''), injection scale (models
''SD''), uniform resistivity (models ''RD''), anomalous resistivity (models
''AD''), and dependence on the guide field (models ''BD'').

Among all parameters of the model we list those which vary, i.e. the strength of
guide field $B_{0z}$, the uniform and anomalous resistivities, $\eta_u$ and
$\eta_a$, respectively, the power of turbulence $P_{inj}$ and its injection
scale $k_{inj}$ with the half-thickness of the injection shell $\Delta k_{inj}$,
the number of perturbed Fourier components of velocity $N_f$ and the amplitude
of perturbation $\tilde{v}_f$ at the injection scale. In addition, we include
some of the most important parameters obtained from the simulations, such as the
peak amplitude at the injection scale $v_l$ obtained from the spectra of
velocity at the final time of simulation (t=12.0), the total Alfv\'en speed
$\tilde{V}_A = \sqrt{\left(B_{0x}^2+B_{0z}^2\right)/\rho_0}$ and the plasma beta
parameter $\beta = p / p_{mag}$. Note that the Alfv\'en speed $V_A =
|B_{0x}|/\sqrt{\rho_0}$ defined by the strength of the antiparallel component of
magnetic field is the same for all models and equals 1.0.

All models presented in this section were calculated with the grid size $\Delta
x \approx 0.004$ corresponding to the resolution 256x512x256. In the process of
constructing our numerical model we have performed number of tests for
convergence of numerical solution during the Sweet-Parker stage. These studies
indicated that the convergence is reached for both HLL and HLLD solvers at this
resolution, but the HLL scheme characterizes by a much higher numerical
dissipation. This is expected since the HLL solver smooths out the rotational
discontinuities resulting from the presence of magnetic field. As will will show
in \S~\ref{ssec:sweet-parker}, the numerical resistivity in the models
calculated with the HLLD scheme has the value of about $6.1\cdot10^{-4}$. The
convergence studies signify that the same models calculated using the HLL scheme
with the same resolution reveal a higher numerical resistivity of order of
$8.7\cdot10^{-4}$, which is over 40\% larger than $\eta_{num}$ in the HLLD
scheme for the same grid size $\Delta x \approx 0.004$. For models with lower
resolution where the grid size $\Delta x \approx 0.008$ we obtained values of
the numerical resistivity $\eta_{num}^{hlld} \approx 6.6\cdot10^{-4}$ and
$\eta_{num}^{hll} \approx 1.12\cdot10^{-3}$ for the HLLD and HLL schemes,
respectively. Here, the difference of numerical resistivity between both
resolutions for the HLLD solver is insignificant, justifying the convergence of
the solution, while the HLL scheme behaves rather poorly showing an increase of
numerical dissipation by over 75\% with respect to the model with smaller grid
size. Thus all our models presented here were calculated using the HLLD scheme
if not indicated otherwise.

\section{Reconnection Rate Measure}
\label{sec:rec_rate}

We measure the reconnection rate by averaging the inflow velocity $V_{in}$
divided by the Alfv\'en speed $V_A$ over the inflow boundaries, i.e.
\begin{equation}
\langle V_{in} / V_A \rangle = \frac{1}{2} \int\limits_{S} {dx dz \left( \left. \frac{v_y}{V_A} \right|_{y=y_{min}} - \left. \frac{v_y}{V_A} \right|_{y=y_{max}} \right)},
\end{equation}
where $S$ defines the area of the XZ inflow boundaries.  Since we have two XZ
boundaries, located at $y=y_{min}$ and $y=y_{max}$, we need to take half of the
resulting integral.  This measure works well for laminar reconnection, when the
system is perfectly stable and where the time derivative of the magnetic flux is
zero.  In the presence of turbulence, however, this time derivative can
fluctuate or the turbulence in the center of the box could affect the flow of
the plasma.  In this way we would get a flow of magnetic flux without the
presence of reconnection.  In order to include all effects contributing to the
change of magnetic flux, we define a new more general measure of the
reconnection rate.

\begin{figure}
\center
\includegraphics[width=0.45\textwidth]{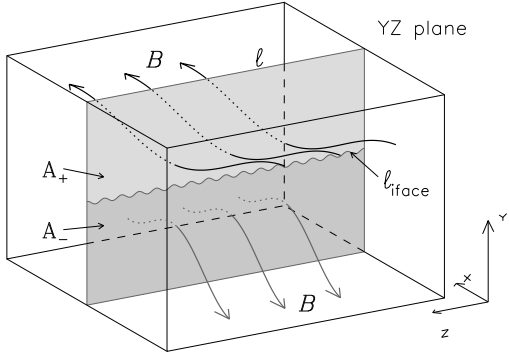}
\caption{Schematic 3D visualization of the reconnection rate evaluation.  $A_+$
and $A_-$ areas are defined by the sign of $B_x$ component.
\label{fig:reconn_measure}}
\end{figure}

We start by considering a conserved quantity; the magnetic flux $\Phi$.  First,
we consider the flux contained within a plane inside the simulation volume (see
Fig.~\ref{fig:reconn_measure}).  If $\hat{x}$ is the direction of the
reconnecting field, then we start by considering the time derivative of the net
flux of $B_x$.  It is
\begin{equation}
\partial_t \Phi = - \oint \vc{E} \cdot d \vc{l} = \oint \left( \vc{v} \times \vc{B} - \eta \vc{j} \right) \cdot d \vc{l}
\end{equation}

This equation is not exactly satisfied in our simulations unless we allow for
numerical resistivity.  This discrepancy can be used to derive an independent
measure of the effective resistivity of the code, which is roughly consistent
with value derived from the speed of laminar reconnection.  Now we split the
area of integration into two pieces, $A_+$ and $A_-$, defined by the sign of
$B_x$ (see Fig.~\ref{fig:reconn_measure}).  Instead of adding two areas we
subtract them, i.e.
\begin{equation}
\partial_t \Phi_{+} - \partial_t \Phi_{-} = \partial_t \int |B_x| dA,
\end{equation}
which we can write explicitly in terms of line integrals around $A_+$ and $A_-$
\begin{eqnarray}
\partial_t \int |B_x| dA & = & \oint{\vc{E} \cdot d \vc{l}_{+}} - \oint{\vc{E} \cdot d \vc{l}_{-}} \\
& = & \oint \mathrm{sign} (B_x) \vc{E} \cdot d \vc{l} + \int 2 \vc{E} \cdot d \vc{l}_{iface}, \nonumber
\end{eqnarray}
where $\vc{l}_{iface}$ is the line separating $A_+$ and $A_-$ (see
Fig.~\ref{fig:reconn_measure}).  The last term describes the mutual annihilation
of positive and negative $B_x$ along the line separating them and by definition,
this is the reconnection rate.  Note that this includes the motion of already
reconnected flux lines through the plane of integration.  Rather than try to
calculate it numerically, we define the interface term as $-2 V_{rec}
|B_{x,\infty}| L_z$, where $|B_{x,\infty}|$ is the asymptotic absolute value of
$B_x$, and $L_z$ is the width of the box.  We can then calculate the other terms
which do not involve trying to find the interface and the parallel component of
the electric field.  The end result, which is the new measure of reconnection
rate, is
\begin{equation}
V_\mathrm{rec} = \frac{1}{2 |B_{x,\infty}| L_z} \left[ \oint{\mathrm{sign} (B_x) \vc{E} \cdot d \vc{l}} - \partial_t \int {|B_x| dA} \right]
\end{equation}

The electric field $\vc{v} \times \vc{B} - \eta \vc{j}$ can be further divided
into an advection term $\vc{v} \times B_x \hat{x}$, a shear term $\vc{v} \times
\left( B_y \hat{y} + B_z \hat{z} \right)$, and a resistive term $- \eta \vc{j}$.
With this in mind the line integral can be rewritten as
\begin{eqnarray}
\oint{\mathrm{sign}\left( B_x \right) \vc{E} \cdot d \vc{l}} = \oint{|B_x| \left( \vc{v}_\perp \times \hat{x} \right) \cdot d \vc{l}} & & \\
+ \oint{\mathrm{sign} \left( B_x \right) v_x \left( \hat{x} \times \vc{B}_\perp \right) \cdot d \vc{l}} & - & \oint{ \eta \vc{j} \cdot d \vc{l}} . \nonumber
\end{eqnarray}

This new reconnection measure contains the time derivative of the absolute value
of $B_x$, and a number of boundary terms, such as advection of $B_x$ across the
boundary and the boundary integral of the resistive term $\eta \vc{j}$.  The
additional terms include all processes contributing the time change of $|B_x|$.
In particular, they can have non-zero values.

\section{Results}
\label{sec:results}

In this section we describe the results obtained from our three dimensional
simulations of magnetic reconnection in the presence of turbulence.  First, we
investigate Sweet-Parker reconnection, the stage before we inject turbulence.  A
full understanding of this stage is required in order to perform further
analysis of reconnection in the presence of turbulence.

\subsection{Sweet-Parker Reconnection}
\label{ssec:sweet-parker}

As we described in \S\ref{sec:initial}, Sweet-Parker reconnection develops in
our models as a result of an initial vector potential perturbation.  In order to
reliably study the influence of turbulence on the evolution of such systems, we
need to reach the stationary Sweet-Parker reconnection before we start injecting
energy.

\begin{figure}
\center
\includegraphics[width=0.45\textwidth]{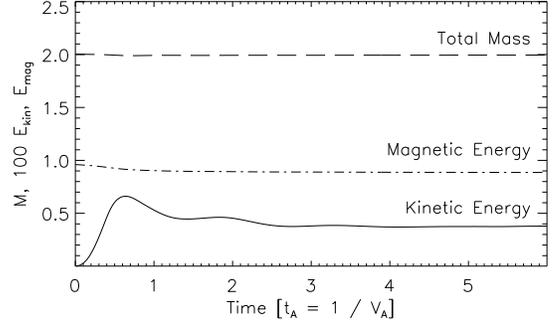}
\caption{Evolution of the total mass $M$, and kinetic and magnetic energies,
$E_{kin}$ and $E_{mag}$, respectively, during the Sweet-Parker stage.  The
kinetic energy $E_{kin}$ has been amplified by a factor of 100 to visualize its
evolution more clearly.  The resistivity in this model is set to
$\eta_{u}=10^{-3}$ and the guide field $B_{0z} = 0.1$. \label{fig:conservation}}
\end{figure}

Figure~\ref{fig:conservation} shows the evolution of total mass, kinetic and
magnetic energies until we start injecting the turbulence (i.e. $t=7$).  All
quantities, after some initial adaptation, reach almost constant values.  We
remind the reader, that the system evolves in the presence of open boundary
conditions, which do not guarantee perfect conservation of mass and total
energy.  Nevertheless, conservation of these quantities is well satisfied during
the Sweet-Parker stage in our models.

\begin{figure}
\center
\includegraphics[width=0.45\textwidth]{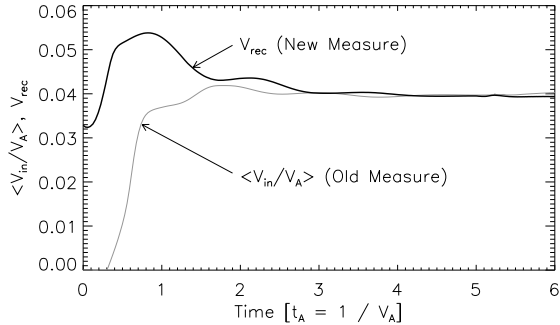}
\caption{Evolution of the reconnection rates, $\langle V_{in}/V_A \rangle$ (old)
and $V_{rec}$ (new), for the same model as in Fig.~\ref{fig:conservation}.  The
reconnection rate grows initially until it reaches the stationary solution.
Note, that both reconnection measures coincide during the later
evolution. \label{fig:rate_sp}}
\end{figure}

The reconnection rate, shown in Figure~\ref{fig:rate_sp} (solid thick line
representing the new reconnection rate measure), also confirms that we have
reached a stationary state.  Initially, the reconnection rate $V_{rec}$ grows
until time $t \approx 1.0$, when it reaches a maximum value of $\approx 0.05$.
Later on, it drops a bit approaching a value of $0.04$.  During the last period
of about three Alfv\'en time units, the change of the reconnection rate is very
small.  We assume that these conditions guarantee a nearly steady state
evolution of the system.  In this plot we also show the evolution of old
reconnection measure $\langle V_{in}/V_A \rangle$ which do not include time
variation of the flux.  We see, that after the initial discrepancy they coincide
and show the same value of reconnection speed.

\begin{figure}
\center
\includegraphics[width=0.45\textwidth]{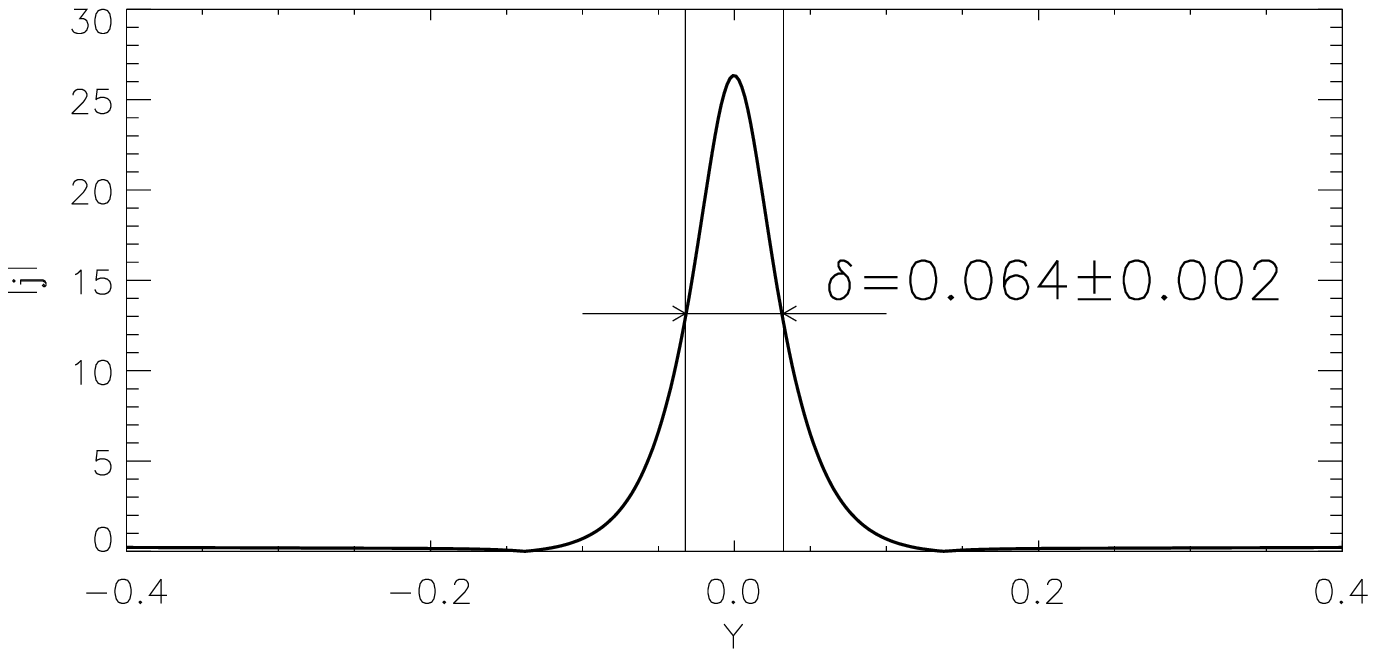}
\includegraphics[width=0.45\textwidth]{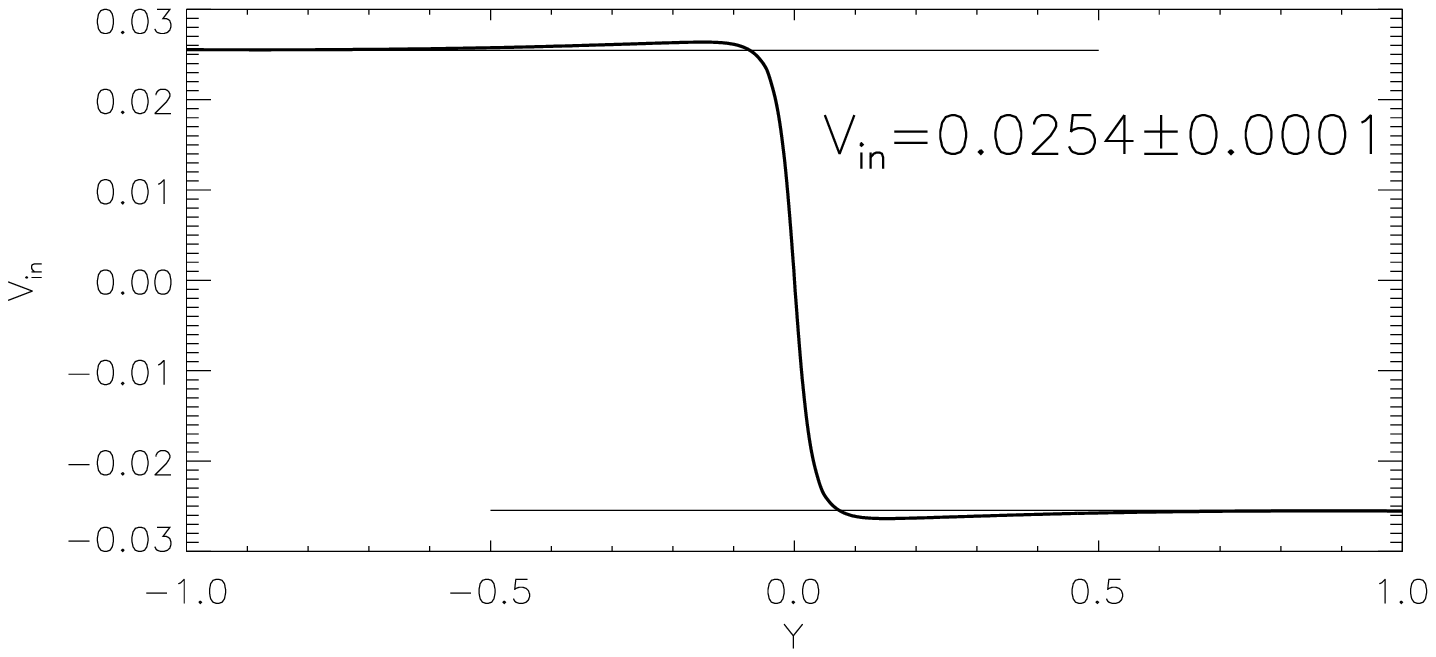}
\includegraphics[width=0.45\textwidth]{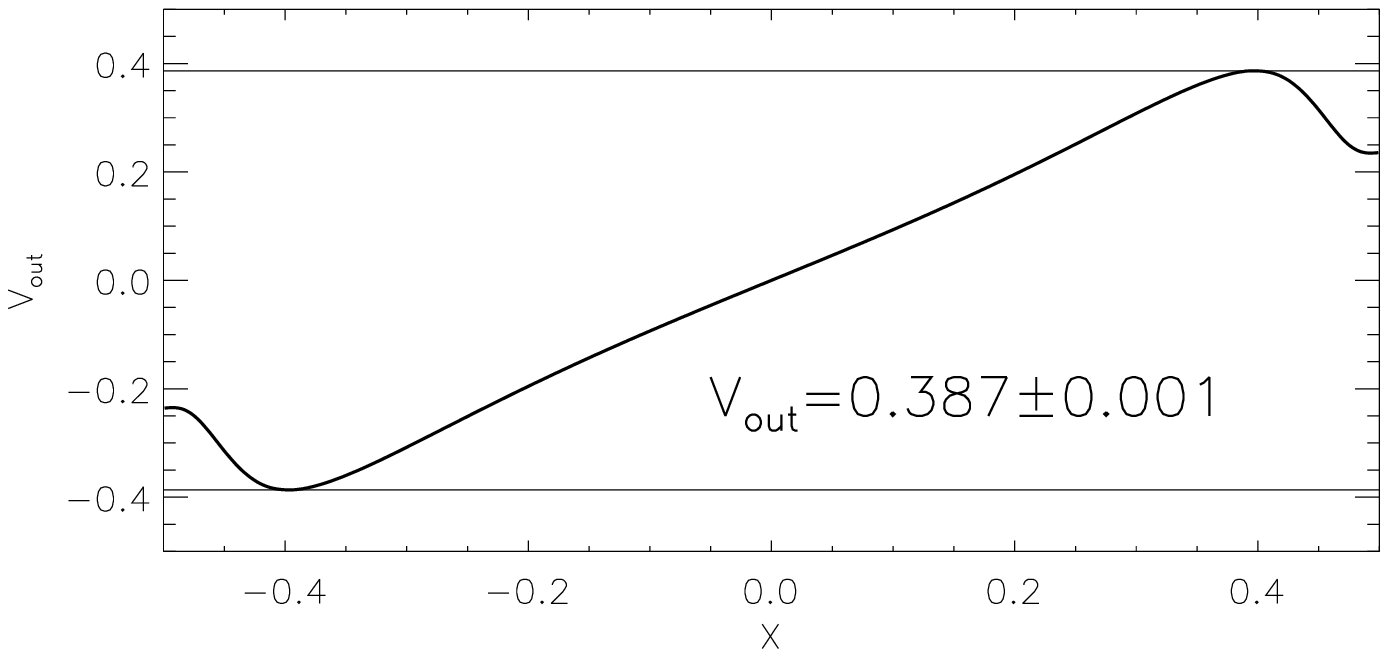}
\caption{Estimates of the current sheet thickness $\delta$ (top), the inflow
(middle) and outflow (bottom) speeds $V_{in}$ and $V_{out}$, respectively at the
time $t=7$ for the model with $\eta_u = 10^{-3}$ and $B_{0z}=0.1$.
\label{fig:estimates}}
\end{figure}

In order to check the reliability of our code, we estimate characteristic
parameters of the laminar reconnection, such as thickness of current sheet
$\delta$, and the inflow and outflow speeds, $V_{in}$ and $V_{out}$,
respectively (see Fig.~\ref{fig:estimates}), from the profiles of the absolute
value of current density $|\vc{j}|$ along the Y direction and the profiles of
$v_x$ and $v_y$ along the X and Y directions, respectively.  We compare these
estimates to the values obtained from the equations describing Sweet-Parker
model.  We limit this procedure to one model with a resolution of 256x512x256,
uniform resistivity $\eta_u = 10^{-3}$ and shared component of magnetic field
$B_{0z}=0.1$.  We analyze the snapshot obtained at the time $t=7$ before we
start introducing turbulence.

First, using the Ohmic dissipation formula $V_{in} = \eta / \delta$ we estimate
the total resistivity coefficient including the numerical contribution.  We
obtain a value of $\eta = \eta_u + \eta_{num} = V_{in} \delta \approx
1.64\cdot10^{-3}$.  It indicates that the numerical resistivity $\eta_{num}$ has
a value of about 60\% of the explicit uniform resistivity.  We use this value of
resistivity in the subsequent estimates.  Next, using mass conservation, $V_{in}
L = V_{out} \delta$ \citep{parker57,sweet58}, we estimate the outflow speed
$V_{out}$.  Since the size of our box along X direction is $L=1.0$, this gives a
value of $V_{out} \approx 0.394$, which agrees well with the value $V_{out} =
0.395$ obtained from the profile of $v_{x}$.  Since we already know the
characteristic parameters $V_{out}$, $L$, and $\eta$, we can estimate the
Lundquist number $S \equiv L V_{out}/\eta$ for our models.  Substituting
estimated $V_{out}$, $L$, and $\eta$ we obtain a value of $S \approx 236$.  As
expected, our numerical models are far from the conditions observed in typical
astrophysical objects where $S \sim 10^{10}-10^{20}$.  Nevertheless, we can still
use them to determine their dependence on the characteristics of the
environment.  From $V_{out}$ and $S$ we can estimate the Sweet-Parker
reconnection rate $V_{rec} = V_{out} S^{-1/2}$, which gives a value of $\sim
0.0251$, matching almost perfectly the value of $V_{in}$ fit from the model.
These estimates indicate, that our numerical model exhibits Sweet-Parker
reconnection during the stationary phase.

Before presenting our results in the presence of turbulence, we describe the
topology of the flow and magnetic field in the laminar case.  In
Figure~\ref{fig:top_sp} we present the velocity and magnetic field configuration
of the steady state.  In the left panel we show the topology of the velocity
field as textures.  The brightness of texture corresponds to the amplitude of
the field, while the texture itself shows the direction of the field lines.  The
topology of the velocity field is mainly characterized by strong outflow regions
along the midplane.  This outflow is produced by the constant reconnection
process at the diffusion region near the center and the ejection of the
reconnected magnetic flux through the left and right X boundaries.  The system
is in a steady state when the flux, which reconnects, is counterbalanced by the
incoming unreconnected flux.  The inflow is much slower than the outflow, but
its direction is still apparent in the left plot of Figure~\ref{fig:top_sp}.

The topology of the magnetic field is shown in the middle panel of
Figure~\ref{fig:top_sp}.  We recognize the antiparallel configuration of the
magnetic lines with uniform strength out of the midplane region.  Near the
midplane, the horizontal magnetic lines are reconnected generating the Y
component, which is ejected by a strong outflow.  In addition, we show the
absolute value of current density in the right panel of Figure~\ref{fig:top_sp}.
We see an elongated diffusion region in the middle of the box, where the
reconnection process takes place.  The maximum of $|\vec{j}|$ does not exceed a
value of 25.  Evidently our boundary conditions allow for physically realistic
inflow and outflow around a central current sheet.

\subsection{Effects of Turbulence}
\label{ssec:turbulent}

\subsubsection{Evolution in the Presence of Turbulence}
\label{ssec:turb_evolution}

The essential part of our studies covers the effects of turbulence on
reconnection.  Our goal was to achieve a stationary state of Sweet-Parker
reconnection, described in the previous subsection, and then introduce
turbulence at a given injection scale $l_{inj} \propto k_{inj}^{-1}$, gradually
increasing its strength to the desired amplitude corresponding to a turbulent
power $P_{inj}$.  We inject turbulence in a region surrounding the midplane and
extending to the distance of around one quarter of the size of the box.  This,
naturally, limits the injection scale to $k_{f} \lesssim 3$.  The transition
period, during which we increase the strength of turbulence has a length of one
Alfv\'enic time unit in all models and in the presented model starts at $t=6$.
This means that from $t=7$ we inject turbulence with the maximum power
$P_{inj}$.

\begin{figure}
\center
\includegraphics[width=0.45\textwidth]{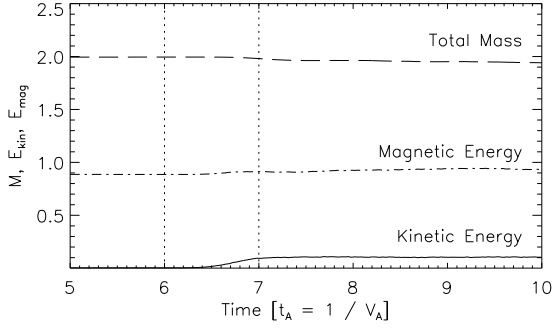}
\caption{Evolution of total mass $M$, and kinetic and magnetic energies,
$E_{kin}$ and $E_{mag}$, respectively for the stage with turbulence.  The
kinetic energy $E_{kin}$ is not amplified in this plot.  Two dotted vertical
lines bound the period of gradually increasing turbulence.  The resistivity in
this model is set to $\eta=10^{-3}$ and the shared component of magnetic field
$B_{0z} = 0.1$. \label{fig:energies_lv}}
\end{figure}

In Figure~\ref{fig:energies_lv} we present an example of the evolution of total
mass, and kinetic and magnetic energies in a model with $P_{in} = 1.0$,
$k_{f}=8$, and $\eta_u=10^{-3}$.  We inject turbulence, gradually increasing its
strength from $t=6$ to $t=7$.  This period is marked by two dotted vertical
lines in Figure~\ref{fig:energies_lv}.  We see an increase of kinetic energy
during this period due to the injection and after $t=7$ all quantities saturate
and do not vary much.  This means that even in the presence of turbulence and
with open boundary conditions our system conserves total mass and energy
relatively well.

\begin{figure}
\center
\includegraphics[width=0.45\textwidth]{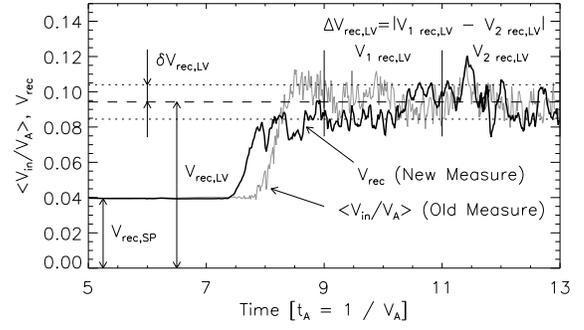}
\caption{Evolution of the reconnection rates, $\langle V_{in}/V_A \rangle$ (old)
and $V_{rec}$ (new), for the same model as in Fig.~\ref{fig:conservation}.  In
this plot we present the measured rates of the Sweet-Parker reconnection
$V_{rec,SP}$ and during the presence of turbulence, $V_{rec,LV}$.  Symbol
$\delta V_{rec,LV}$ is the time variance.  $\Delta V_{rec,LV}$ is the estimated
uncertainty of the measure. \label{fig:rate_lv}}
\end{figure}

In Figure~\ref{fig:rate_lv} we show the evolution of reconnection rates $\langle
V_{in}/V_A \rangle$ and more advanced $V_{rec}$, both described in
\S\ref{sec:rec_rate}.  In this plot we recognize an increase of both rates
during the introduction of turbulence.  After the initial period between $t=6$
and $t=8$, during which the system is adjusting to the new state, we see that
both measures coincide and even though they are fluctuating, they reach a
stationary state characterized by faster reconnection.  A more detailed
comparison of the simple and advanced reconnection rate measures is presented in
\S\ref{sec:comparison}.

In Figure~\ref{fig:rate_lv} we also show how we measure the rates of
Sweet-Parker reconnection $V_{rec,SP}$ and LV99 model $V_{rec,LV}$.  Because the
reconnection rates fluctuate in the presence of turbulence we also measure their
time variance $\delta V_{rec,LV}$ using standard deviation.  In addition to the
time variance of $V_{rec}$, we measure their errors by splitting the averaging
region into two subregions and after averaging the rates $V_{1 rec}$ and $V_{2
rec}$ over each subregion (see Fig.~\ref{fig:rate_lv}), we take the absolute
value of their difference $\Delta V_{rec} = V_{1 rec} - V_{2 rec}$.  This
difference corresponds to the error of $V_{rec}$, i.e. it is different from zero
if the rate is not constant in time.  In all further analysis and presented
plots of dependencies we use values estimated in this way.

\begin{figure*}
\center
\includegraphics[width=0.3\textwidth]{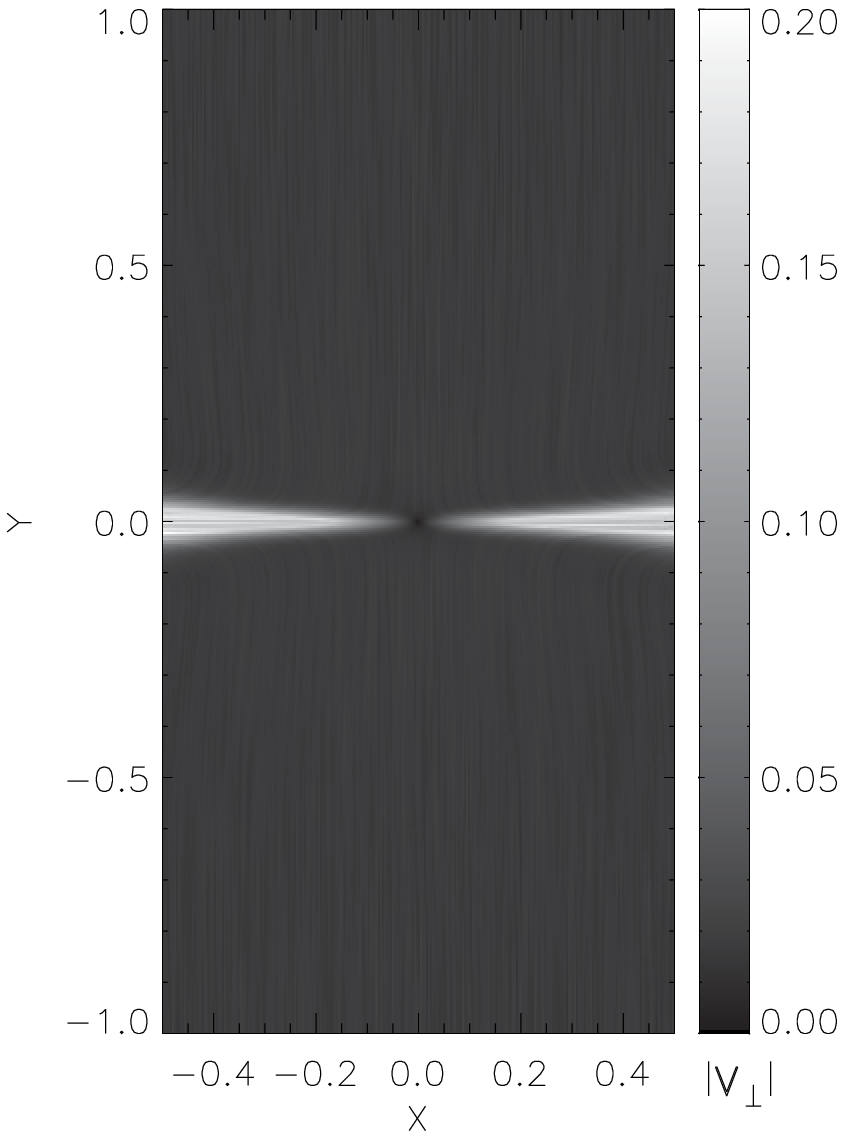}
\includegraphics[width=0.3\textwidth]{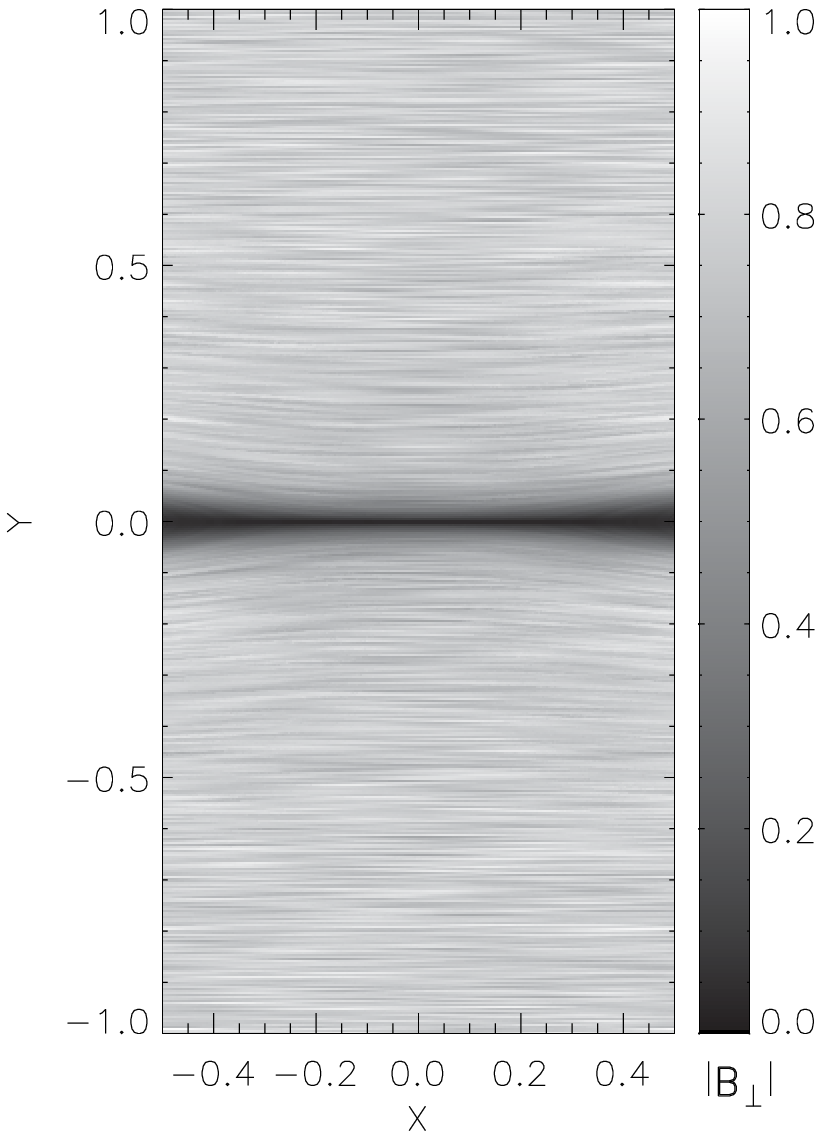}
\includegraphics[width=0.3\textwidth]{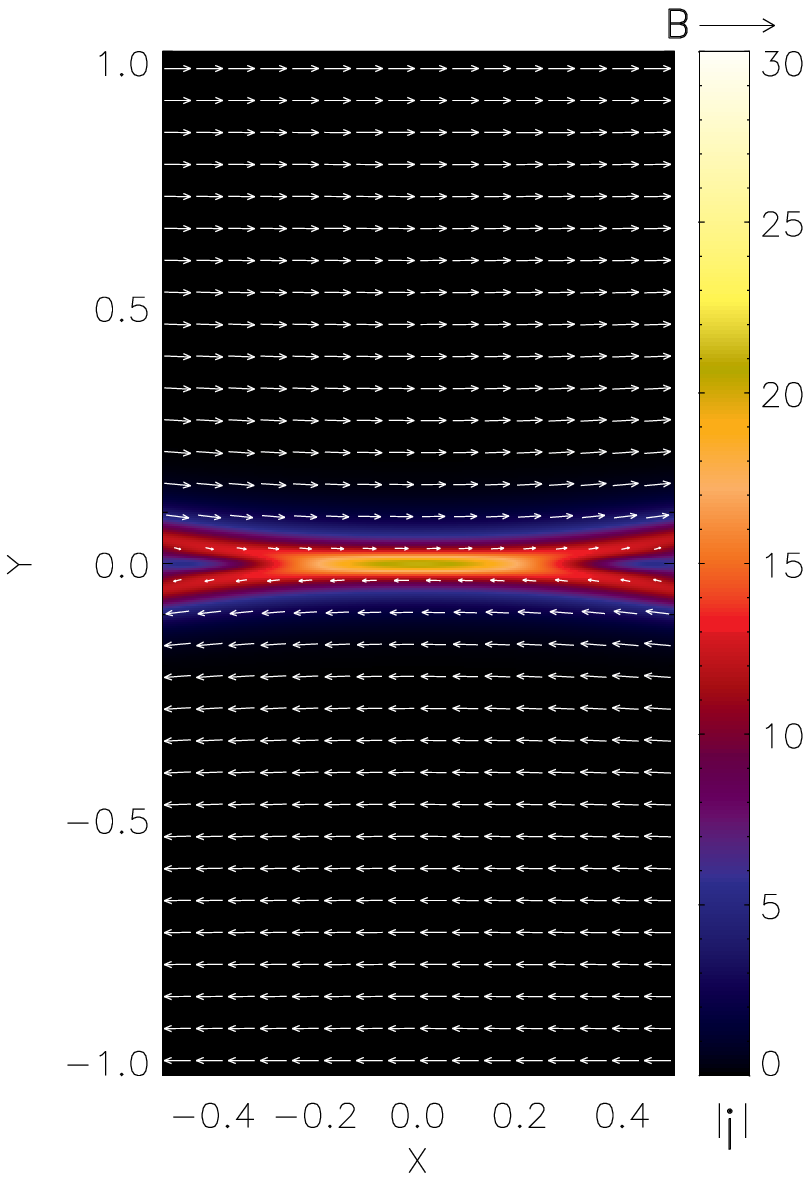}
\caption{Topology and strength of the velocity field (left panel) and magnetic
field (middle panel) during the Sweet-Parker reconnection at $t=7$.  The
strength is calculated from the components of $\vc{V}$ and $\vc{B}$
perpendicular to the normal vector of the XY-plane.  In the right panel we show
the absolute value of current density $|\vec{j}|$ overlapped with the magnetic
vectors.  The images show the XY-cut through the domain at $Z=0$ at time $t=7$
for a model with $B_{0z} = 0.1$, $\eta_u=10^{-3}$, $\eta_a=0.0$, and the
resolution 256x512x256. \label{fig:top_sp}}
\end{figure*}
\begin{figure*}
\center
\includegraphics[width=0.3\textwidth]{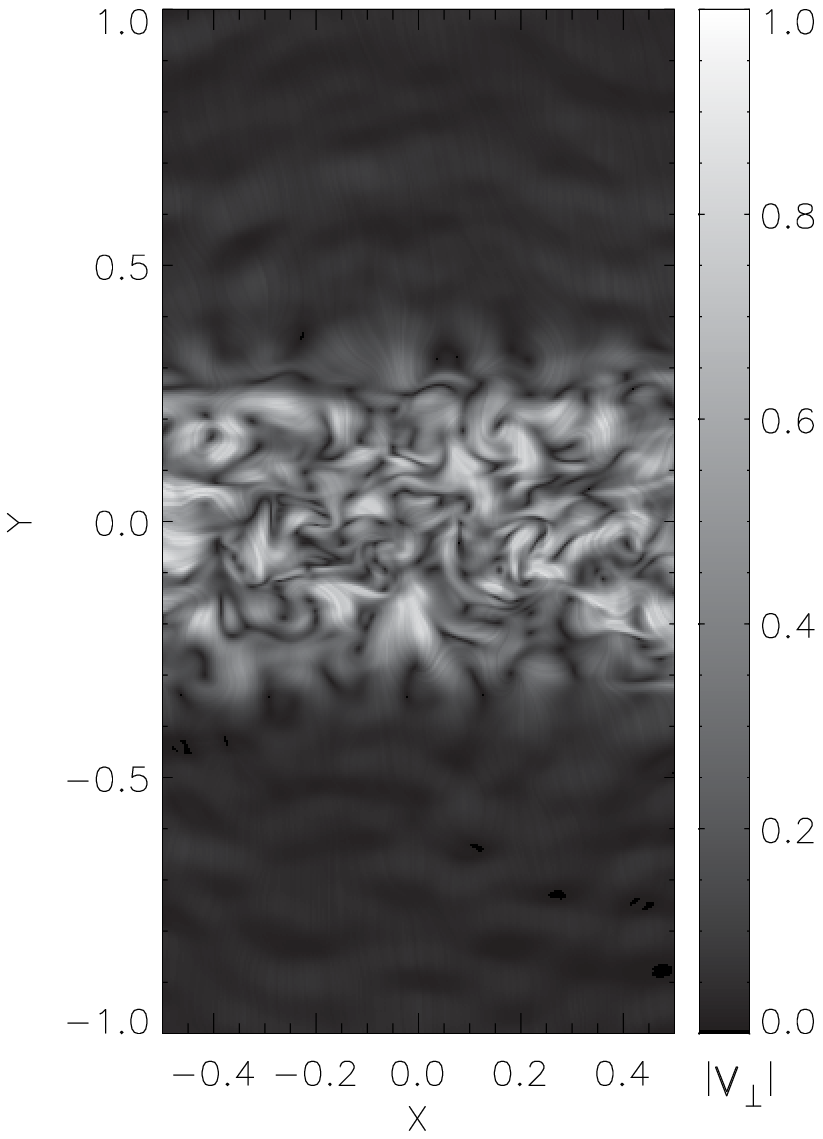}
\includegraphics[width=0.3\textwidth]{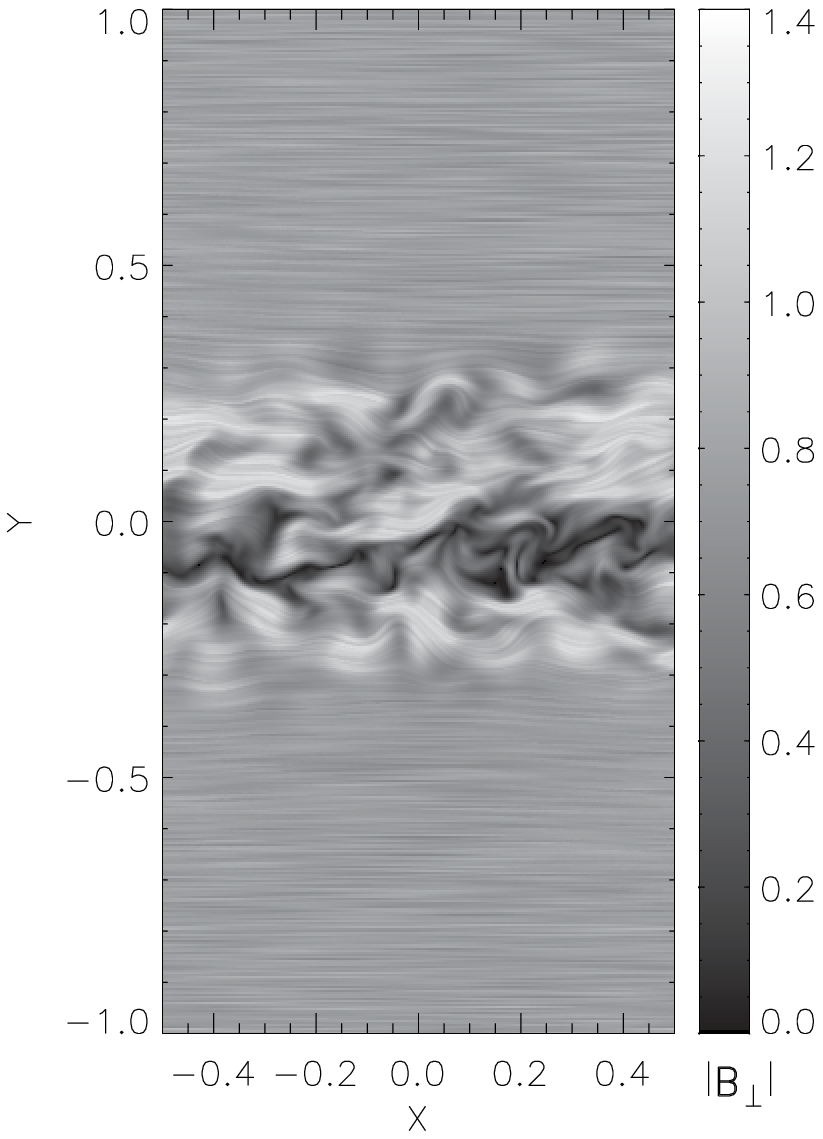}
\includegraphics[width=0.3\textwidth]{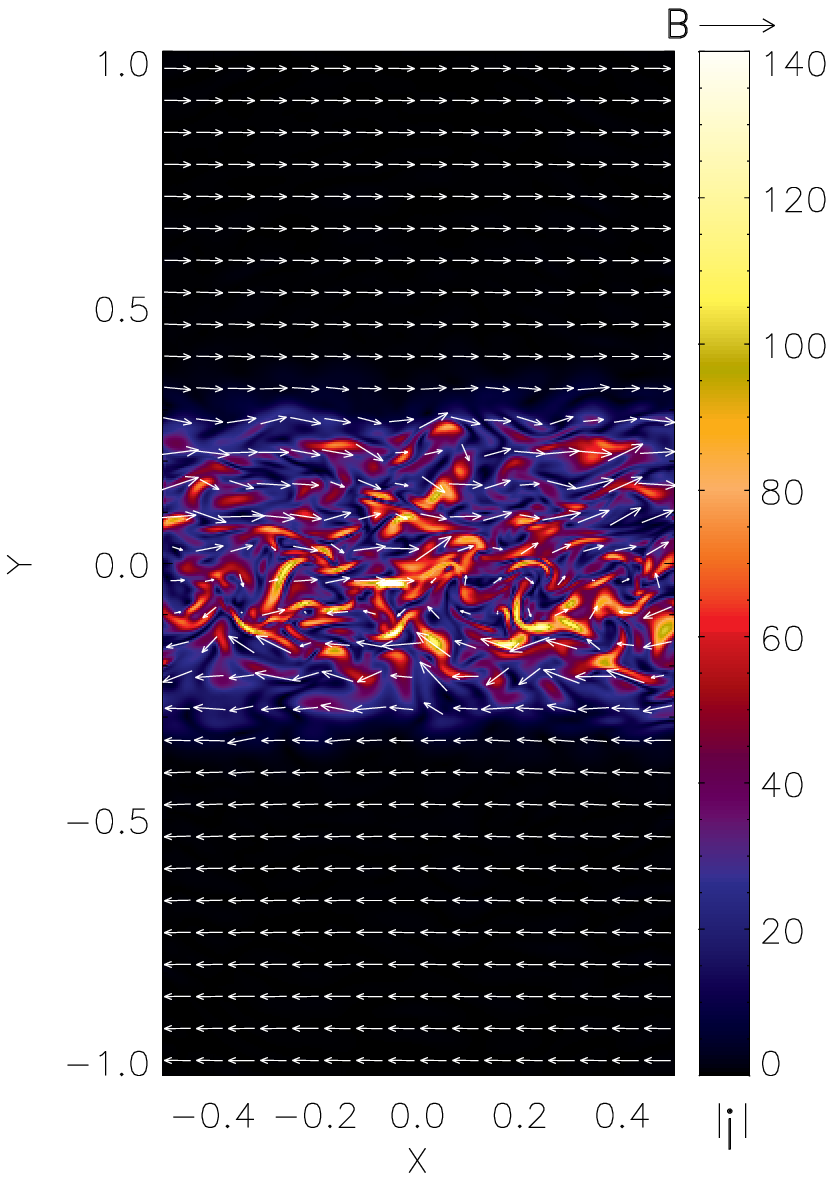}
\includegraphics[width=0.3\textwidth]{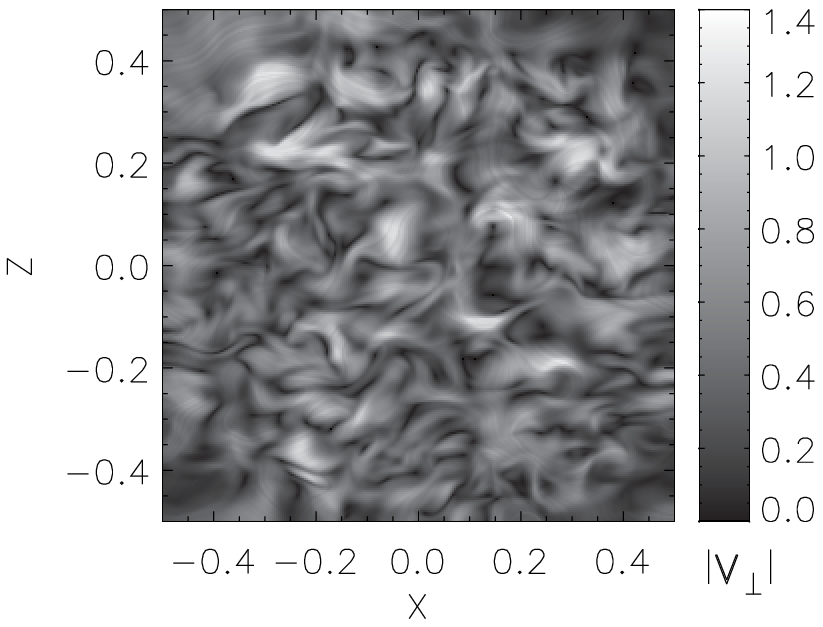}
\includegraphics[width=0.3\textwidth]{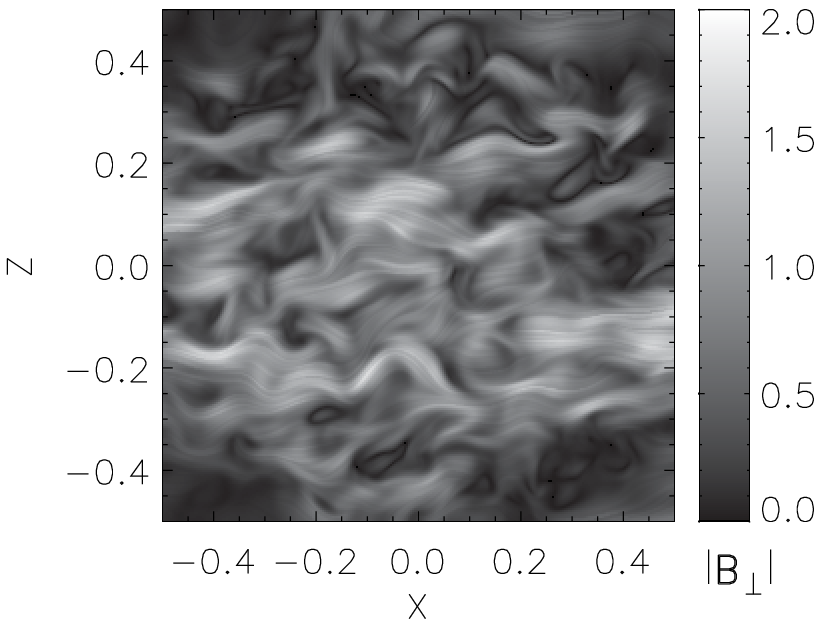}
\includegraphics[width=0.3\textwidth]{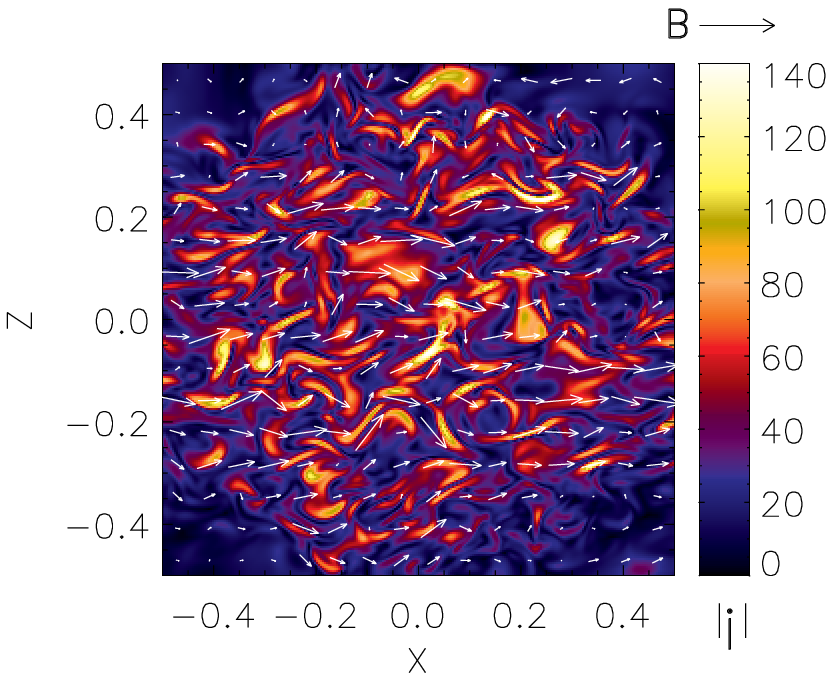}
\caption{Topology and strength of the velocity field (left panel) and magnetic
field (middle panel) in the presence of fully developed turbulence at time
$t=12$.  In the right panel we show distribution of the absolute value of
current density $|\vec{J}|$ overlapped with the magnetic vectors.  The images
show the XY-cut (upper row) and XZ-cut (lower row) of the domain at the midplane
of the computational box.  Turbulence is injected with power $P_{inj}=1$ at
scale $k_{inj}=8$.  Magnetic field reversals observed are due to magnetic
reconnection rather than driving of turbulence, which is subAlfv\'enic.
\label{fig:top_turb}}
\end{figure*}

Before we discuss the main results obtained in these studies, we present and
describe the configuration and topology of the velocity and magnetic fields in
the presence of turbulence.  In Figure~\ref{fig:top_turb} we show examples of
XY-cuts (upper row) and XZ-cuts (lower row) through the box of the velocity
(left panel) and magnetic field (middle panel) topologies with the intensities
corresponding respectively to the amplitude of perpendicular components of
velocity and magnetic field to a normal vector defining the plotted plane.

The first noticeable difference compared to the Sweet-Parker configuration is a
significant change of the velocity and magnetic field topologies.  Velocity has
a very complex and mixed structure near the midplane, since we constantly inject
turbulence in this region (see the left panel in Fig.~\ref{fig:top_turb}).
Although the structure is very complex, most of the velocity fluctuations are
perpendicular to the mean magnetic field.  This is because we are in the nearly
incompressible regime of turbulence (large plasma $\beta$, see
Tab.~\ref{tab:models}) and most of the fluctuations propagate as Alfv\'en waves
along the mean magnetic field.  Slow and fast waves, whose strengths are
significantly reduced, are allowed to propagate in directions perpendicular to
the mean field as well.  As a result most of the turbulent kinetic energy leaves
the box along the magnetic lines.  We observe, however, the efficient bending of
magnetic lines at the midplane (see the upper middle plot in
Fig.~\ref{fig:top_turb}).  This is not result of a driving, but result of
reconnection.  In general the interface between positively and negatively
directed magnetic lines is much more complex than in the case of Sweet-Parker
reconnection.  This complexity favors creation of enhanced current density
regions, where the local reconnection works faster since the current density
reaches higher values (see the right panel of Fig.~\ref{fig:top_turb}).  Since
we observe multiple reconnection events happening at the same time (compare the
right panel of Fig.~\ref{fig:top_turb} to the Sweet-Parker case in
Fig.~\ref{fig:top_sp}), the total reconnection rate should be significantly
enhanced.

\subsubsection{Dependence on Turbulence Strength}
\label{ssec:power}

\begin{figure*}
\center
\includegraphics[width=0.45\textwidth]{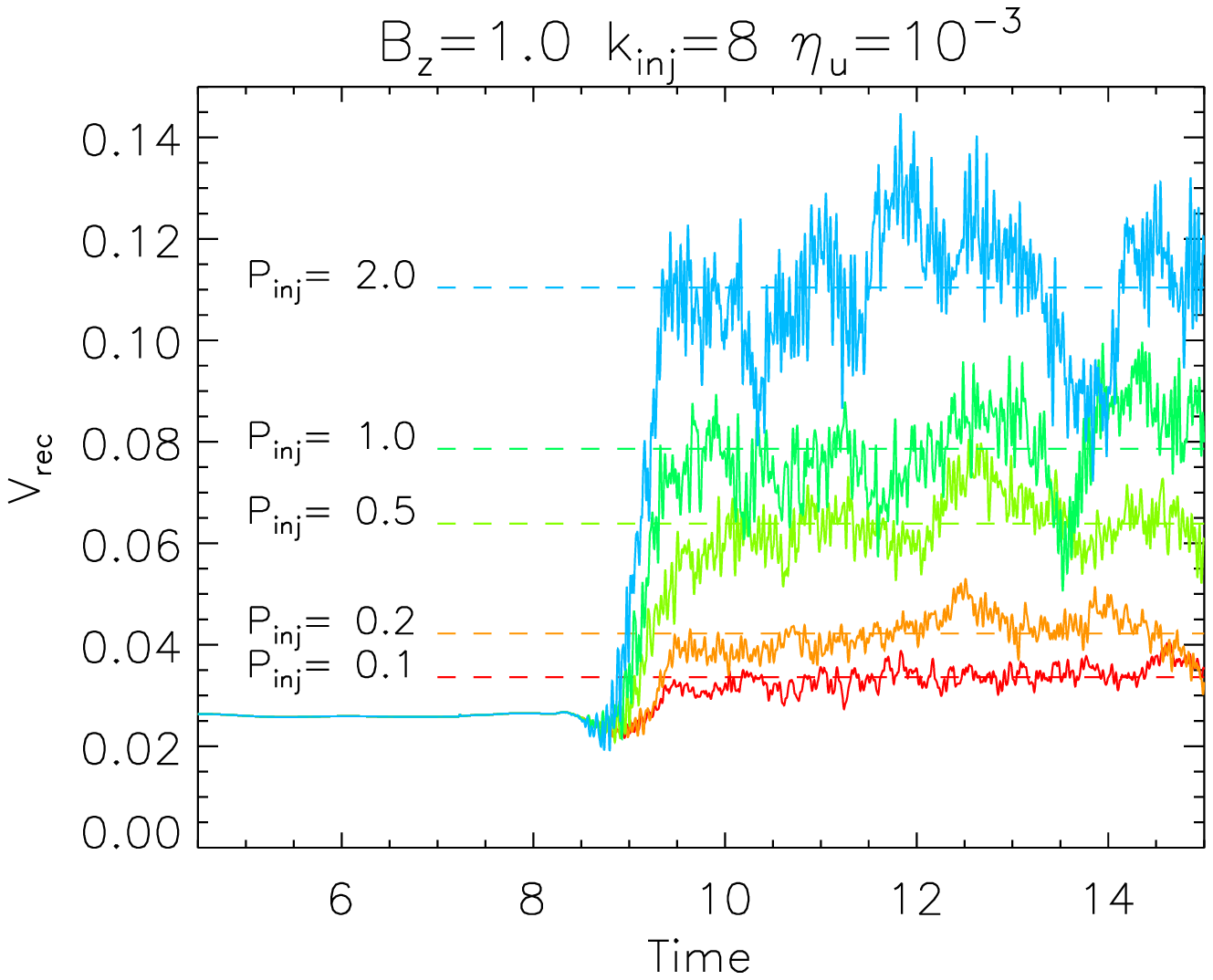}
\includegraphics[width=0.45\textwidth]{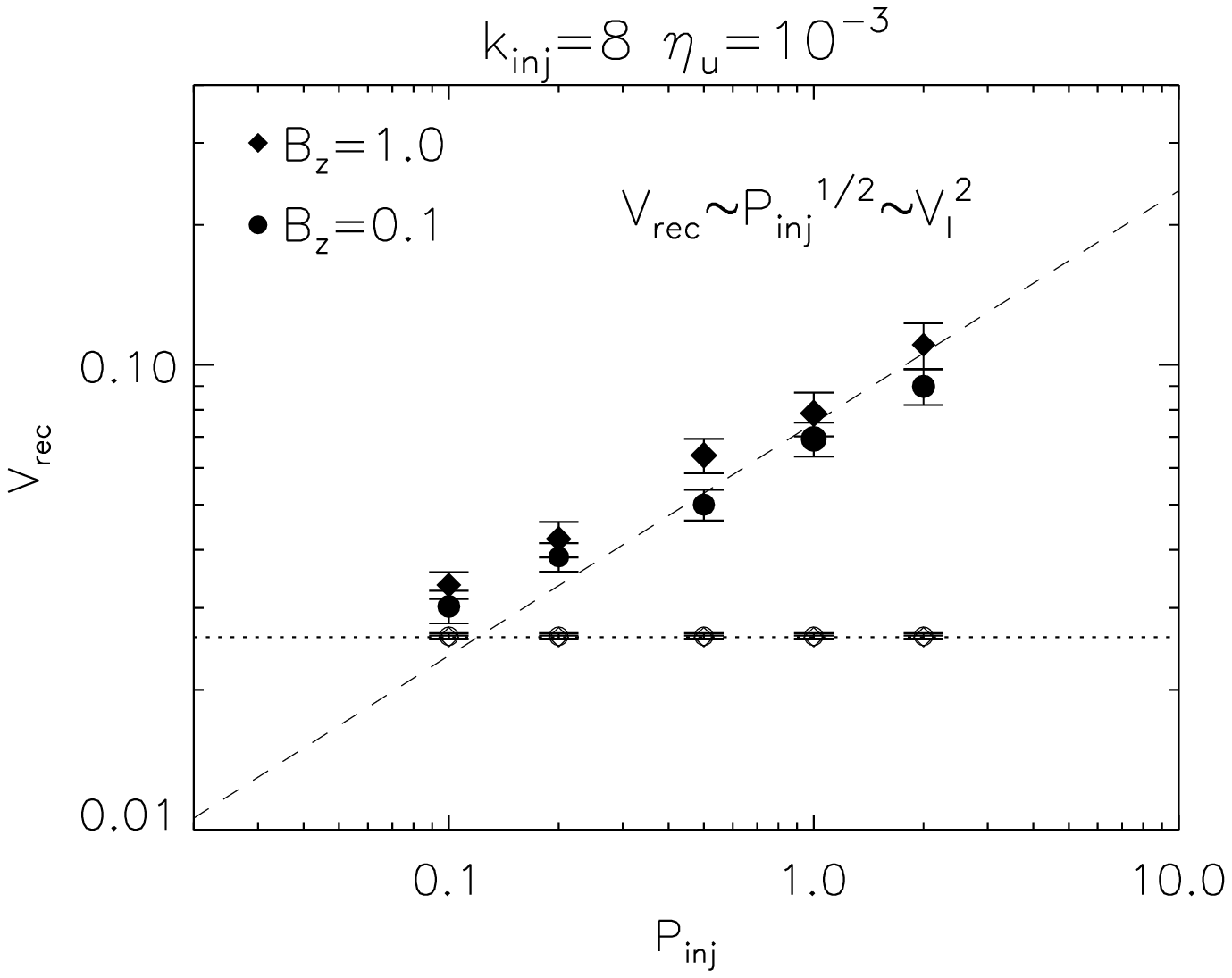}
\caption{{\em Left:} Time evolution of the reconnection speed $V_{rec}$ for
models PD (see Table~\ref{tab:models}) with different powers of turbulence
$P_{inj}$ (see legend).
{\em Right:} The dependence of the reconnection speed $V_{rec}$ on $P_{inj}$.
Error bars represent the time variance of $V_{rec}$.  The size of symbols
corresponds to the error of $V_{rec}$ (the way we calculate errors is described
in \S\ref{ssec:turbulent}).
\label{fig:pow_dep}}
\end{figure*}

\begin{figure*}
\center
\includegraphics[width=0.45\textwidth]{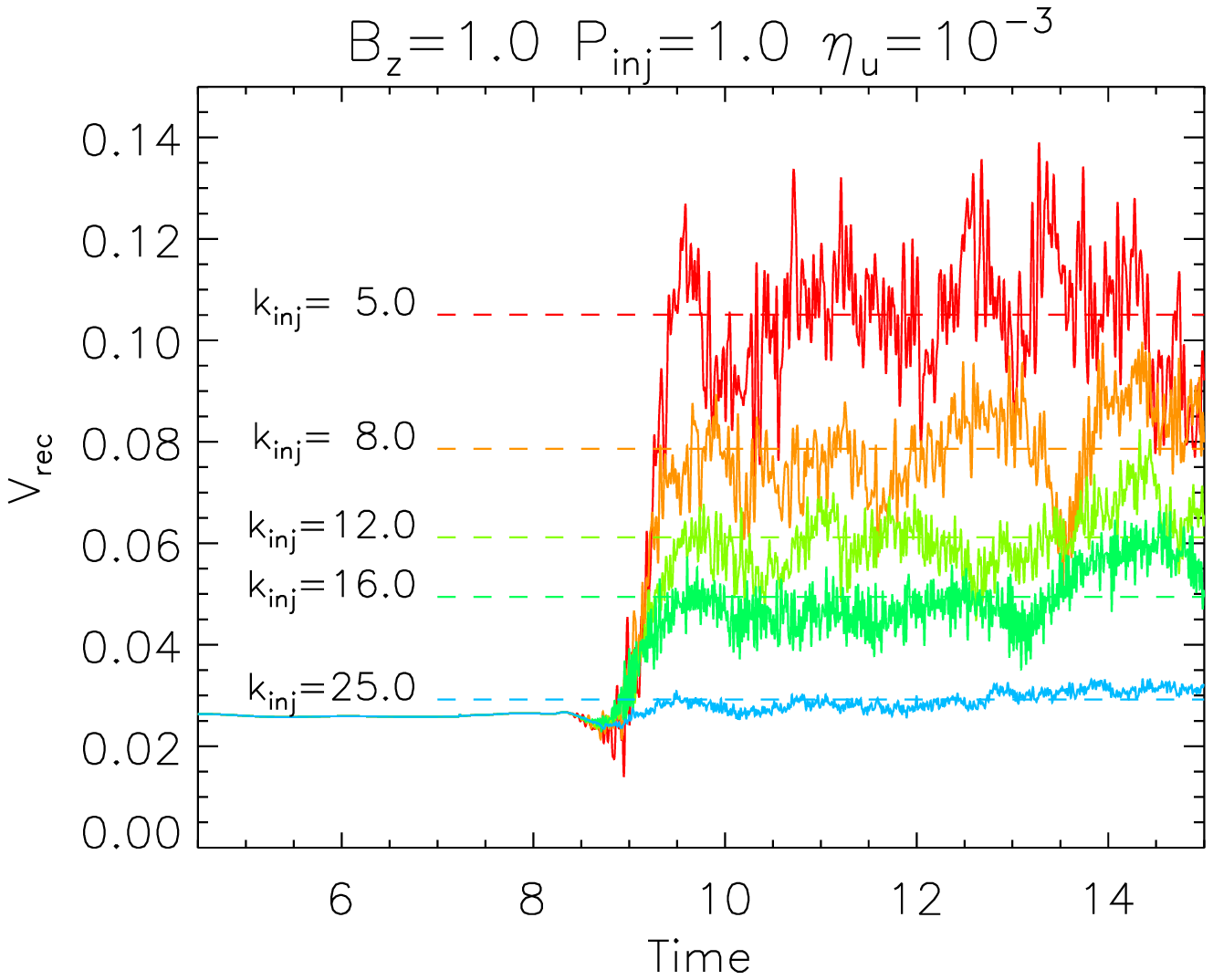}
\includegraphics[width=0.45\textwidth]{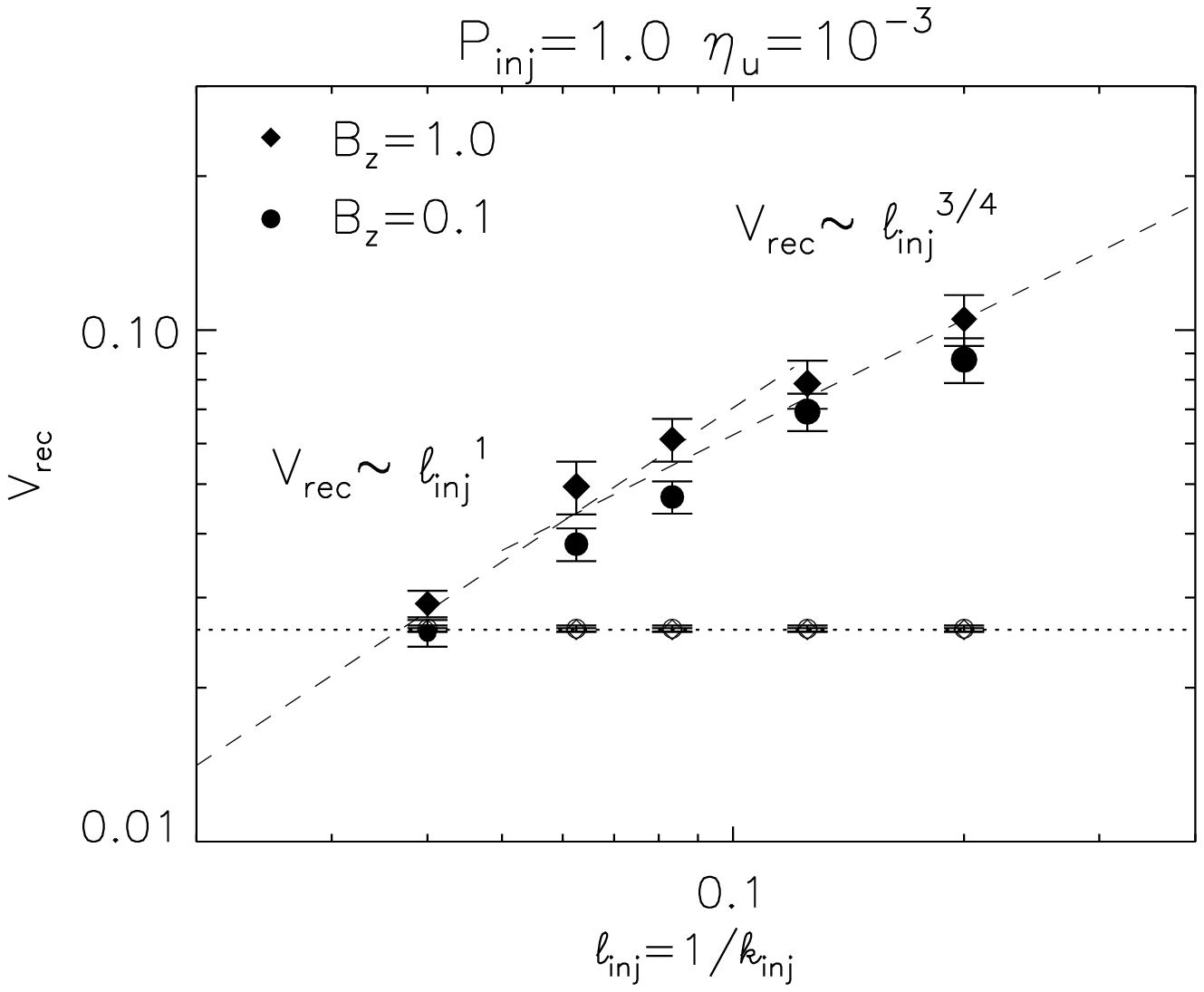}
\caption{{\em Left:} Time evolution of the reconnection speed $V_{rec}$ for
models SD (see Table~\ref{tab:models}) with different injection scale $l_{inj}$
(see legend).
{\em Right:} The dependence of the reconnection speed $V_{rec}$ on $l_{inj}$.
Error bars and the size of symbols have the same meaning as in
Fig.~\ref{fig:pow_dep}. \label{fig:sca_dep}}
\end{figure*}

\begin{figure*}
\center
\includegraphics[width=0.45\textwidth]{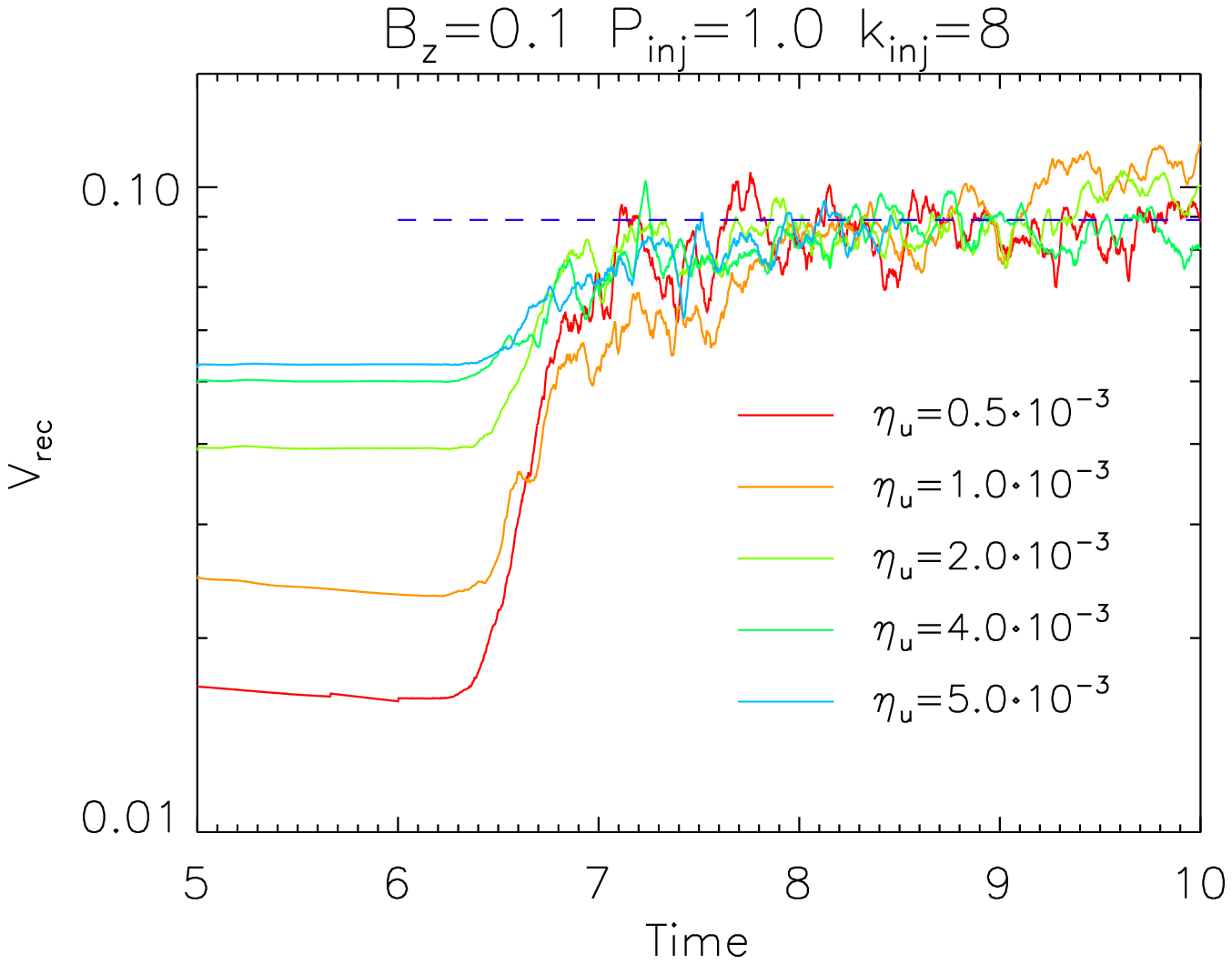}
\includegraphics[width=0.45\textwidth]{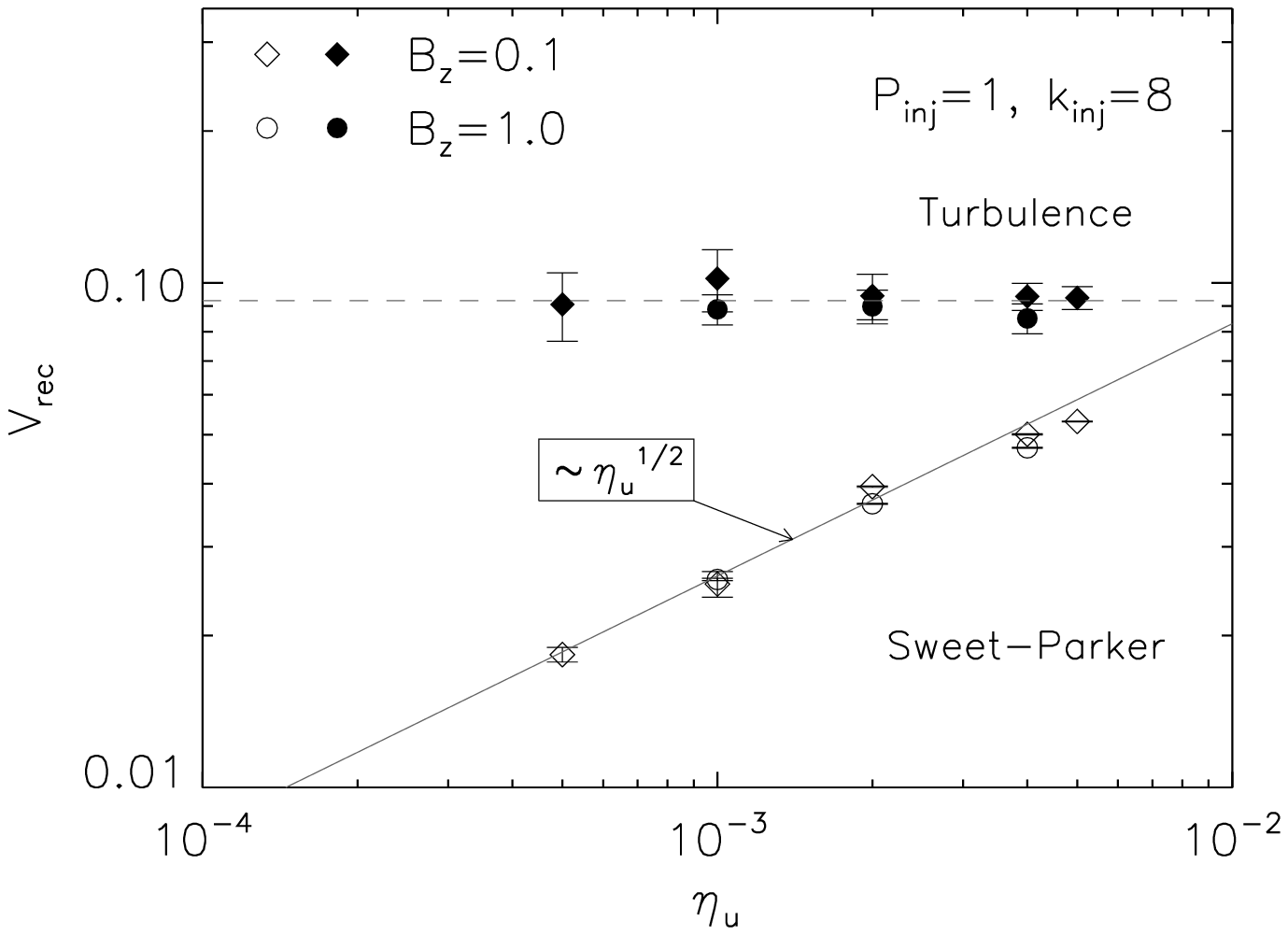}
\caption{{\em Left:} Time evolution of the reconnection speed $V_{rec}$ for
models RD (see Table~\ref{tab:models}) with different resistivity coefficients
$\eta_u$ (see legend).
{\em Right:} The dependence of the reconnection rate $V_{rec}$ on the uniform
resistivity $\eta_u$.  The solid line shows the theoretical dependence in
Sweet-Parker model, $V_{rec,SP} \sim \eta_u^{1/2}$.  Open symbols show
Sweet-Parker reconnection rate for different models.  Filled symbols show
reconnection rates $V_{rec}$ in the presence of turbulence.  Dashed line shows
the mean value of reconnection rate averaged over all models during the
turbulence stage (all filled symbols).  Error bars show the measure of time
variance of $V_{rec}$.
\label{fig:ueta_dep}}
\end{figure*}

We ran several models (models PD in Tab.~\ref{tab:models}) with varying powers
of turbulence.  All other parameters were kept the same, allowing us to estimate
the dependence of the reconnection rate $V_{rec}$ on the power of injected
turbulence $P_{inj}$.

Figure~\ref{fig:pow_dep} (left panel) shows the evolution of reconnection speed
in models with turbulent power $P_{inj}$ varying in range by more than one order
of magnitude, from 0.1 to 2.0.  The evolution of $V_{rec}$ reaches stationarity
in a relatively short period, about one Alfv\'en time unit, as estimated from
the plot.  We averaged $V_{rec}$ from $t=8.2$ to $t=12$.  From this plot we see
that the reconnection rates are relatively stable.  In the right panel of
Figure~\ref{fig:pow_dep} we plot the averaged reconnection speed over the
strength of turbulence.  Filled symbols represent the averaged reconnection rate
in the presence of turbulence.  Open symbols represent the reconnection rate
during the Sweet-Parker process, i.e. without turbulence.  The error bars show
the time variance of $V_{rec}$.  The size of symbols indicates the uncertainty
in our estimate of the reconnection speed $\Delta V_{rec, LV}$ normalized to the
uncertainty in the reconnection speed during the Sweet-Parker evolution $\Delta
V_{rec, SP}$ using formula $size = 2.0 - \ln \Delta V_{rec, LV} / \ln \Delta
V_{rec, SP}$.  Meaning that if $\Delta V_{rec, LV}$ is of order of $\Delta
V_{rec, SP}$ their symbols have the same size.

Fitting a power law to the calculated points indicates that the reconnection
speed $V_{rec}$ scales as $\sim P_{inj}^{1/2}$, as expected from the LV99
prediction (see Eq.~\ref{eq:scaling}).

\subsubsection{Dependence on Injection Scale}
\label{ssec:injection_scale}

We performed similar studies to determine the dependence of the reconnection
speed $V_{rec}$ on the scale at which we inject turbulence, $l_{inj}$.  Keeping
the same power of turbulence for all models SD (see Tab.~\ref{tab:models}) we
inject turbulence at several scales, from $k_{inj} = 5$ to $k_{inj} = 25$.  At
the upper end of this range the turbulence barely broadens the Sweet-Parker
current sheet. At the lower end the turbulent eddies are barely contained within
the volume within which we excite turbulent motions.

In Figure~\ref{fig:sca_dep} (left panel) we present the results for this series
of models.  From the plot we clearly see a strong dependence of the reconnection
rate on the injection scale.  After the injection of turbulence, models with
smaller power injection scales reach smaller values of the reconnection rate
$V_{rec}$.  In the model with the largest injection scale $k_{inj}=5$,
reconnection is 5 times faster than the Sweet-Parker rate.  After averaging over
time, we plot the dependence of $V_{rec}$ on the injection scale in the right
plot of Figure~\ref{fig:sca_dep}.

The prediction of LV99 model is that the reconnection rate should scale as $\sim
l_{inj}$.  In the case of a strong guide field we see that this relation is
observed for small injection scales.  However, when $l_{inj}$ is comparable with
length of the volume of driven turbulence the dependence becomes flatter.  For a
weak guide field the dependence is flatter from the very beginning and is better
fit by $V_{rec} \sim l_{inj}^{3/4}$.  This difference requires further
investigation.

We see several possible sources for the discrepancy.  For instance, the
existence of a turbulent inverse cascade can modify the effective $l_{inj}$.  In
addition, reconnection can also modify the characteristics of turbulence, such
as the power spectrum and anisotropy.  We shall study this issue elsewhere, but
within this work get satisfied with the qualitative agreement of the predictions
and the numerical simulations.

\subsubsection{Dependence on Resistivity}
\label{sec:resistivity}

In the global constraint on the reconnection rate (Eq.~\ref{eq:constraint})
derived in LV99, there is no explicit dependence on the resistivity.  In order
to test this, we performed another set of models (RD in Tab.~\ref{tab:models})
in which we change the uniform resistivity $\eta_u$ only.  We expect that in
these models we should see at early times the theoretical dependence of the
Sweet-Parker reconnection rate ($V_{rec,SP} \sim \eta_u^{1/2}$), and at late
times, after we introduce turbulence, a complete lack of any dependence on
resistivity.

In the left panel of Figure~\ref{fig:ueta_dep} we show the evolution of the
reconnection rate for a subset of RD models with $B_{0z} = 0.1$ and with
$\eta_u$ varying from $5 \cdot 10^{-4}$ to $5 \cdot 10^{-3}$.  In this plot we
can recognize a strong dependence of the Sweet-Parker reconnection on the
uniform resistivity, while in the presence of turbulence the reconnection speed,
even though it fluctuates, reaches roughly the same level for all models.  This
indicates that, as expected, there is little or no dependence of the
reconnection rate on resistivity in the presence of turbulence.

In the right panel of Figure~\ref{fig:ueta_dep} we show reconnection rates
obtained from the set RD of models with a weak and strong guide fields, $B_{0z}
= 0.1$ and $B_{0z} = 1.0$, respectively.  The open symbols show the rates of
laminar reconnection $V_{rec, SP}$.  They follow precisely the theoretical
relation $V_{rec,SP} \sim \eta_u^{1/2}$.  Evolving the models further in the
presence of turbulence allowed us to test the dependence of the reconnection
rates on resistivity in the presence of turbulence.  We plot these results using
filled symbols.  We see that there is virtually no dependence on $\eta_u$.
Moreover, we see that this property does not depend on the strength of $B_{0z}$.

\begin{figure}
\center
\includegraphics[width=0.45\textwidth]{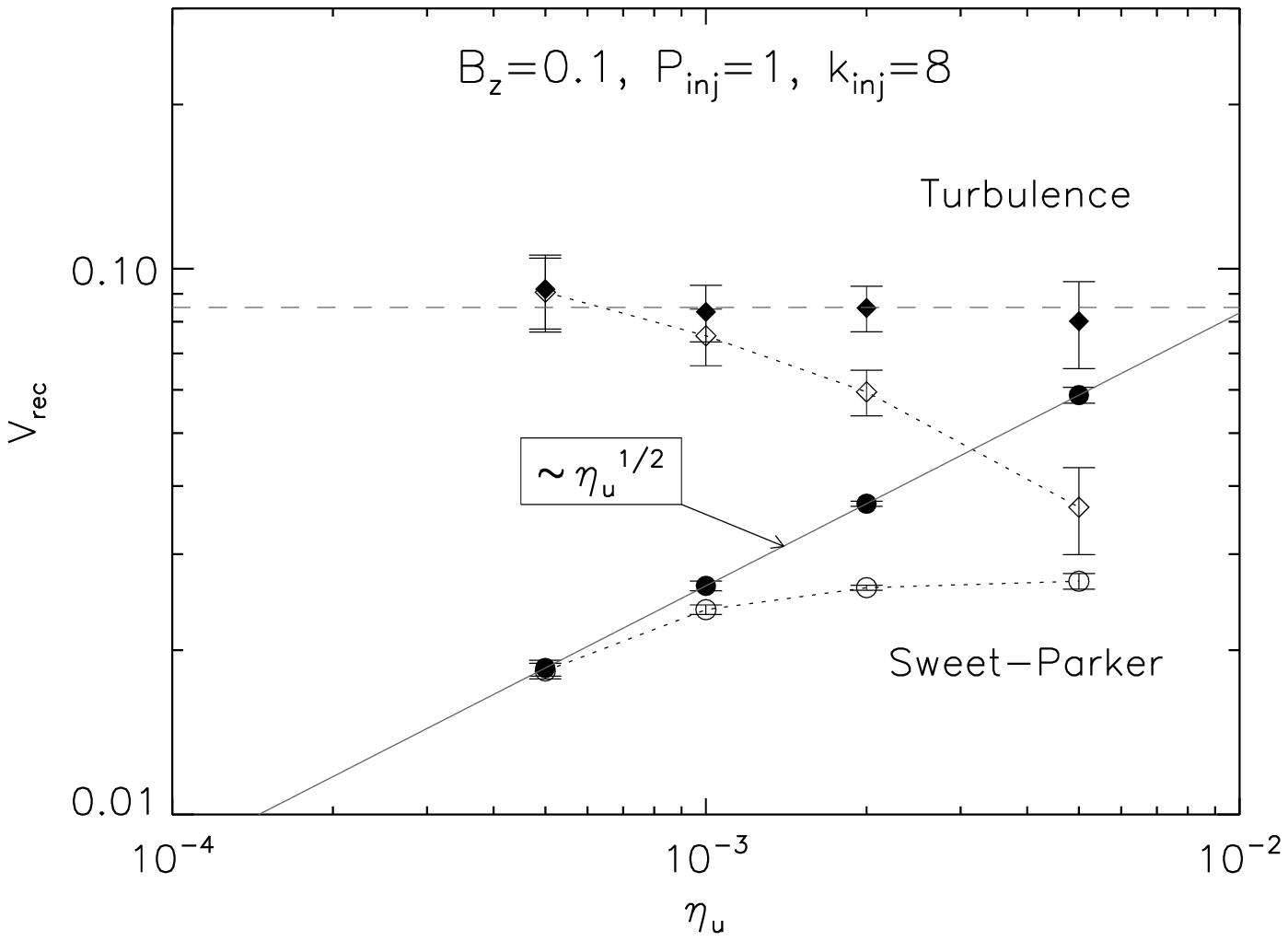}
\caption{Dependence of the reconnection rate $V_{rec}$ on the uniform
resistivity $\eta_u$ without the decaying resistivity zone.  The solid line
shows the theoretical dependence in Sweet-Parker model, $V_{rec,SP} \sim
\eta_u^{1/2}$.  Open circles show Sweet-Parker reconnection rate for different
models.  Open diamonds show reconnection rates $V_{rec}$ in the presence of
turbulence.  Filled symbols show values corrected by the factor calculated from
the theoretical dependence (see text). \label{fig:ueta_dep_corr}}
\end{figure}

These results were obtained in models with a decaying zone of resistivity near
the X boundary.  In order to exhibit the importance of properly treating the
resistive terms at the outflow boundary conditions we show in
Figure~\ref{fig:ueta_dep_corr} values obtained in models with regular open
boundary conditions without a zone of decaying resistivity.  Here, the open
circles show the rates of laminar reconnection with no turbulence.  We clearly
see a growing departure of these results from the theoretical prediction,
$V_{rec,SP} \sim \eta_u^{1/2}$ (solid line) for models with large values of
uniform resistivity.  For $\eta_u \gtrsim 2 \cdot 10^{-3}$, laminar reconnection
seems to be insensitive to the value of resistivity.  This effect is purely
numerical and caused by improper handling of boundary conditions when the
resistive term in the induction equation starts to dominate.  Open diamonds in
Figure~\ref{fig:ueta_dep_corr} show the corresponding rates for models with
turbulence.  Even though initially, for smaller values of resistivity, the
reconnection rate seems to be independent of $\eta_u$, later on, when the
resistivity is larger, it starts to decay with $\eta_u$.  The problem of
boundaries affects the turbulent case as well.

We can test our ability to compensate for the affect of large resistivity on the
boundary conditions using the theoretical dependence $V_{rec, SP} \sim
\eta^{1/2}$.  If we calculate correction factors by taking the ratio of
$\eta^{1/2} / V_{rec, SP}$, we can use these coefficients to correct
reconnection rate during the turbulent stage.  The result is plotted with the
filled symbols.  The relation signifies virtually no dependence of the $V_{rec}$
on $\eta$.

\begin{figure}[t]
\center
\includegraphics[width=0.45\textwidth]{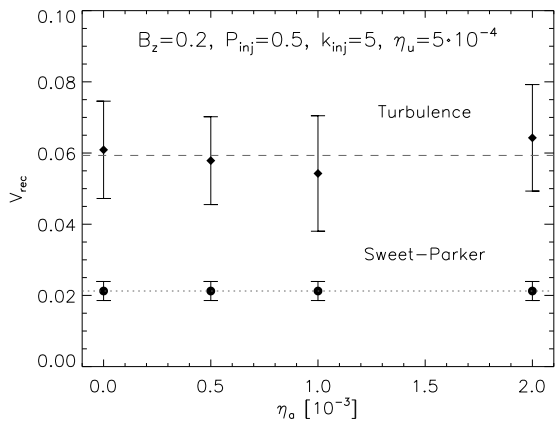}
\caption{Dependence of the reconnection rate $V_{rec}$ on the anomalous
resistivity $\eta_a$.  Two horizontal lines show the mean reconnection rate
averaged over all plotted models during the Sweet-Parker stage and in the
presence of turbulence (dotted and dashed lines, respectively).  The critical
value of the current density $j_{crit}$ is set to 25.0 in all models plotted
here. \label{fig:aeta_dep}}
\end{figure}

In addition to the uniform resistivity dependence, we have studied the
dependence on anomalous effects as well.  The results of these studies are
presented in Figure~\ref{fig:aeta_dep} and show four models with the same
uniform resistivity $\eta_u = 5 \cdot 10^{-4}$ and the critical current density
$j_{crit} = 25.0$ but with the anomalous resistivity parameter $\eta_a$ varying
between $0.0$ and $2 \cdot 10^{-3}$ (see models AD in Tab.~\ref{tab:models}).  In
Figure~\ref{fig:aeta_dep} we plot the dependence of the reconnection rate on the
anomalous resistivity parameter $\eta_a$.  We see that also the reconnection
speed $V_{rec}$ is insensitive to the value of $\eta_a$ to within the variations
of the reconnection rate in each model (see the error bars).

In this section we shown that the results of our resistivity studies agree with
the LV99 model and indicate no sensitivity to Ohmic resistivity in the presence
of turbulence.  This is important for two reasons.  First, it supports the
stochastic reconnection model proposed in LV99.  Second, it gives reason to
believe that simulations of reconnection and MHD turbulence in astrophysical
objects can be successfully applied to real objects, despite the enormous
difference between the magnetic Reynolds numbers that can be simulated and the
values that actually obtain in stars, accretion disks, and galaxies.

\subsection{Role of the Guide Field}
\label{sec:guide}

The robustness of the LV99 model is determined by its application to any
configuration of the reconnecting field.  Usually textbooks describe the
Sweet-Parker model as having opossitely directed field lines, which undergo the
reconnection process in the diffusion region and are perfectly antiparallel,
since this model is strictly two dimensional.  This situation is very
particular.  In reality, even in the absence of turbulence, magnetic field lines
can enter the diffusion region at different angles $\alpha$ (see right panel of
Fig.~\ref{fig:setup_planes}), or in other words, an uniform component parallel
to the reconnecting field could be present.  This component is called a guide
field and in our simulations is determined by the value of $B_{0z}$.

Direct simulations of the LV99 model with a strong guide field of order of the
reconnecting component strength encounter problems when the Z boundaries are
open.  In our picture, the simulation domain is a box embedded in a large scale
configuration of magnetic field around the interface of two volumes
characterized by the different direction of magnetic lines.  In this picture,
magnetic lines crossing our domain extend to infinity.  This means that any
force acting on a field line in the domain will feel the tension of that line,
so the general configuration of the field line will not change.  It only
encounters a small local perturbation.  When the open boundary conditions are
applied, the force acting on a magnetic line is not bounded by the magnetic
tension anymore, and in the presence of shear flows generated by turbulence,
magnetic line ends attached to the boundaries can slip along them changing the
global topology of magnetic field.  Moreover, the change of direction of the
field transforms $B_z$ into $B_x$ component, which suppresses the inflow of
fresh unreconnected flux.  This is an artificial and undesirable situation.

There are two solutions to this problem.  First, we can construct a special type
of open boundary conditions, which fixes magnetic lines at some distance far
from the box.  Then, at the boundary, we would have to extrapolate the magnetic
field lines and calculate the required tension from this condition.  This
approach, however, seems to be complex and requires additional conditions
assumed {\em a priori} at the boundaries.  Thus, we have chosen another approach
by applying periodic boundary conditions along the Z direction.  This is much
simpler and does not limit our model, i.e. it does not influence the inflow and
outflow of the magnetic flux.

\begin{figure}
\center
\includegraphics[width=0.45\textwidth]{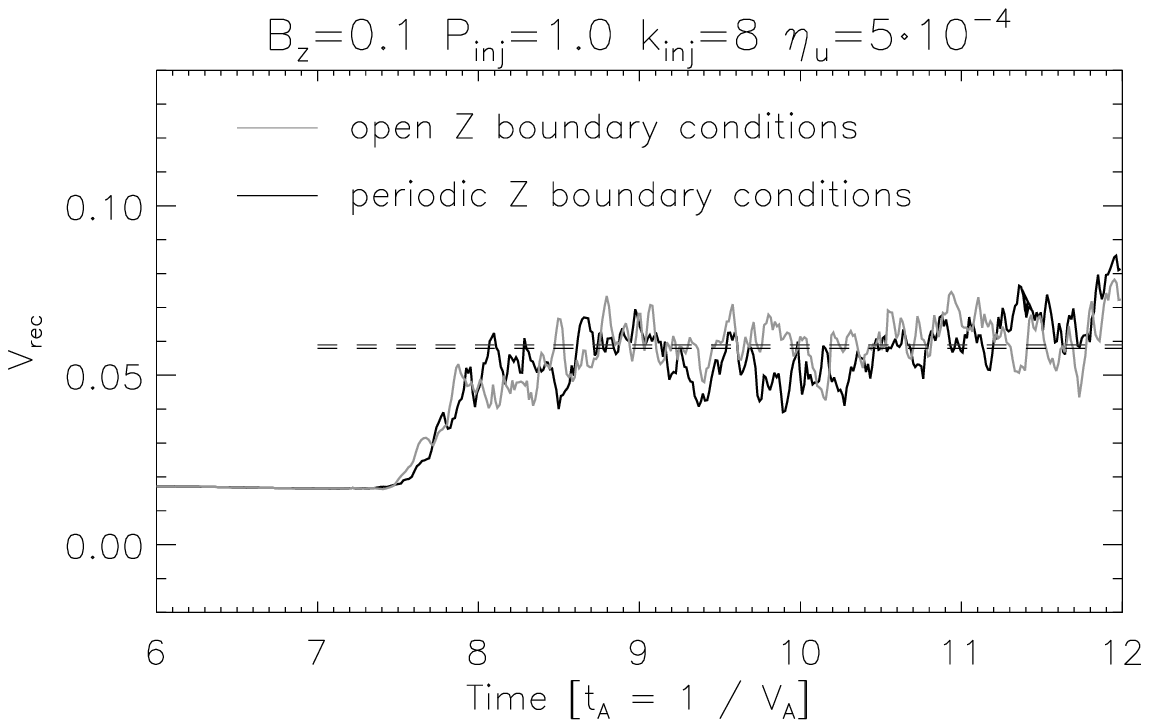}
\includegraphics[width=0.45\textwidth]{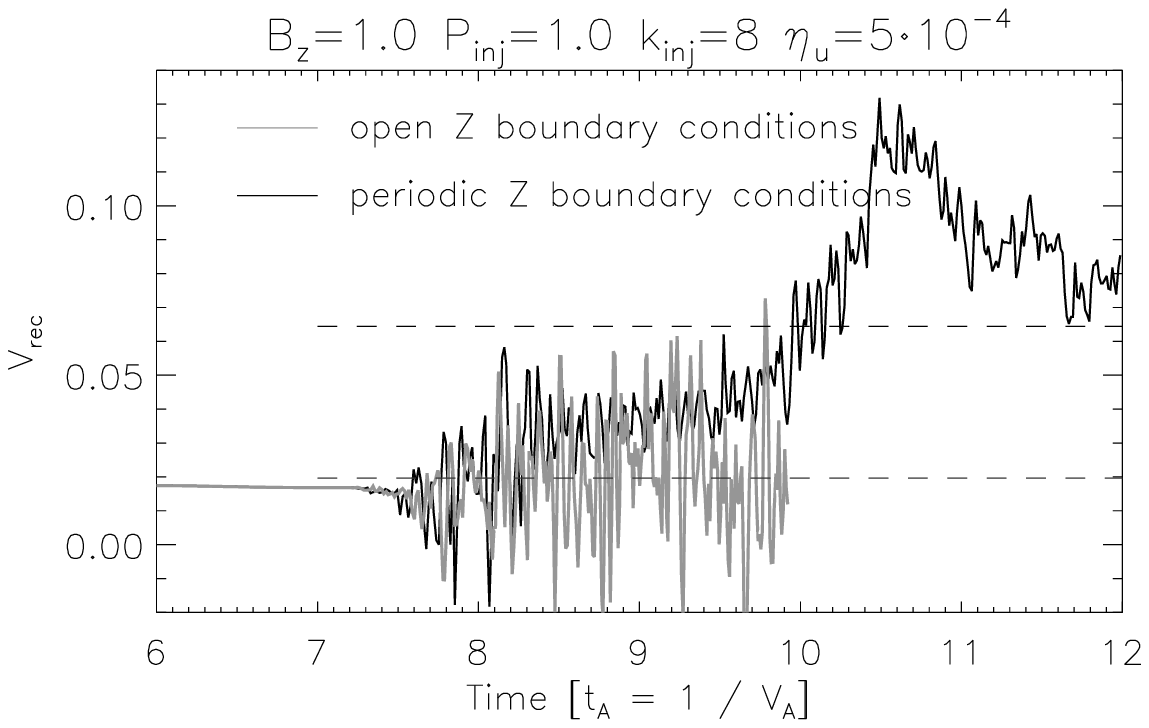}
\caption{Demonstration of the importance of boundary condition along
Z-direction.
{\em Upper:} Comparison of two models with the same conditions but different
types of Z boundaries for a weak $B_{0z}$.
{\em Lower:} Comparison of two models with the same conditions but different
types of Z boundaries for a strong $B_{0z}$. \label{fig:bz_open_vs_per}}
\end{figure}

In Figure~\ref{fig:bz_open_vs_per} we show the comparison of $V_{rec}$ for
models with a weak (top) and strong (bottom) guide fields $B_{0z}$.  Each plot
contains two otherwise identical models, but one with open boundaries along Z
direction and another with periodic boundaries.  As we see, in the case of weak
$B_{0z}$ there is no difference if we apply open or periodic boundaries.  In
both models, mean values of reconnection speed during the presence of turbulence
are almost identical (see black and grey dashed lines in top panel of
Fig.~\ref{fig:bz_open_vs_per}).  However, once we start increasing the strength
of the guide field, our choice of the boundary conditions starts to be
important.  In the case of open boundaries along the Z direction, the mean
reconnection rate is suppressed by the strong turbulent shear of $B_z$, which is
clearly seen in the bottom plot of Figure~\ref{fig:bz_open_vs_per}.  Applying
periodic boundaries along the Z direction we can restore the fast reconnection
rate, but the system needs more time to adjust to a steady state.  This is
caused by the fact that after starting injection of turbulence, part of the
injected kinetic energy is converted into magnetic energy, thus the relaxation
of the system takes place more slowly than in the case of weak $B_z$.  This
demonstration shows that the choice of applied boundary conditions along the Z
direction is important and has a crucial influence on the reconnection rate, at
least in the case of strong magnetic guide field.

\begin{figure}
\center
\includegraphics[width=0.45\textwidth]{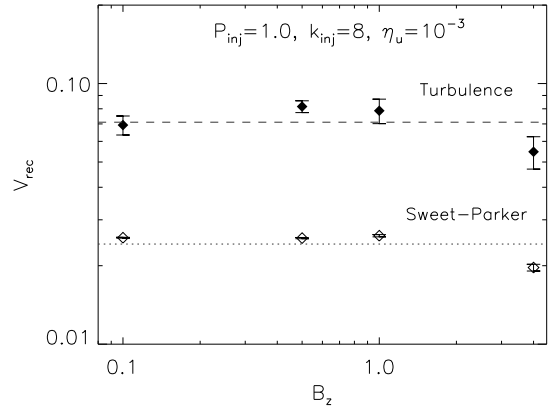}
\caption{Dependence of the reconnection rate $V_{rec}$ on the strength of the
guide field $B_{0z}$. \label{fig:bz_dep}}
\end{figure}

In Figure~\ref{fig:bz_dep} we show four models (set BD in Tab.~\ref{tab:models})
with different strengths of the guide field $B_{0z}$.  This plot shows that the
reconnection speed is comparable over a broad range of $B_{0z}$ (and the angle
$\alpha$), which means that reconnection will work with similar efficiency for
any configuration of magnetic field in which the lines of the magnetic field
enter the diffusion region at different angles.  The studied models with $0.1
\le B_{0z} \le 1.0$ cover range of $\alpha$ from 0$^\circ$ to 90$^\circ$.  For a
given power input a larger guide field is expected to correspond to a smaller
amount of field wandering and therefore a smaller reconnection speed (see the
model with $B_{0z} = 4.0$ corresponding to $\alpha \approx 152^\circ$).  We,
however, do not see a prominent dependence on the shared component of magnetic
field, which calls for further studies of the effects of the strong guide
field.

\subsection{Comparison of the Old and New Measures}
\label{sec:comparison}

\begin{figure*}
\center
\includegraphics[width=0.45\textwidth]{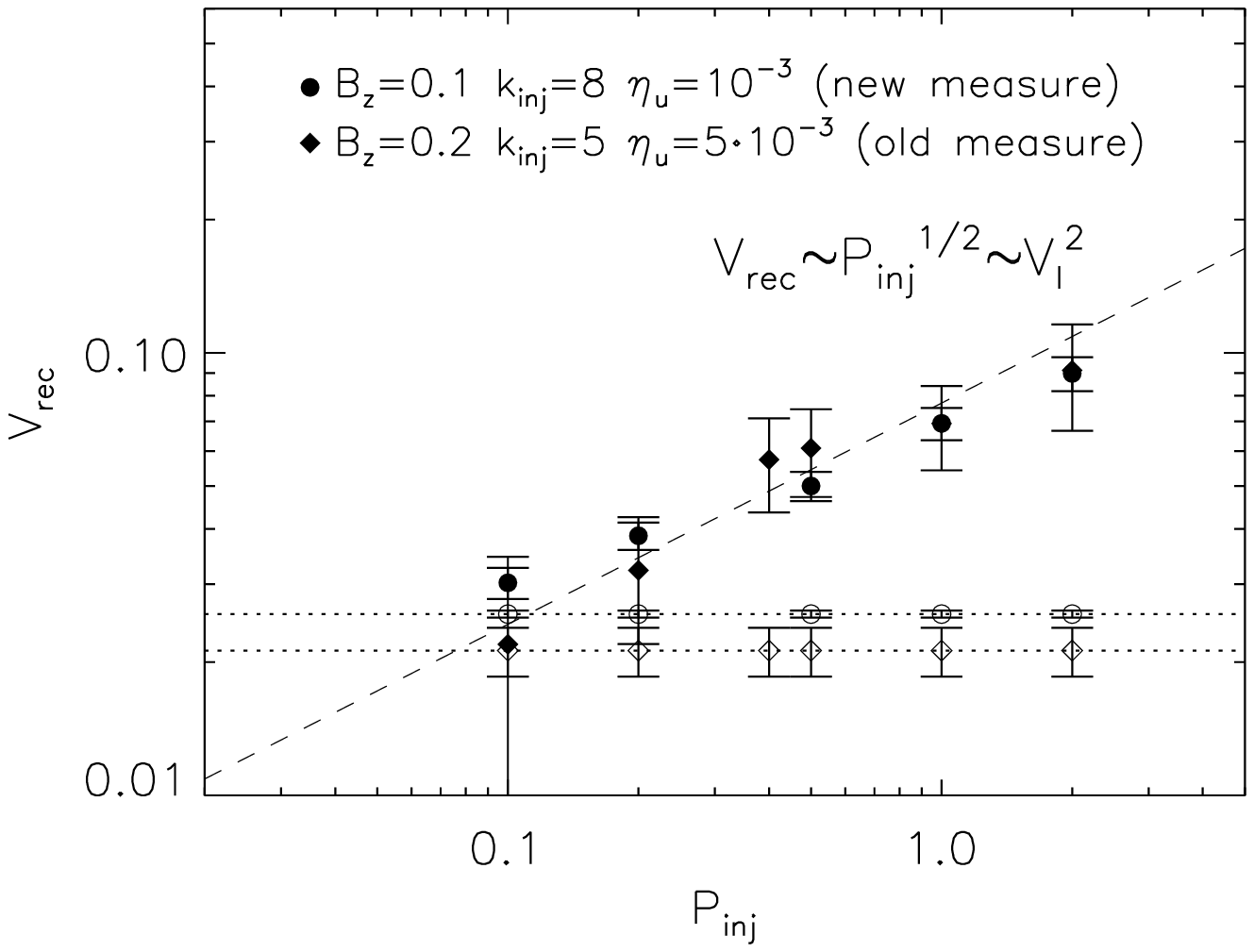}
\includegraphics[width=0.45\textwidth]{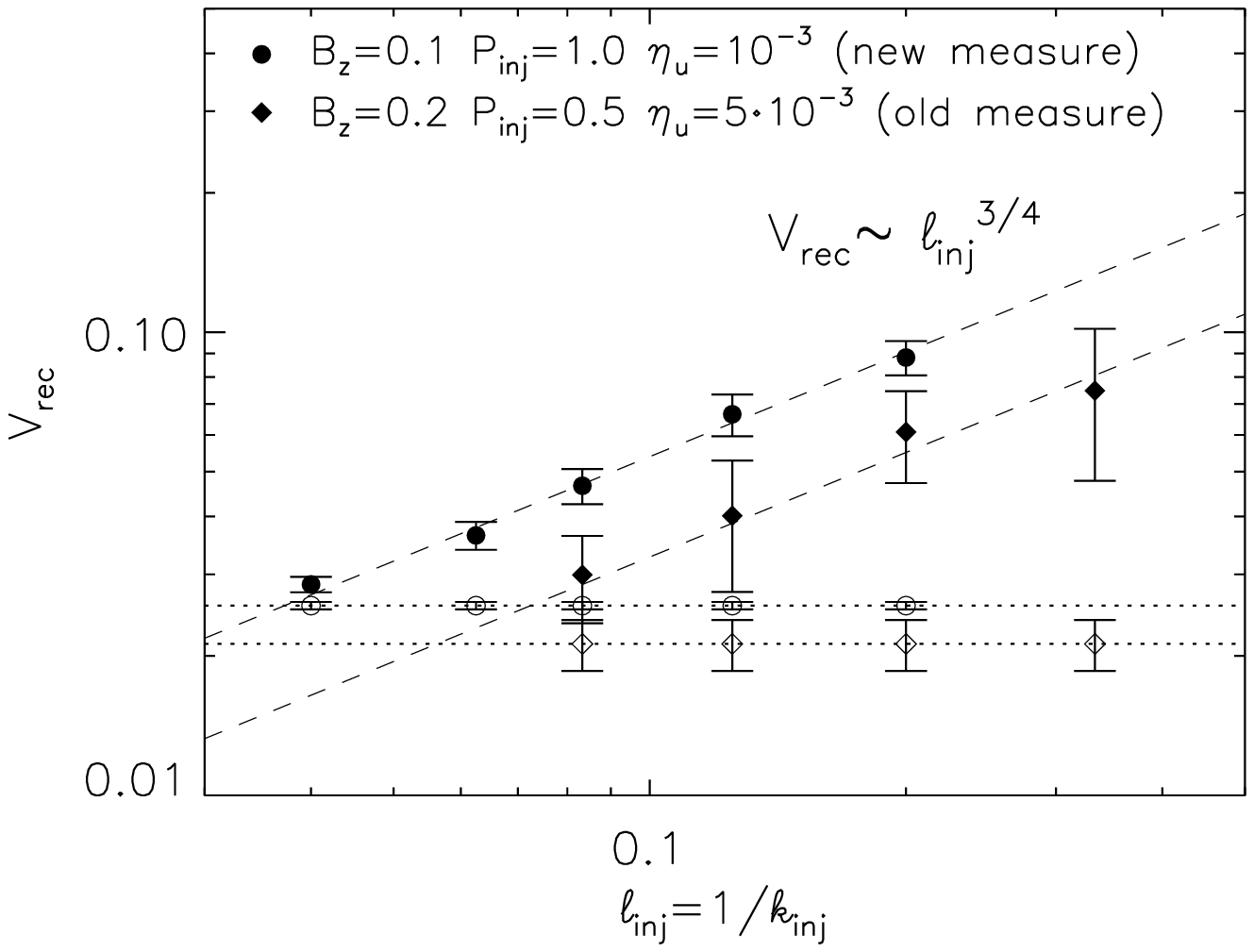}
\caption{Comparison of the old and new reconnection rate estimates.  In the left
plot we show the dependence of the reconnection rate $V_{rec}$ on the power of
turbulence $P_{inj}$ (diamonds) using both methods.  The corresponding
Sweet-Parker rates, without the presence of turbulence, are shown using
x-symbols.  In the right plot we show the dependence of the reconnection rate
$V_{rec}$ on the injection scale $l_{inj}$ (diamonds).  Again, the Sweet-Parker
rates are shown using x-symbols. \label{fig:deps}}
\end{figure*}

In Figure~\ref{fig:deps} we show a comparison of the reconnection rate obtained
using the old and new measures as a function of the power of turbulence (left
plot) and the injection scale (right plot).  In the case of the turbulent power
dependence we used two different sets of models for each method.  For the old
method of $V_{rec}$ estimation the guide field $B_z=0.2$, and the uniform
resistivity $\eta=5 \cdot 10^{-4}$.  The guide field in new models is set to
0.1.  The injection scales are $k_{inj}=5$ and 8, for the old and new models,
respectively.  Dashed line shows the LV99 dependence $V_{rec} \sim V_T^2$.

The old measure models were calculated with a more dissipative scheme based on
the HLL solver instead of HLLD, with a higher value of numerical resistivity.
During the Sweet-Parker stage we note some difference in the values of
reconnection rate between two sets of models resulting from the different values
of uniform resistivity $\eta_u$ in the models and other parameters, and
different type of schemes used to evolve the numerical models.  After the
injection of turbulence, as we shown in Figures~\ref{fig:ueta_dep} and
\ref{fig:aeta_dep}, during the stage of the presence of turbulence, the
reconnection rate is insensitive to the value of resistivity.  In
Figure~\ref{fig:deps} we see that both relations, using the old and new methods,
show the same dependence, even though they were fitted to two different sets of
models.  This means that the reconnection rate dependence on the injected power
is not sensitive to the strength of the guide field $B_z$, the injection scale
or the value of uniform resistivity and the numerical diffusivity of the method
used in solving the MHD equations.

In the right plot of Figure~\ref{fig:deps} we see a comparison of the old and
new measures of $V_{rec}$ as a function of the injection scale $l_{inj}$.
Similar to the dependence on the power of turbulence, this dependence does not
change and in both cases is $V_{rec} \sim l_{inj}^{3/4}$.  This confirms again
that the dependence of the reconnection rate $V_{rec}$ on the properties of the
turbulence do not change its character.  The only difference seen in both
relations is the amplitude of the reconnection rate.  This difference comes from
the fact, that in the old measure models we used weaker turbulence
($P_{inj}=0.5$ vs. $P_{inj}=1.0$).  Another difference could result from the
fact that we include more terms in the elaboration of the new reconnection rate,
such as the boundary shear, the boundary resistive terms, the time derivative of
the absolute value of $|B_x|$.  These terms can have non-zero time averages,
although we expect them to have at least stationary values.

The results showed here confirms that the new and old methods of estimating the
reconnection rate reveal the same dependence on the power of turbulence and
injection scale.  It signifies that the old measure $\langle V_{in} / V_A
\rangle$ and new one $V_{rec}$ are both adequate in determining the speed of
reconnection and more importantly, that they measure precisely the reconnection
rate in our system excluding all other processes, which could contribute to the
plasma flow, but not to the reconnection process itself.  It is also important
that these results are independent of the numerical scheme or the initial
conditions.

\subsection{Non-Reconnection Case}
\label{sec:non-reconn}

The results presented so far indicate an agreement with the LV99 model.  We see
that the presence of turbulence near current sheets enhances reconnection
significantly.  However, since our testing is limited to numerical models, which
have explicitly defined boundaries, our results may be sensitive to interactions
between turbulence and the boundary conditions implemented in the numerical
model.  These interactions could stimulate a flow through the box resulting in a
false, non-zero reconnection rate without the presence of reconnection itself.
Moreover, the measures of reconnection rate that we applied here may indicate
the existence of other processes contributing to the plasma flow and related to
the turbulence itself rather than reconnection.  In order to test the
reliability of our numerical model and the methods of reconnection rate
estimation we present below a comparison of two models with the same set of
parameters, but with different initial configurations of magnetic field, namely,
one with the antiparallel $B_x$ as used in all numerical models presented here
and another with the uniform $B_x$ where is no laminar reconnection before we
start injecting turbulence.  From the latter model one would obviously expect
the lack of reconnection and zero reconnection rate $V_{rec}$.

\begin{figure}
\center
\includegraphics[width=0.45\textwidth]{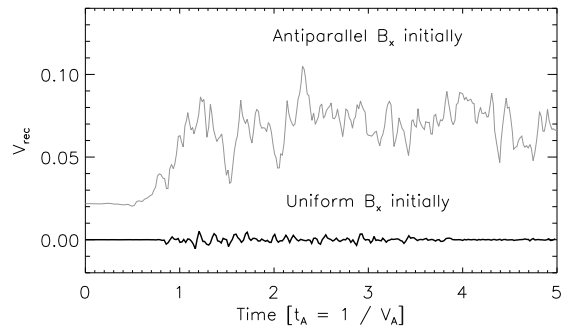}
\caption{Comparison of the reconnection rate $V_{rec}$ for two models with the
same set of parameters, $P_{inj}=1.0$, $k_{f}=8$, $B_{0z}=0.1$,
$\eta_{u}=10^{-3}$, but different initial configurations of magnetic field:
reversed $B_x$ (grey) and uniform $B_x$ (black).  Both models were run with the
same resolution 128x256x128. \label{fig:rec_vs_uni}}
\end{figure}

In Figure~\ref{fig:rec_vs_uni} we show the evolution of reconnection rate
$V_{rec}$ for two models with the same set of parameters, $P_{inj}=1.0$,
$k_{inj}=8$, $B_{0z}=0.1$, and $\eta_u=10^{-3}$.  Models were started from
different initial configurations of magnetic field.  The initial configuration
of $\vc{B}$ in the first model is taken from the steady-state laminar
reconnection of Sweet-Parker type.  In the second model we set uniform magnetic
field $B_{x}=1.0$ initially.  Starting the evolution from these conditions we
inject turbulence in the same way in both models, increasing its strength
gradually through the period of one Alfv\'enic time unit.  Then, turbulence is
injected with the power of constant value of $P_{int}=1.0$.

Figure~\ref{fig:rec_vs_uni} shows a clear difference between the reconnection
rates in both models, i.e., in the case of initial Sweet-Parker configuration
further evolution of $V_{rec}$ is similar to the results presented in the
previous sections.  The reconnection rate grows from a value corresponding to
the Sweet-Parker rate of about 0.02 to higher values when it saturates at time
$t \approx 1.4$ at the level of $V_{rec} \approx 0.07$.  Comparing this to the
evolution of $V_{rec}$ in the model with the initial uniform field we see only
small fluctuations of reconnection rate around zero, what indicates the absence
of global reconnection in the system.

Results presented in this subsection indicate essentially a lack of reconnection
in the case of uniform initial field, what means that the enhancement of
reconnection rate in the LV99 model is resulting from an influence of turbulence
on the laminar reconnection process and not a result of turbulence itself.

\section{Discussion}
\label{sec:discussion}

%
%
\subsection{Relation to Other Studies of Turbulent Reconnection}

Our simulations presented in this paper test the LV99 model of reconnection.  At
the same time, the notion that magnetic field stochasticity might affect current
sheet structures is not unprecedented.   For instance, some papers have
concentrated on the effects that turbulence induces on the microphysical level.
In particular, \cite{speiser70} showed that in collisionless plasmas the
electron collision time should be replaced with the electron retention time in
the current sheet.  Also \cite{jacobson84} proposed that the current diffusivity
should be modified to include the diffusion of electrons across the mean field
due to small scale stochasticity.  Our simulations are performed at the MHD
level and do not take these effects into account\footnote{These effects will
usually be small compared to effect of a broad outflow zone containing both
plasma and ejected shared magnetic flux.  Moreover, while both of these effects
will affect reconnection rates, they are not sufficient to produce reconnection
speeds comparable to the Alfv\'en speed in most astrophysical environments.}.

"Hyper-resistivity" \citep{strauss86,bhattacharjee86,hameiri87,diamond03} is a
more subtle attempt to derive fast reconnection from turbulence within the
context of mean-field resistive MHD. The form of the parallel electric field can
be derived from magnetic helicity conservation.  Integrating by parts one
obtains a term which looks like an effective resistivity proportional to the
magnetic helicity current.  There are several assumptions implicit in this
derivation.  The most important objection to this approach is that by adopting a
mean-field approximation, one is already assuming some sort of small-scale
smearing effect, equivalent to fast reconnection.  Furthermore, the integration
by parts involves assuming a large scale magnetic helicity flux through the
boundaries of the exact form required to drive fast reconnection. As we are not
aware of any attempts to test the idea of "hyper-resistivity" numerically, we
shall not discuss this concept further.

\cite{strauss88} examined the enhancement of reconnection through the effect of
tearing mode instabilities within current sheets.   However, the resulting
reconnection speed enhancement is roughly what one would expect based simply on
the broadening of the current sheets due to internal mixing.  \cite{shibata01}
extended the concept suggesting that tearing may result in fractal reconnection
taking place on very small scales.  \cite{waelbroeck89} considered not the
tearing mode, but the resistive kink mode to accelerate reconnection.

We do not expect either tearing or kink modes to allow us to evade the
constraints on the global plasma flow that leads to slow reconnection
speeds\footnote{Sometimes the growth rate of the resistive modes is erroneously
identified with the reconnection speeds.  This does not account for the global
outflow of the matter constraint that should be satisfied for reconnection of
magnetic field.  When this is done, the results are different.  For instance,
LV99 equated the shearing rates arising from the global outflow and the rate of
the tearing mode growth at the scale of the reconnection region.  They obtained
the reconnection speed $V_A R_m^{3/10}$, which was faster than the Sweet-Parker
rate, but incredibly slow for most astrophysical systems.}.  The effects of
tearing modes may be important for the initiation of turbulence in the situation
when the initial configuration is laminar
\cite[see][]{dahlburg92,dahlburg94,dahlburg97,ferraro04,lazarian08}.  We plan to
study this effect elsewhere.  However, once turbulence is initiated reconnection
should proceed independently of the tearing mode.  Straus' idea is closely
related to recent attempts to explain the existence of the observed thick
reconnection region in the context of numerical models of collisionless
reconnection \citep{ciaravella08} and predictions of particle acceleration
within these regions \citep{drake06a}.  Here we do not appeal to collisionless
effects, but demonstrate both thick reconnection regions and fast reconnection
in the presence of turbulence.

The closest study to ours was done by \cite{matthaeus85} \cite[see
also][]{matthaeus86}.  The authors studied 2D magnetic reconnection in the
presence of external turbulence.  An enhancement of the reconnection rate was
reported, but the numerical setup precluded the calculation of a long term
average reconnection rate.  A more recent study along the approach in
\cite{matthaeus85} is one in \cite{watson07}, where the effects of small scale
turbulence on 2D reconnection were studied and no significant effects of
turbulence on reconnection were reported for the setup chosen by the authors.
The major differences from the present study stem from the fact that we test a
3D model of reconnection, as the LV99 depends on effects, e.g.  field wandering,
that happen only in 3D.  Thus within the present study we do not address the
question of what is happening to 2D reconnection in the presence of turbulence,
appealing instead to the 3D nature of astrophysical reality (see more in
\S\ref{ssec:ionized_gas}).

Finally, we would like to stress that our calculations do not attempt to
validate the concept of turbulent diffusivity frequently employed in the mean
field dynamo \cite[see][]{moffat78,blackman08}.  This concept appeals to the
alleged ability of hydrodynamic motions to mix magnetic lines on very small
scales, as if the magnetic field were a passive scalar.  Naturally, this
assumption is wrong for any astrophysical field of dynamical importance.  In our
calculations, on the contrary, the magnetic field energy exceeds the kinetic
energy and thus, the turbulence is weak.  Instead of mixing magnetic lines, as
in the case of "magnetic turbulent diffusivity" creating reversals of magnetic
fields on the smallest scales, our driving is only able to induce wandering of
the magnetic field lines within a well-defined mean direction of the magnetic
field.

In general, it is important not to confuse stochastic reconnection with the {\it
ill motivated} concept of turbulent diffusivity.  For instance, \cite{kim01}
addressed the problem of stochastic reconnection by calculating the turbulent
diffusion rate for magnetic flux inside a current sheet.  They obtained similar
turbulent diffusion rates for both two dimensional and reduced three dimensional
MHD (2D with the Z-components of velocity and magnetic field).  In both cases
the presence of turbulence had a negligible effect on the flux transport.  The
authors pointed out that this would prevent the anomalous transport of magnetic
flux within the current sheet and concluded that both 2D and 3D stochastic
reconnection proceed at the Sweet-Parker rate, even if the individual small
scale reconnection events happen quickly, which apparently contradicts our
calculations in the paper.

We would like to stress that the turbulent diffusion rates within the current
sheet are irrelevant to the process of stochastic reconnection (see discussion
in LVC04).  The basic claim in LV99 is that the realistic magnetic field
topologies allow multiple connections between the current sheet and the exterior
environment, which would persist even if the stochastic magnetic field lines are
stationary ("frozen in time") before reconnection happens.  This leads to global
outflow constraints which are weak and do not depend on the properties of the
current sheet.  In particular, the analysis in LV99 assumed that the current
sheet thickness is determined purely by Ohmic dissipation and that turbulent
diffusion of the magnetic field is negligible inside and outside of the current
sheet.

\subsection{2D versus 3D reconnection}

The fact that our study is in 3D is essential, as the LV99 model is
intrinsically three dimensional.  The general picture is of tangled field lines,
with reconnection taking place via a series of "Y-points" or modified
Sweet-Parker sheets distributed in some fractal way throughout the turbulence.
A large scale Sweet-Parker sheet will be replaced by a more fractured surface,
but the current sheets will occupy a vanishingly small fraction of the total
volume and the field reversal will remain relatively well localized.  The model
predicts that the reconnection speed would be approximately equal to the strong
turbulent velocity with a modest dependence on the ratio of the eddy length to
the current sheet length.  There should be no dependence on resistivity.  The
major results contained in our figures showing the dependence of the
reconnection speed on resistivity, input power and input scale agree with the
quantitative predictions of the LV99 model.  We are not aware of any competing
models to compare our simulations with.

In the absence of quantitative model to be tested, simulations aimed at studying
reconnection speed have been done in 2D.  This allowed to achieve higher
resolutions (compared to that contemporary available in 3D) but substantially
constrained magnetic field dynamics.  For instance, the 2D study in
\cite{matthaeus86} stresses the importance of turbulence for modifying the
character of magnetic reconnection and specifies heating and transport as the
effect of particular significance, as well as formation of Petscheck-type
"X-points" in 2D turbulence.

As we mentioned earlier, LV99 appeals to "Y-points" and heating does play a role
in it. Our simulations are also isothermal.  \cite{kim01} showed that the
transport of magnetic flux is not enhanced to the reconnection zone.  At the
same time, field  wandering is the essential feature of 3D reconnection, which
is absent in the 2D case.  In our forthcoming paper \citep{kulpa09} we shall
demonstrate the difference of 2D and 3D reconnection using direct numerical
simulations\footnote{To our best knowledge no quantitative predictions are
currently available for the 2D reconnection.}.

\subsection{Reconnection in collisionless and collisional plasma}

The LV99 model of reconnection is applicable to the collisional medium.  For
instance, it is applicable to the interstellar medium (known to be both
turbulent and magnetized), for which one cannot apply the Hall-MHD reconnection
\citep{yamada07}.  This is a relieving news for the interstellar medium, star
formation and Solar simulations \citep{ostriker01,mckee07}, as numerical
reconnection in MHD codes is fast.  Note, that the requirement of being {\it
collisionless} in terms of magnetic reconnection is different from the usual use
of the term in astrophysics.  For instance, for the diffuse ISM the
collisionallity parameter is $\omega_c \tau_e$, where $\omega_c$ is cyclotron
frequency of electrons and $\tau_e$ is the collisional time for an electron.
However, for Hall-MHD reconnection to be applicable the criterion is different.
It is required that the Sweet-Parker current sheet $\delta_{SP}$ width be less
than the ion inertial length $d_i$.  Thus the "reconnection criterion for media
to be collisionless" is $(L/d_i)^{1/2}/(\omega_c \tau_e)<1$, which presents a
much severe constraint on the possible rate of collisions.  As a result,
magnetic reconnection happens mediated by the Hall-MHD only if the extend of the
contact region $L$ (see Fig.~\ref{fig:lv99model}) does not exceed $10^{12}$~cm.
Magnetic fields in the ISM should interact over much larger scales.

Astrophysical environments also contain media to which the Hall-MHD reconnection
is applicable, e.g. Solar corona, interplanetary medium.  Is LV99 model
applicable to such environments?  The answer is positive if the level of
turbulence is high enough.  Indeed, the reconnection on microscales can happen
fast, i.e. in the Hall-MHD fashion.  However, this may not change the global
reconnection rate.  Indeed, the LV99 model shows that even with relatively slow
Sweet-Parker reconnection at microscales the global reconnection is limited not
by Ohmic resistivity, but the rate of magnetic field wondering.  In fact,
anomalous resistivity that we use in some of our numerical runs is a proxy for
the Hall-MHD effect on reconnection and we do not see any dependence of the
reconnection rate on the anomalous resistivity (see Fig.~\ref{fig:aeta_dep}).
We believe that the Hall-MHD {\it local} reconnection of magnetic fields is
taking place interplanetary medium, which is being tested by local {\em in-situ}
measurements, while the {\it global} reconnection rates are determined by
magnetic field wandering as prescribed in LV99.  Future dedicated numerical
experiments and space measurements should test this idea.

\subsection{Reconnection in Partially Ionized Gas}
\label{ssec:ionized_gas}

The LV99 model of reconnection deals with either fully ionized plasma or with
plasmas where the neutrals constitute less than 10\% of species.  For higher
percentage of the neutrals, the MHD-type turbulent cascade is truncated by the
viscosity of neutrals.  How does reconnection happen in a partially ionized gas?

A partially ionized plasma fills a substantial volume within our galaxy and the
earlier stages of star formation take place in a largely neutral medium.  This
motivates our study of the effect of neutrals on reconnection.  The role of
ion-neutral collisions is not trivial.  On one hand, they may truncate the
turbulent cascade, reducing the small scale stochasticity and decreasing the
reconnection speed.  On the other hand, the ability of neutrals to diffuse
perpendicular to the magnetic lines allows for a broader particle outflow and
enhances reconnection rates.

Reconnection in partially ionized gases before the introduction of the LV99
model looked hopelessly slow.  For instance, in \citet[][henceforth
VL99]{vishniac99} we studied the diffusion of neutrals far from the reconnection
zone assuming the anti-parallel magnetic lines.  The ambipolar reconnection
rates obtained in VL99, although large compared to the Sweet-Parker model, are
insufficient either for fast dynamo models or for the ejection of magnetic flux
prior to star formation.  In fact, the increase in the reconnection speed
stemmed entirely from the compression of ions in the current sheet, with the
consequent enhancement of both recombination and ohmic dissipation. This effect
is small unless the reconnecting magnetic field lines are almost exactly
anti-parallel \cite[VL99, see also][]{heitsch03a,heitsch03b}. Any dynamically
significant shared field component will prevent noticeable plasma compression in
the current sheet, and lead to speeds practically indistinguishable from the
standard Sweet-Parker result.  Since generic reconnection regions will have a
shared field component of the same order as the reversing component, the
implication is that reconnection and ambipolar diffusion do not change
reconnection speeds significantly.

LCV04 presented a model of turbulence in a partially ionized gas.  This model
agrees well with numerical simulations available as the limiting case which can
be characterized by one fluid with a high Prandtl number
\cite[see][]{cho02b,cho03}.  Using this model LVC04 described field wandering,
which is the core of the LV99 model of reconnection.  They showed that the
magnetic reconnection proceeds fast, both in the diffuse interstellar and
molecular cloud partially ionized gas.

In the paper above we employed one-fluid code and therefore we could not test
the predictions in LVC04.  However, LVC04 model at its core is based on the
ideas presented in LV99.  Therefore, our successful testing of LV99 model
provides us with optimism in relation to the LVC04 model.  The latter should be
tested with a two fluid code capitalizing on our present experience with the
boundary conditions, turbulent driving and the ways of measuring of reconnection
rate obtained in this paper.  Needless to say, the numerical confirmation of
fast reconnection in partially ionized gas would open wide prospects for
quantifying processes taking place in molecular clouds, including the enigmatic
processes of star formation.  For instance, \cite{shu06} showed that magnetic
field is removed from star-forming cores faster than allowed by the standard
ambipolar diffusion scenario \cite[see][]{tassis05a,tassis05b}.  \cite{shu07}
proposed a mechanism that required efficient reconnection of magnetic loops
based on the ``hyper-resistivity'' concept.  As we discussed earlier, this
concept is not self-consistent and is therefore problematic.  Fast magnetic
reconnection in the partially ionized gas could serve a similar purpose.

\subsection{The Current Simulations and Remaining Problems}

In this paper we have performed extensive numerical testing of the LV99 model of
magnetic reconnection.  A advantage of this model is that it is very generic and
does not appeal to any particular plasma effects and/or initial boundary
conditions.  In many respects, it is a natural generalization of the
Sweet-Parker model to the case when turbulence is present.  It does not require
the current sheet to open up, as is required in the Petscheck reconnection
model.  The only requirement is the presence of field line wandering, which is a
well documented aspect of magnetic turbulence (see LVC04 and ref.  therein).

Our results indicate that the reconnection speed increases in the presence of
weak turbulence in a manner consistent with the LV99 predictions.  In
particular, the reconnection speed $V_{rec}$ (Eq.~\ref{eq:constraint}) shows a
strong dependence on the characteristics of the turbulence, such as the rate of
energy dissipation and the scale of injection.  It is very important that we
observe no explicit dependence on the fluid resistivity, which is the
requirement for fast magnetic reconnection in astrophysical environments.

To make progress in simulating the weakly stochastic reconnection, we had to
develop a new numerical set up, which included both driving turbulence over a
part of the box, inflow conditions for the incoming flux and outflow conditions
for the reconnected flux.  This enabled us to study both the transitional
regimes of reconnection and the steady state one.  In comparison, simulations in
the periodic box would not be instructive, as stochastic reconnection is dynamic
and does not reach a stationary state in one crossing time.

In addition, we have developed a new way of measuring the reconnected flux (see
\S~\ref{sec:rec_rate}).  This was required as worries were associated with the
accuracy of the "intuitive measure", which is the rate with which the
unreconnected flux enters the computational box at the inflow boundaries.
Potentially, the "intuitive measure" could overestimate the reconnected flux if
the unreconnected magnetic lines escape through the leaky outflow boundaries of
the computation box.  The new measure accounts for such a loss of magnetic flux.

To our satisfaction, we found that the new and old measures provide consistent
results both in the case of no turbulence and with turbulence, which means that
the effect of the outflow of unreconnected field was subdominant for the
situations when the magnetic fluxes were intersecting at an appreciable angle.
For the limiting case, however, when the large scale reconnection was absent,
i.e. when the corresponding angle between the two fluxes was zero and without
turbulence magnetic fields were all parallel, the "intuitive measure" was a
non-zero value, while the new measure of reconnection gave zero reconnection
rate.  This shows the new measure is more reliable.

Our results confirm that the reconnection of magnetic field in the presence of
turbulence is independent of Ohmic resistivity, allowing for fast reconnection.
In addition, we showed that the reconnection rate does not depend on the
anomalous resistivity, which proves that the LV99 model of reconnection is also
applicable to collisionless plasmas and may dominate reconnection in such
systems if the level of turbulence is efficiently high.  Note, that according to
\cite{lazarian08} turbulence in a system may be successfully generated by the
release of magnetic energy through reconnection. Thus, we anticipate that the
initial reconnection in a collisionless plasma may be induced through the Hall
reconnection, but later the resulting MHD turbulence will determine the
reconnection rate.

Apart from the prediction in LV99 that the reconnection rate should be
independent of resistivity, we have also tested the quantitative scalings of the
reconnection rates presented in LV99.  In particular, we tested the dependence
of the reconnection rate on the turbulent injection power.  Our numerical
results confirm the LV99 prediction $V_{rec} \sim V_l^2$ which, as we explained
earlier, translates into $V_{rec} \sim P_{inj}^{1/2}$ (see
Eq.~\ref{eq:scaling}).

We do not expect to see the dependencies of the reconnection rate on viscosity
until the Prandtl number is less than unity since the cascade can proceed to
scales smaller than the current sheet thickness affecting the rate of local
reconnection.  This is the case we are studying here.  For large Prandtl numbers
the reconnection may be suppressed as magnetic field becomes more laminar at
sufficiently large scales.  However, two main factors of the LV99 model
contributing to the enhancement of the reconnection rate in the presence of
turbulence are that the turbulence allows for multiple reconnection events
acting simultaneously and secondly, the turbulence should broaden the outflow
region to the widths corresponding to the scales much larger than the
dissipation scale and comparable to the injection scale allowing to remove more
reconnected flux.  While the first factor could be affected by the large
viscosity, the second is still valid if the injection scale is much larger than
the dissipation scale.  The case when the effective Prandtl numbers are larger
than unity due to the presence of neutrals is discussed in \cite{lazarian04}.
We plan to study the dependence on the Prandtl number in our next publication.

The LV99 model predicts that the reconnection speed should grow with the
injection scale as $V_{rec} \sim l_{inj}$.  However, some of our numerical tests
show a weaker dependence, $V_{rec} \sim l_{inj}^{3/4}$.  The origin of this
discrepancy is unclear.  Our model of how weak isotropic turbulence evolves
cascades into strong turbulence may be too naive.  In particular, the
approximation that the parallel wavelength is unchanged during this process may
be simplistic.  In addition, we may simply not have sufficient dynamic range
between the dimensions of the subvolume where turbulence is excited and the
thickness of the Sweet-Parker layer.  Corrections at either end could result in
the appearance of a shallower dependence on the driving scale.  The true nature
of this discrepancy should be explored in future work.

Even though these numerical simulations allow us to study reconnection in the
presence of turbulence for a limited range of magnetic Reynolds numbers (in this
paper $R_m < 10^3$), the results provide good testing of the relations derived
by LV99.  The strong dependence of $V_{rec}$ on the injection scale, when scaled
to the real conditions of the interstellar medium, shows a dramatic enhancement
of the reconnection speed, even in the situation of an almost perfectly frozen
magnetic field in the medium.  This allows for fast reconnection with the
characteristic time comparable to one Alfv\'en time unit.

\subsection{Implications of the LV99 Model}

Reconnection is one of the most fundamental processes involving magnetic fields
in conducting fluids or plasmas.  Therefore, the identification of a robust
process responsible for reconnection has many astrophysically important
consequences.  Below we shall mention a few selected implications of the
successful validation of the LV99 model.

Following \cite{zweibel09} we note that solar flares inspired much of the
earlier research on reconnection \cite[see][]{pneuman81,bastian98,hudson08}.  As
the plasma involved is substantially rarefied, the restrictive conditions for
the collisionless reconnection are satisfied in this particular environment.
\cite{cassak05} stated that bistable Hall reconnection can be important in this
case.  Stochastic reconnection, as we discuss below, provides an alternative
explanation.

Indeed, an important prediction of the LV99 model is related to the {\it
reconnection instability} that arises in the situation when the initial
structure of the flux prior to reconnection is laminar.  Reconnection at the
Sweet-Parker rate is negligible.  This allows magnetic flux to accumulate.
However, when the degree of stochasticity exceeds a threshold value, the
reconnection itself should excite more turbulence, creating a positive feedback
resulting in a flare \cite[see][]{lazarian08}.  The instability is a generic
property of laminar field reconnection in both collisionless and collisional
environments.  Referring to the Sun, one may speculate that the difference
between gradual and eruptive flares arises from the original state of magnetic
field prior to the flare.  In the case when the magnetic field is sufficiently
turbulent the accumulation of magnetic flux does not happen and the flare is
gradual.  Similarly, the observed spatial spread of energy release during solar
flares may be due to the spread of the region of turbulent fields once
reconnection is initiated at one place.  Further research is necessary for
testing these ideas.

In the Sweet-Parker model reconnection can accelerate charged particles, e.g.
due to the electric field in the reconnection region \cite[see][]{litvinenko03}.
 However, the speed of Sweet-Parker reconnection is negligible for most
astrophysical environments, thus the transfer of energy from the magnetic field
to particles is absolutely negligible, if reconnection follows the Sweet-Parker
predictions.

It is interesting to notice that the first-order Fermi acceleration process is
intrinsic to the LV99 model of reconnection.  Consider a particle entrained on a
reconnected magnetic field line (see Fig.\ref{fig:lv99model}).  This particle
may bounce back and forth between magnetic mirrors formed by oppositely directed
magnetic fluxes moving towards each other with the velocity $V_R$.  Each bounce
will increase the energy of a particle in a way consistent with the requirements
of the first-order Fermi process\footnote{Another way of understanding the
acceleration of energetic particles in the reconnection process above is to take
into account that the length of magnetic field lines is decreasing during
reconnection.  As a result, the physical volume of the energetic particles
entrained on the field lines is shrinking.  Thus, due to Louiville theorem,
their momentum should increase to preserve the constancy of the phase volume.}
\citep{dalpino01,dalpino03,dalpino05,lazarian06}.  This is in contrast to the
second-order Fermi acceleration that is frequently discussed in terms of
accelerating particles by turbulence generated by reconnection \citep{larosa06}.

An interesting property of this acceleration mechanism is that it is potentially
testable observationally, since the resulting spectrum of accelerated particles
is different from that arising from a shock.  \cite{dalpino01,dalpino05} used
this mechanism of particle acceleration to explain the synchrotron power-law
spectrum arising from the flares of the microquasar GRS 1915+105.  The mechanism
is similar to the acceleration mechanism that was discussed later by
\cite{drake06b}.  \cite{drake06b} considered the acceleration of electrons and,
similarly, to the \cite{matthaeus84}, assumed that the acceleration happens
within 2D contracting loops.

To enable sustainable dynamo action and, for example, generate a galactic
magnetic field, it is necessary to reconnect and rearrange magnetic flux on a
scale similar to a galactic disc thickness within roughly a galactic eddy
turnover time ($\sim 10^8$~years).  This implies that reconnection must occur at
a substantial fraction of the Alfv\'en velocity.  The preceding arguments
indicate that such reconnection velocities should be attainable if we allow for
a realistic magnetic field structure, one that includes both random and regular
fields \cite[see][e.g.]{hanasz04}.  This does solve one part of the problem of
dynamo.  The other part is related to magnetic helicity conservation
\cite[see][]{vishniac03}.

Interestingly enough, while earlier on we criticized the concept of turbulent
diffusivity as ill founded, we should mention that the robust reconnection of
turbulent fields can allow parcels of magnetized fluid to move being less
constrained by the surrounding magnetic fields.  This allows matter and heat to
transfer with the rates determined by a simple turbulent diffusion coefficient
$\sim v_{turb} l_{turb}$ which for typical parameters of plasma in galaxy
clusters exceed the rates of thermal diffusion of electrons \cite{lazarian06}.
An important distinction remains however.  Reconnection happens in thin current
sheets, involving a negligible fraction of the magnetic field.  Unlike a large
effective resistivity, fast reconnection cannot violate magnetic helicity
conservation.

Finally, LV99 showed that fast reconnection of stochastic magnetic field makes
the models of strong MHD turbulence self-consistent.  Indeed, critical balance
in the GS95 model requires the existence of eddy-type motions perpendicular to
the magnetic field.  In the absence of reconnection this would result in
unresolved knots that should drain energy from the cascade.  The estimates in
LV99 showed that the rates of reconnection predicted by the model are sufficient
to resolve magnetic knots within one period.

\subsection{Final Remarks}

In this paper we have considered reconnection for weak MHD turbulence.  What
would happen if the turbulence is strong, i.e. when the injection of turbulent
energy happens with superAlfv\'enic velocities?  In this situation we expect the
turbulent cascade to proceed initially in a way similar to ordinary
hydrodynamics with strong hydrodynamic motions easily bending magnetic field
lines.  However, at some small scale we expect the magnetic field to be
dynamically important and resist bending.  Our reconnection results should be
applicable to turbulence starting at this scale.

The successful numerical testing of the turbulent reconnection model presented
in this paper appeals for more studies in this direction.  It is important to
stress that, unlike brute force simulations, we were able to test the LV99 model
of three dimensional reconnection giving support to the idea that we could
reliably extend our results to much higher Lundquist numbers.  The problems for
future studies cover the importance of compressibility in this type of
reconnection, since a substantial fraction of observed interstellar medium is
supersonic.  Another interesting question is the ability of turbulent
reconnection to self-sustaining or to generate turbulence by
itself\footnote{While, as we argued earlier, turbulence is a natural state of
most of the astrophysical fluids, it is important to know whether the
reconnection is determined by the preexisting level of turbulence, or it can get
self-accelerated.}.  These are problems we hope to address in the future.

\section{Summary}
\label{sec:summary}

In this article we put to the test the LV99 model of reconnection by
investigating the influence of weak turbulence on the reconnection process using
3D numerical simulations.  The turbulence is weak in the sense that the
injection velocities are less than the Alfv\'en speed and the magnetic field
direction is only weakly perturbed by turbulence.  It is strong in the sense of
producing a strongly nonlinear cascade of energy.  In our study we experimented
with different boundary conditions and with different ways of measuring the
reconnection speed.  We analyzed the dependence of the reconnection speed on the
turbulence injection power, on the injection scale, as well as on Ohmic and
anomalous resistivity.  We found that:

\begin{itemize}

\item Turbulence in 3D drastically changes the topology of the magnetic field
near the interface of oppositely directed magnetic field lines.  These changes
include the fragmentation of the current sheet, favoring multiple simultaneous
reconnection events, as well as a substantial increase in the thickness of the
outflow of reconnected magnetic flux and matter.

\item The intuitive measure of reconnection defined as the inflow of magnetic
field velocity corresponds well, in the condition of stationary reconnection, to
a more rigorously defined reconnection measure that we introduced.  The two
measures show the same dependencies of $V_{rec}$ on the power $P_{inj}$ and
injection scale $l_{inj}$ of the turbulence, regarding the different scheme used
to solve MHD equations and different sets of initial parameters.

\item The reconnection rate is determined by the thickness of the outflow
region.  For large scale turbulence, the reconnection rate depends on the
amplitude of fluctuations and injection scale as $V_{rec} \sim P_{inj}^{1/2}
\sim V_l^2$ which corresponds to LV99 predictions.

\item The reconnection rate grows with the turbulence injection scale, which
qualitatively corresponds to the LV99 predictions.  However, in some cases the
rate of growth is better approximated by $V_{rec} \sim l_{inj}^{3/4}$ scaling
rather than $V_{rec}\sim l_{inj}$ predicted in LV99.  The difference may stem
from details of the initial weak cascade of energy, or from limitations in the
dynamic range available for study.

\item Reconnection in the presence of weak turbulence is not sensitive to Ohmic
resistivity, which corresponds to the LV99 prediction that the reconnection of
weakly stochastic magnetic fields are generically fast.

\item The introduction of anomalous resistivity does not change the rate of
reconnection of a weakly stochastic field either, which supports the assertion
in LV99 that in the presence of magnetic field stochasticity, which is generic
for most of astrophysical environments, plasma effects, e.g.  collisionless
effects, are irrelevant in determining reconnection speeds.

\item The strength of the guide field does not change reconnection rates for
similar rates of turbulent energy injection.  Thus fast reconnection is possible
for generic configurations when magnetic bundles intersect each other at
arbitrary angles.
\end{itemize}

\acknowledgments
The research of GK and AL is supported by the Center for Magnetic
Self-Organization in Laboratory and Astrophysical Plasmas and NSF Grants
ATM-06-48699 and AST-08-08118.  The work of ETV is supported by the National
Science and Engineering Research Council of Canada.  Part of this work was made
possible by the facilities of the Shared Hierarchical Academic Research
Computing Network (SHARCNET:www.sharcnet.ca).  This research also was supported
by the National Science Foundation project TG-AST090078 through TeraGrid
resources provided by Texas Advanced Computing Center
(TACC:www.tacc.utexas.edu).   We would like to thank Mel Goldstein, Paul Cassak,
Bill Matthaeus, Dmitri Uzdenski, Ellen Zweibel and Masaaki Yamada for fruitful
discussions.  Helpful suggestions of the anonymous referee are acknowledged.


\end{document}